\documentclass[aoas]{imsart}

\usepackage{amsmath}
\usepackage{enumerate}
\usepackage{natbib}
\usepackage{url} 
\usepackage[skip=5pt]{caption}
\usepackage[normalem]{ulem}
\usepackage{amsthm}
\usepackage{natbib}
\usepackage{xr}
\usepackage[colorlinks,citecolor=blue,urlcolor=blue,filecolor=blue,backref=page]{hyperref}
\usepackage{graphicx}
\usepackage{algorithm}
\usepackage{algpseudocode}
\usepackage{algorithmicx}
\usepackage{amssymb}
\usepackage{amsbsy}
\usepackage{float}
\usepackage{comment}
\usepackage{mathtools}
\usepackage{color,soul}
\usepackage{blkarray}
\usepackage[dvipsnames]{xcolor}
\usepackage{color, colortbl}
\usepackage{mathbbol}
\usepackage[T1]{fontenc}
\usepackage{mathrsfs}
\usepackage{hhline}
\usepackage{caption,subcaption}
\usepackage{afterpage}
\usepackage{multirow}
\usepackage{adjustbox}
\usepackage{rotating}
\usepackage{tabularx}
\usepackage{booktabs}
\usepackage{anyfontsize}
\usepackage{appendix}
\usepackage[shortlabels]{enumitem}

\makeatletter
\newenvironment{notoprulealgorithm}
  {\let\old@fs@ruled\@fs@ruled
   \def\@fs@ruled{\hrule height 0pt \old@fs@ruled}%
   \@float{algorithm}}
  {\end@float}
\makeatother

\makeatletter
\newenvironment{noendalgorithm}
  {\let\oldendalg\endalgorithmic
   \def\endalgorithmic{\hrule height 0pt \oldendalg}%
   \@float{algorithm}}
  {\end@float}
\makeatother

\renewcommand{\tilde}[1]{\widetilde{#1}}

\newcommand{\sigmasq}{\sigma^2}

\usepackage{bbm} 

\newcommand{\bolds}[1]{\boldsymbol{#1}}

\newcommand{\calV}{{\cal V}}

\newcommand{\Cov}{\text{Cov}}

\newcommand{\ba}{\bolds{a}}
\newcommand{\bA}{\bolds{A}}

\newcommand{\bB}{\bolds{B}}

\newcommand{\bD}{\bolds{D}}

\newcommand{\bbf}{\bolds{f}}

\newcommand{\bh}{\bolds{h}}
\newcommand{\bH}{\bolds{H}}

\newcommand{\bI}{\bolds{I}}

\newcommand{\bM}{\bolds{M}}

\newcommand{\bO}{\bolds{O}}
\newcommand{\bp}{\bolds{p}}

\newcommand{\bQ}{\bolds{Q}}

\newcommand{\bR}{\bolds{R}}

\newcommand{\bu}{\bolds{u}}

\newcommand{\bV}{\bolds{V}}

\newcommand{\bW}{\bolds{W}}
\newcommand{\bx}{\bolds{x}}

\newcommand{\by}{\bolds{y}}

\newcommand{\bz}{\bolds{z}}
\newcommand{\bZ}{\bolds{Z}}

\newcommand{\balpha}{\bolds{\alpha}}
\newcommand{\bbeta}{\bolds{\beta}}
\newcommand{\bdelta}{\bolds{\delta}}
\newcommand{\bet}{\bolds{\eta}}

\newcommand{\btheta}{\bolds{\theta}}
\newcommand{\bSigma}{\bolds{\Sigma}}
\newcommand{\bgamma}{\bolds{\gamma}}

\newcommand{\bmu}{\bolds{\mu}}

\newcommand{\bLambda}{\bolds{\Lambda}}

\newcommand{\bphi}{\bolds{\phi}}

\newcommand{\brho}{\bolds{\rho}}

\newcommand{\bPsi}{\bolds{\Psi}}
\newcommand{\bzeta}{\bolds{\zeta}}

\newcommand{\IdentityMat}{\bI}
\newcommand{\Norm}{\mathcal{N}}
\newcommand{\bzero}{\mathbf{0}}

\newcommand{\lrnd}{\left(}
\newcommand{\rrnd}{\right)}
\newcommand{\lsq}{\left[}
\newcommand{\rsq}{\right]}
\newcommand{\lcur}{\left\lbrace}
\newcommand{\rcur}{\right\rbrace}
 \newcommand{\given}{\,|\,}
\newcommand{\T}{\mathrm{\scriptscriptstyle T}}

\startlocaldefs

\theoremstyle{remark}

\endlocaldefs

\begin{document}

\begin{frontmatter}
\title{Detecting spatial health disparities using disease maps}
\runtitle{Spatial health disparities}

\begin{aug}
\author[A]{\fnms{Luca Aiello} \snm{}\ead[label=e1, mark]{l.aiello4@campus.unimib.it}}
\and
\author[D]{\fnms{Sudipto Banerjee} \snm{}\ead[label=e2]{sudipto@ucla.edu}}
\address[A]{Department of Economics, Management and Statistics, University of Milano-Bicocca.
\printead{e1}}

\address[D]{Department of Biostatistics, 
University of California, Los Angeles, 
\printead{e2}}

\end{aug}

\begin{abstract}

Epidemiologists commonly use regional aggregates of health outcomes to map mortality or incidence rates and identify geographic disparities. However, to detect health disparities across regions, it is necessary to identify ``difference boundaries” that separate neighboring regions with significantly different spatial effects. This can be particularly challenging when dealing with multiple outcomes for each unit and accounting for dependence among diseases and across areal units. In this study, we address the issue of multivariate difference boundary detection for correlated diseases by formulating the problem in terms of Bayesian pairwise multiple comparisons by extending it through the introduction of adjacency modeling and disease graph dependencies. Specifically, we seek the posterior probabilities of neighboring spatial effects being different. To accomplish this, we adopt a class of multivariate areally referenced Dirichlet process models that accommodate spatial and interdisease dependence by endowing the spatial random effects with a discrete probability law. Our method is evaluated through simulation studies and applied to detect difference boundaries for multiple cancers using data from the Surveillance, Epidemiology, and End Results Program of the National Cancer Institute. 

\end{abstract}

\begin{keyword}
\kwd{Bayesian hierarchical models}
\kwd{Difference Boundary Analysis}
\kwd{Disease mapping}
\kwd{False discovery rates}
\kwd{Spatial Dirichlet Processes}
\kwd{Spatial epidemiology}
\end{keyword}

\end{frontmatter}

\section{Introduction}\label{sec: intro}

Health disparities, or inequities, broadly refer to sections of the population being deprived of fair and equal opportunities to seek healthcare \citep[see, e.g., the recent text by][]{rao2023}. Spatial disparities in health manifest in the form of variations in health outcomes over geographic regions and are often visually represented using disease maps \citep{koch2005cartographies}. While the role of spatial data analysis in epidemiological investigations has been extensively documented \citep[see, e.g.,][and further references therein]{waller2004applied, waller2010handbook, lawson2013statistical, lawson2016handbook}, a specialized exercise of detecting ``boundaries'' or zones of abrupt change on disease maps is directly relevant to statistical detection of health disparities. In public health, such \emph{difference boundaries}, or \emph{wombling boundaries} \citep[so called after a seminal paper by][]{womble1951differential}, indicate significantly different disease mortality and incidence across regions, thereby improving decision-making for disease prevention and control, geographic allocation of health care resources, and so on \citep[see][for some algorithmic approaches with applications]{jacquez2003local, jacquez2003geographic, lu2005bayesian, li2011mining, ma2007bayesian, fitzpatrick2010}. Model-based approaches aiming for full probabilistic uncertainty quantification have also received much attention and include, but are not limited to, developments in \cite{lu2007bayesian, ma2010hierarchical, li2015bayesian, banerjee2012bayesian, hanson2015spatial, corpas-burgos2020serra} and \cite{gao2022spatial}.

This manuscript builds on the above developments in what may be described as still a fledgling area. We devise a probabilistic learning mechanism for adjacency boundaries on a map when such information may become available. An administrative or political boundary separating two adjacent regions need not delineate them in terms of the health outcomes measured there. In fact, in disease mapping and spatial smoothing of areal data the customary approach is to assume that neighboring regions are similar to each other. This assumption runs counter to detecting health disparities, where we seek out differences and not similarities. Therefore, modeling spatial disparities needs to balance underlying processes generating similarities based upon geographic proximity with processes generating significant differences between neighbors. We specifically pursue an effective manner of modeling the adjacency relations on a map. Our contribution is summarized in the context of the two broad themes. The first builds stochastic models for the adjacency matrices \citep{lu2007bayesian, ma2010hierarchical, lee2012boundary, corpas-burgos2020serra} that, in principle, can accommodate information from explanatory variables in ascertaining the presence of edges. Based upon the complexity of the resulting models and numerical difficulties in estimating such edges, a different approach forgoes introducing variables in the adjacency, but detects edges using multiple comparisons by estimating the spatial random effects and, subsequently, testing how many such pairs are significantly different. A Bayesian approach \citep{li2015bayesian, banerjee2012bayesian, gao2022spatial} holds appeal as the posterior distribution of spatial effects produce an exact Bayesian false discovery rate (FDR) \citep{muller2004optimal} without recourse to asymptotic assumptions that are inappropriate in areal settings.     

We offer a framework that melds the two themes. We build upon the recent developments in \cite{gao2022spatial} by embedding directed acyclic graphical autoregression (DAGAR) models \citep{datta2019spatial} within a hierarchical Bayesian nonparametric model. This nonparametric specification is critical as it introduces discrete probability masses on the spatial random effects and allows us to meaningfully calculate the posterior probabilities of two neighboring spatial effects being equal. This sets up boundary detection using Bayesian FDR. We further enrich the model by allowing the DAGAR adjacency matrix to learn from explanatory variables that may carry information regarding whether two neighboring regions are similar. Thus, each element of a binary adjacency matrix is modeled using an exponential threshold devised by \cite{lee2012boundary}. We develop joint models for multiple diseases (cancers), where the diseases are dependent and accommodate disease-specific learning of adjacency relationships. We allow difference boundaries to vary by health outcomes and explain the impact of explanatory variables on such boundaries. Here, we consider unstructured disease dependencies and those posited by conditional dependencies in graphical models.

The data analytic merits of our proposed approach are best appreciated in comparison to approaches that adhere to one, but not both, of the boundary detection themes discussed above. Methods that allow uncertainty through purely parametric adjacency relations (including regression models for adjacency matrices) within widely used conditional autoregression models often encounter numerical difficulties in convergence of iterative estimation algorithms, such as Markov chain Monte Carlo (MCMC). On the other hand, methods offering FDR based boundary detection without allowing adjacency learning from explanatory variables can lead to inflated detection rates of boundaries by failing to account for additional uncertainties in adjacency relations. We demonstrate this phenomenon in the cancer boundary detection rates reported recently by \cite{gao2022spatial}.      

The balance of our article proceeds as follows. Section~\ref{sec: data} describes the data we subsequently analyze. This data, extracted from the National Cancer Institute's (NCI) Surveillance Epidemiology and End Results (SEER) database records occurrences of four different cancer types across the counties of California together with explanatory variables for modeling outcomes and boundaries. Section~\ref{sec: models} develops our proposed multivariate Bayesian modeling framework using an areal stick-breaking process with a DAGAR spatial distribution with unknown boundaries as its baseline measure. Section~\ref{sec: simulations} presents a suite of simulation experiments to elicit insights into the performance of our proposed model using various metrics. Section~\ref{sec: cancer_analysis} applies spatial boundary detection to the data set described in Section~\ref{sec: data} and we conclude with some remarks in Section~\ref{sec: discussion}. Additional materials are provided in the Appendix.       

\section{Data}\label{sec: data}

We explore an areal data set extracted from the SEER$^*$Stat database using the SEER$^*$Stat statistical software \citep{seer}. The data records the occurrence of four potentially interrelated types of cancers: lung, esophageal, larynx, and colorectal. The occurences of these cancers are aggregated from January 2012 through December 2016 and presented as counts in each of the $58$ counties of California. Previous research has shown that lung and esophageal cancers share common risk factors \citep{agrawal2018risk} and metabolic mechanisms \citep{shi2004frequencies}. Furthermore, it has been found that lung cancer is one of the most frequent second primary cancers in patients with colon cancer \citep{kurishima2018lung}. Patients with laryngeal cancer are also at a high risk of developing second primary lung cancer \citep{akhtar2010second}. Hence, we seek to account for the dependence between the cancers due to non-spatial factors.

Let  $(y_{id})$ be the observed counts for cancer type $d$ ($d = 1, 2, 3, 4$) in county $i$ ($i = 1, 2, \dots , 58$). Our model, developed in Section \ref{sec: models}, accounts for the expected number of cases ($E_{id}$) by adjusting for age-sex demographics in each county. Specifically, we calculate the expected age-sex adjusted number of cases in county $i$ for cancer $d$ as $E_{id} = \sum_{k=1}^m c_{d}^k N_{i}^k$ \citep{jin2005generalized}, where $c_{d}^k = (\sum_{i=1}^{58} y_{id}^k) / (\sum_{i=1}^{58} N_{i}^k)$ is the age-sex specific incidence rate in age-sex group $k$ for cancer $d$ across all California counties; $y_{id}^k$ represents the incidence count in age-sex group $k$ of county $i$ for cancer $d$, and  $N_i^k$ represents the population in age-sex group $k$ of county $i$. The age groups have been determined based on $5$-year intervals up to $85$ years or older. These age intervals are as follows: less than $1$ year, $1$–$4$ years, $5$–$9$ years, $10$–$14$ years, and so on, up to $80$–$84$ years and $85+$ years. This results in a total of $m = 19 \times 2 = 38$ age-sex groups. 

\begin{figure}[t]
    \centering
    \includegraphics[width = 0.66\textwidth]{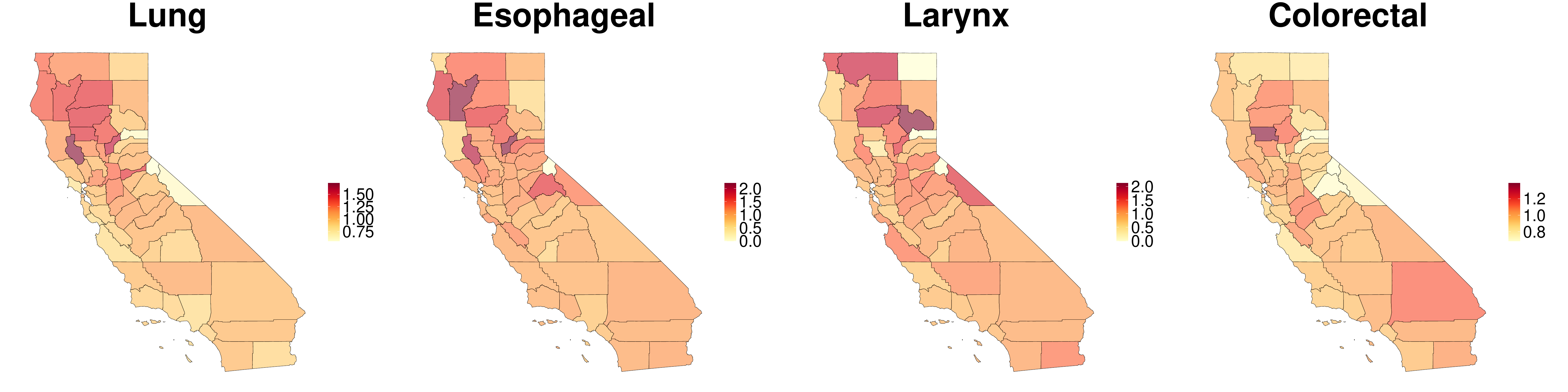}
    \caption{Maps of age-sex adjusted standardized incidence ratios (SIR) for lung, esophageal, larynx, and colorectal cancer in California, 2012–2016.}
    \label{fig:data}
\end{figure}

Figure \ref{fig:data} displays a map of California's counties, illustrating the age-sex adjusted standardized incidence ratios ($SIR_{id} = y_{id}/E_{id}$) for different types of cancer. The SIR values are classified into quintiles, and distinct thresholds are established for each level of SIRs. The map reveals a notable pattern, where regions exhibiting comparable SIRs tend to cluster together geographically, forming distinct groups. This clustering is evident in the group of regions with the highest incidence rates for each cancer.  The high SIR values for all of the cancer types are concentrated in northern counties, with the only exception being that of a high SIR for colorectal cancer in a southern county (San Bernardino). Additionally, in northern counties, we notice that several counties with high SIRs are situated next to areas with low SIRs. Our proposed model, which detects boundaries, provides a robust, adaptable, interpretable and clear analytical tool for identifying these geographical clusters.

\begin{figure}[t]
    \centering
    \includegraphics[width = 0.75\textwidth]{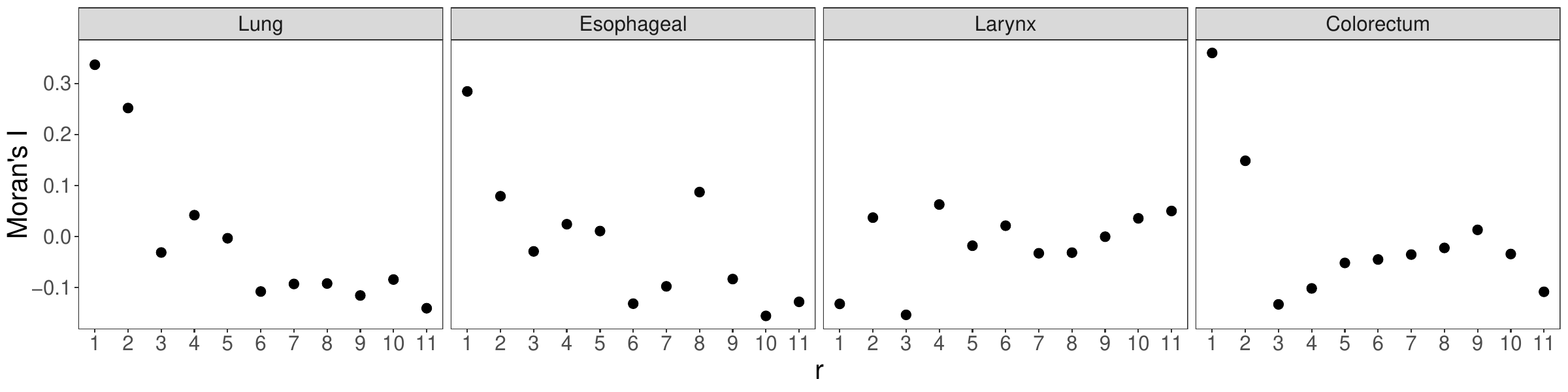}
    \caption{Moran’s I of $r$-th order neighbors for lung, esophageal, larynx, and colorectal cancer.}
    \label{fig:morans}
\end{figure}

We used Pearson's correlation coefficient as an exploratory tool to evaluate the relationships between various types of cancers. In a preliminary analysis, we treated the SIRs from different counties as independent samples. Our findings indicate a significant association in the incidence of lung cancer with each of esophageal (correlation coefficient 0.58), larynx (0.40), and colorectal (0.50) cancers. We also measured a correlation coefficient of $0.42$ between esophageal and larynx cancer. Moreover, to assess the spatial relationships of each cancer type, we employed the Moran's I statistic, which was computed based on the $r$-th order neighbors for each cancer. These calculations were then used to generate an areal correlogram, as described by \cite{banerjee2014hierarchical}. In our analysis, we defined distance intervals as $(0, d_1], (d_1, d_2], (d_2, d_3], \dots$, with the $r$-th order neighbors referring to units that fall within the distance range of $(d_r-1, d_r]$, and indicating that they are separated by a distance greater than $d_r-1$ but within a distance of $d_r$. The distance measure we used was the Euclidean distance derived from an Albers map projection of California. Figure~\ref{fig:morans} presents the results of our correlation analysis, which shows the spatial associations for lung, esophageal, colorectal, and larynx cancers. It is clear that as the order of neighbors ($r$) increases, the spatial associations diminish notably for lung, esophageal, and colorectal cancers. However, the observed pattern is less pronounced and clear for larynx cancer.

\begin{figure}[t]
    \centering
    \includegraphics[width = \textwidth]{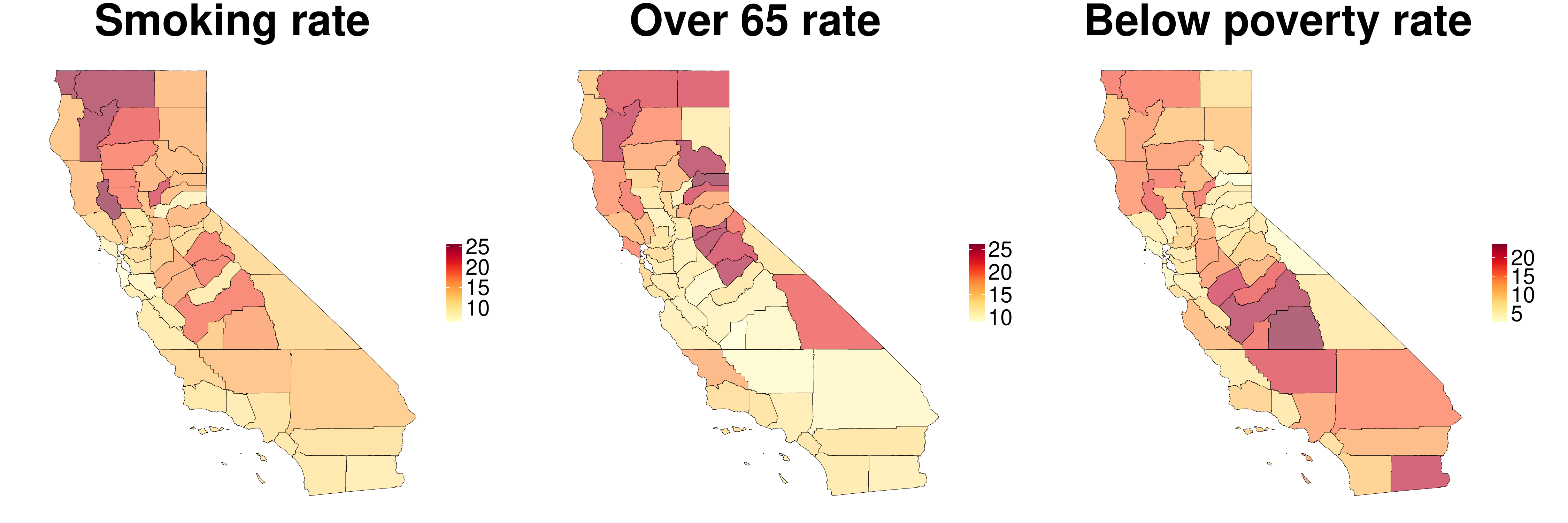}
    \caption{Maps of county-level covariates}
    \label{fig:covariates_plot}
\end{figure}

To provide a comprehensive analysis and explain the impact of risk factors, we incorporate covariates into our model to capture both the mean structure and identify neighboring areas with significant differences in rates. A detailed analysis is presented in Section~\ref{sec: cancer_analysis}. In Figure~\ref{fig:covariates_plot}, we display the smoking, over $65$, and below poverty threshold rates for each county, arranged from left to right. These plots reveal notable patterns: the northern counties exhibit a higher concentration of smoking rates, with a few exceptions in the central counties; both northern and eastern counties have higher rates of individuals over $65$; and the northern and central counties have higher rates below the poverty threshold. Conversely, the coastal regions demonstrate relatively lower rates across all the covariates, indicating a distinct contrast between different geographical areas \citep[see][for similar findings]{doll2005mortality}. Interestingly, central and southern counties exhibit lower rates specifically in the over $65$ variable. The significance and implications of these findings will be discussed in greater detail in Section~\ref{sec: cancer_analysis}.

\section{Model development}\label{sec: models}

We construct a generalized linear mixed model for the disease outcome, $y_{id}$, using a distribution from the exponential family with a canonical link,
\begin{equation}\label{eq:glm_link_mean}
    g(E(y_{id})) = \bx_{id}^{\T}\bbeta_d + \phi_{id},
\end{equation}
for each areal unit $i = 1,\dots,n$ and disease $d = 1,\dots, q$. The large scale trend is modeled using a regression, where $\bx_{ij}$ is a vector of explanatory variables and $\bbeta_d$ is the corresponding vector of slopes, while $\phi_{id}$ is the spatial random effect for disease $d$ in region $i$. For detecting difference boundaries, we seek $P(\phi_{id} \neq \phi_{jd} \given \by, i \sim j)$, where $\by$ is the collection of observed $y_{id}$'s and $i \sim j$ means that regions $i$ and $j$ are neighbors. This raises an issue about the prior distribution for the $\phi_{id}$'s. A continuous probability distribution is inappropriate because the resulting posterior probability is $P(\phi_{id} \neq \phi_{jd} \given \by, i \sim j) =1$, which leads to all $n$ boundaries being detected. This limitation highlights the need for a model that accounts for spatial dependence while assigning discrete probability masses on spatial random effects. Such a model enables including non-zero probabilities for the random effects being equal. Indeed, our approach will ensure non-null probabilities of equality between the random effects, resulting in a smaller number of boundaries detected compared to $n$. The Dirichlet Process (DP) offers a natural way to cluster regions, as it ensures that $P(\phi_{id} = \phi_{jd} \given \by, i \sim j) > 0$. Moreover, it presents a versatile framework for capturing the spatial effects represented by $\phi_{id}$. The model below belongs to a subclass of stick-breaking process priors and includes the DP as a special instance.

\subsection{Spatial random effects}
Let $\bphi = (\bphi_1^\T,\dots,\bphi_q^\T)^\T$ be the $N\times 1$ vector, where $N=nq$ and $\bphi_d = (\phi_{1d},\dots,\phi_{nd})^\T$ for each $d=1,\ldots,q$. Following  \cite{gao2022spatial} we let 
$\{1, \dots , n, \dots , (q-1)n + 1, \dots ,N\}$ be an enumeration of the pairwise $(i,d)$ indices corresponding to the ordering of $\bphi$, thereby dealing with a unique $N\times 1$ vector $\bphi$. We let $\bphi \sim G_N$, where $G_N$ is an unknown distribution further specified by $G_N = \sum_{u_1,\dots,u_N} \pi_{u_1,\dots,u_N} \delta_{\theta_{u_1}, \dots, \theta_{u_N}}$, where 
\begin{equation}\label{eq:DP}
    \begin{split}
        \pi_{u_1,\dots,u_N} &= P \left(\sum_{k=1}^{u_1-1} p_k < F^{(1)}(\gamma_1) < \sum_{k=1}^{u_1} p_k, \ldots,\right. \\
        &\qquad \qquad \qquad \quad \left. \ldots\sum_{k=1}^{u_N-1} p_k < F^{(N)}(\gamma_N) < \sum_{k=1}^{u_N} p_k\right); \quad 
        \bgamma \sim \Norm_N(\mathbf{0},\bSigma_{\bgamma}),
    \end{split}
\end{equation}
where $\theta_{k} \given \tau_s \overset{iid}{\sim} \Norm\left(0, 1/\tau_s\right)$ for $k=1,\dots,K$. The motivation behind the choice of the independent and identically distributed assumption for the atoms base distribution stems from the necessity of establishing ties across both regions and cancers. For instance, if we were to assign different means to different cancers, we could still accommodate ties within each cancer, wherein different regions' random effects assume the same value. However, we would be unable to establish ties across cancers, as the underlying distribution from which the atoms are sampled would differ.

Priors for model parameters are specified as $\bbeta_d \overset{ind}{\sim} \Norm(\bzero, \bSigma_{\bbeta_d})$, while the precision (inverse variance) parameter $1/\tau_s \sim IG(a_s, b_s)$, where $a_s$ is shape and $b_s$ is scale. We specify the stick-breaking weights by $p_1 = V_1$ and $p_j = V_j \prod_{k<j}(1-V_k)$ for $j=2,\dots,K$, where $V_k \overset{iid}{\sim} Beta(1, \alpha)$. Moreover, $u_1,\dots,u_N$ are indices of the elements of $\btheta = (\theta_1,\ldots,\theta_N)^{\T}$ sampled for the $N$ observations, $\delta_{\btheta}$ is the Dirac measure located at $\btheta$, and $F^{(1)}(\cdot),\dots,F^{(N)}(\cdot)$ are cumulative distribution functions of the marginal distribution of the corresponding $\bgamma$ elements, indexed in accordance with the elements of $\bphi$.

The weights in the distribution $G_N$ are constructed by utilizing the marginal cumulative distribution function (CDF) of the elements of $\bgamma$. This construction combines two approaches: the hybrid Dirichlet process \citep{petrone2009hybrid}, which involves the use of point-referenced continuous spatial copulas to model weight dependency, and the latent stick-breaking process \citep{rodriguez2010latent}, which employs ordered atoms. Marginally, each $F^{(i)}(\gamma_{i})$ is $\mathcal{U}(0,1)$ but they are dependent through the $\gamma_{i}$, and the distribution of each spatial random effect $\phi_{i}$ is a regular univariate DP, with the spatial dependence between these DPs introduced using a copula representation for the weights. In theory, $K=\infty$ yields a fully non-parametric model, but we follow the standard practice of replacing the infinite sum with the sum of the initial $K$ ($K \leq n$) terms, as the probability masses diminish rapidly. Should concerns about truncation bias arise, an alternative method called slice-sampling \citep[described in][]{kalli2011slice} is available for exact sampling. In particular, for areal data the value of $K$ is naturally constrained by the number of areal units rendering the infinite representation redundant.

The marginal covariance between two spatial random effects $\phi_i$ and $\phi_j$ where, in this case, $i,j = 1,\dots,N$, can be expressed as
\begin{equation}\label{eq:rnd_cov}
    \Cov{(\phi_i,\phi_j)} = \frac{b_s}{a_s - 1} \sum_{k=1}^K \pi_{kk}^{(i,j)}
\end{equation}
where 
\begin{equation}\label{eq: rnd_cov2}
    \pi_{kk}^{(i,j)} = P \lrnd \sum_{t=1}^{k-1} p_t < F^{(i)}(\gamma_i) < \sum_{t=1}^{k} p_t,\sum_{t=1}^{k-1} p_t < F^{(j)}(\gamma_j) < \sum_{t=1}^{k} p_t \rrnd 
\end{equation}
Equation~\eqref{eq:rnd_cov}~and~\eqref{eq: rnd_cov2} evinces how the spatially dependent elements of $\bgamma$ induce associations among the elements of $\boldsymbol{\phi}$. The covariance of $\bphi$ also depends on the hyperparameters of the precision parameter prior of the base distribution. Higher values of the probabilities, i.e., $\pi_{kk}^{(i,j)}$, produce higher correlation between the random effects. Unsurprisingly, a higher likelihood of $\phi_i$ and $\phi_j$ exhibiting similar values implies a higher covariance in \eqref{eq:rnd_cov}. We next attend to modeling the spatial covariance matrix $\boldsymbol{\Sigma}_{\boldsymbol{\gamma}}$.

\subsection{DAGAR with adjacency modeling}\label{subsec: dagar_adj_modeling}

DAGAR uses a fixed set of arbitrarily ordered regions yielding a \emph{topologically ordered} set of vertices, $\calV = \{1,\dots,n\}$, and a set ${\cal E}$ of directed edges that encode ``neighbors'' of every region. Rather than defining all geographically adjacent regions as neighbors, DAGAR defines neighbors of a given region $i$ as regions that (i) are geographically adjacent to $i$; AND (ii) precede $i$ in the topological order. The directed graph ${\cal G} = ({\cal V}, {\cal E})$ is constructed from this definition. The precision matrix of a DAGAR random variable is $\bQ(\rho)=(\IdentityMat - \bB)^\T \bLambda (\IdentityMat - \bB)$, where $\bB = (b_{ij})$ is $n \times n$ with elements $b_{ij} = \frac{\rho}{1 + (n_{<i} - 1)\rho^2}$ if there is a directed edge $(i,j) \in {\cal E}$ for $i \geq 2$, where $n_{<i}$ is the number of neighboring regions preceding $i$, and $b_{ij} = 0$ if $(i,j) \notin {\cal E}$. In DAGAR, $(i,j) \in {\cal E}$ if $j<i$ in the topological order and regions $i$ and $j$ are geographically adjacent. This implies that $b_{ii}=0$ and $b_{ij}=0$ for all $i \leq j$ so $\bB$ is strictly lower triangular. The $n\times n$ diagonal matrix $\bLambda = \mbox{diag}(\lambda_{i})$ has elements $\lambda_i = \frac{1 + (n_{<i} - 1)\rho^2}{1 - \rho^2}$ along its diagonal for $i=1,\dots,k$, where $n_{<1} = 0$. \cite{datta2019spatial} show that $\rho$ is a direct indicator of spatial autocorrelation. DAGAR holds further appeal as it accounts for all edges without any omissions, while still yielding a computationally efficient precision matrix. 

Two extensions to the DAGAR are relevant to our current inferential objectives. First, as articulated in Section~\ref{sec: intro}, we intend to model the adjacency matrix rather than specifying it purely from the map. Geographic adjacency, by itself, need not be enough to infer similarities or disparities and we accommodate learning about the adjacency relations using risk factors or other explanatory variables. We thereby recognize that inferring about spatial disparities relies upon effectively estimating difference boundaries. We exploit the available data on spatially oriented risk factors and explanatory variables to drive inference on difference boundaries. Second, while extending the model to jointly model dependent cancer occurrences, we allow the impact of the factors informing about the edges to vary by the diseases under considerations. This results in an effective amalgamation of the approach in \cite{lee2012boundary} with that in \cite{gao2022spatial}.  

\subsubsection{Adjacency modeling}\label{subsubsec: adj_modeling} 
Unlike in more traditional spatial autoregression models, such as simultaneous and conditional autoregression (SAR and CAR, respectively), where the precision matrix includes an adjacency matrix whose elements can be modeled using explanatory variables \citep{lu2007bayesian}, the elements of the DAGAR precision matrix, $b_{ij}$, are specified in a manner that renders $\rho$ as a spatial correlation parameter. It is, therefore, unapparent how information from explanatory variables, predictors or risk factors can be introduced in the adjacency. In order to achieve this, we construct a strictly lower triangular $n \times n$ adjacency matrix, $\bW = (w_{ij})$ based upon the fixed topological order of regions to assimilate information from preceding neighboring regions. This results in $w_{ij} = 0$ for $i\leq j$ or $i \not\sim j$, where $i\sim j$ indicates ``neighbors'' as defined in the preceding section. Note that $\bW$ encodes all of the information contained in the original DAGAR. Thus, for each region $i$ its preceding neighbors are contained in $\lcur j: j<i, j \sim i \rcur \equiv \lcur j: w_{ij} = 1 \rcur$. Subsequently, we introduce $\bW$ into the DAGAR precision matrix as
\begin{equation}\label{eq: dagar_modified1}
        \bQ(\rho,\bW) = (\IdentityMat - \bB)^\T \bLambda (\IdentityMat - \bB);\quad \mbox{ where }\quad \bB = \tilde{\bB} \circ \bW\;,
\end{equation}
where $\tilde{\bB} = (\tilde{b}_{ij})$ with elements $\tilde{b}_{ij} = \frac{\rho}{1 + (n_{<i} - 1)\rho^2}$, $\lambda_i = \frac{1 + (n_{<i} - 1)\rho^2}{1 - \rho^2}$, $ n_{<i} = \sum_{j=1}^n w_{ij}$ and $\circ$ denotes the Hadamard (or elementwise) product.

The matrix $\tilde{\bB}$ is dense with all its elements being possibly non-zero. However, $\tilde{\bB}\circ \bW$ retains the sparse lower-triangular structure characteristic of DAGAR.  
Instead of directly setting the non-null elements of $\bW$ to fixed values, one possibility is to model each element of $\bW$ as a separate random quantity. For instance, \cite{lu2007bayesian} employed a logistic model for the elements of a symmetric adjacency matrix in a CAR model while considering boundary-specific risk factors. In our DAGAR setting, a parallel approach would be to model the lower triangular elements $w_{ij} \given p_{ij} \stackrel{ind}{\sim} Bernoulli(p_{ij})$ and introducing a regression model in, for example, a logistic link $\log (p_{ij}/(1-p_{ij}))$. Other possible approaches include adapting random adjacency models, such as in \cite{ma2010hierarchical}, to DAGAR by leveraging an Ising model for elements of $\bW$. However, the information needed to estimate such models is rarely available and require informative prior distributions resulting in excessively parametrized covariance models for $\bgamma$ that inaccurately estimate uncertainties and produce biased inference for spatial boundaries \citep[also see][who argue that effectively modeling geographical adjacency would require a separate parameter for every pair of neighbors resulting in overparametrized models]{li2011mining}.  

In what follows, we adopt a simpler and more effective approach based on \cite{lee2012boundary}, following who we model the adjacency matrix as 
\begin{equation}\label{eq:w_binary}
    w_{ij}(\bet) =
    \begin{cases}
        1 \text{\,\,\, if \,\,\,} \exp{\lrnd - \bz_{ij}^\T\bet \rrnd} \geq 0.5 \text{\,\,\, and \,\,\,} i \sim j \\
        0 \text{\,\,\, otherwise}\;,
    \end{cases}
\end{equation}
where $\bz_{ij}$ consist of explanatory variables specific to pairs of regions $i$ and $j$ and are assumed to be non-negative in each element. In practice, they could represent an absolute measure of discrepancy between the two regions with $\bet$ being the vector of corresponding coefficients. Equation~\eqref{eq:w_binary} encodes geographic neighbors so as to not represent disparities whenever $\bz_{ij}^\T\bet \leq \log 2$. Furthermore, $\bz_{ij}$ excludes an intercept so regions with homogeneous populations, i.e., $\bz_{ij} = \bzero$, are deemed adjacent. The coefficients in $\bet$ act as disparity parameters and are assumed to be nonnegative with each element assigned a Uniform prior distribution between $0$ and a fixed positive constant such that at most $50\%$ of geographic neighbors in the study region are classified as difference boundaries. 
Higher values in $\bet$ assist in exceeding the threshold, which results in a loss of a geographical boundary in $\bW$ to be smoothed over and, hence, indicating a spatial disparity. 

A salient feature of the preceding development is the assimilation of information from explanatory variables $\bx_{id}$ in \eqref{eq:glm_link_mean} and $\bz_{ij}$ in \eqref{eq:w_binary}. The former informs about the observed values of the health outcomes and captures large scale trends, while the latter provides information on disparities among geographic neighbors. The vectors $\bx_{id}$ and $\bz_{ij}$ could include the same variables but need to be indexed differently. For example, one could include the smoking rate for a region $i$ in $\bx_{id}$ and the difference in smoking rates between regions $i$ and $j$ in $\bz_{ij}$. While the slope $\bbeta$ is interpreted as in the customary generalized linear models with exponential families, the coefficients in $\bet$ indicate whether the corresponding variable in $\bz_{ij}$ signifies a disparity or not. In particular, a smaller value of a coefficient $\eta_k$ indicates the $k$-th variable in $\bz_{ij}$. For example, they can be map-based, such as the distance between the centroids of regions $i$ and $j$ or the discrepancies in the percentage of common boundaries shared by the two regions compared to their total geographical boundaries. Auxiliary topological data may be relevant, such as the presence of challenging natural features like mountain ranges or rivers that hinder travel between the regions. Our approach assimilates sociodemographic information into $\bz_{ij}$. This involves comparing the percentages of urban area between the two regions or calculating the standardized absolute difference in a specific regional characteristic, such as the proportion of residents who smoke or even the expected age-adjusted disease count for the region. By including these covariates in $\bz_{ij}$, the model can account for various factors that may influence the neighboring relationship between regions $i$ and $j$.

In this way the covariance structure of $\bgamma$ exhibits a dual advantage: it is fully automatic and parsimonious, and yet it enables the data to determine the number and locations of any adjacency, rather than relying on the investigator's prior knowledge. Furthermore, we are introducing an exploratory tool in the proposed model, that assists investigators in comprehending the localized spatial patterns of disease risk. For instance, in cases where a adjacency is identified despite an observable discontinuity in the data, researchers can delve deeper into the areas in question to investigate other potential factors contributing to this discontinuity.

\subsection{Multivariate framework}\label{subsec: multivariate}

Recalling that our current data analytic goals involve accounting for dependence among multiple diseases and spatial dependence for each cancer, we turn to a rich literature on multivariate areal models \citep{mardia1988multi, gelfand2003proper, carlin2003hierarchical, jin2005generalized, jin2007order, zhang2009smoothed, banerjee2016multivariate, mcnab2018, gao2022hierarchical}. Following recent developments in \cite{gao2022hierarchical} and \cite{gao2022spatial}, who demonstrated advantages of the DAGAR model \citep{datta2019spatial} with regard to overall model performance and interpretation of spatial correlation, we let $\bgamma$ follow a multivariate DAGAR (MDAGAR) model. 

\begin{figure}[t]
    \centering
    \begin{subfigure}[b]{0.32\textwidth}
        \includegraphics[width=\textwidth]{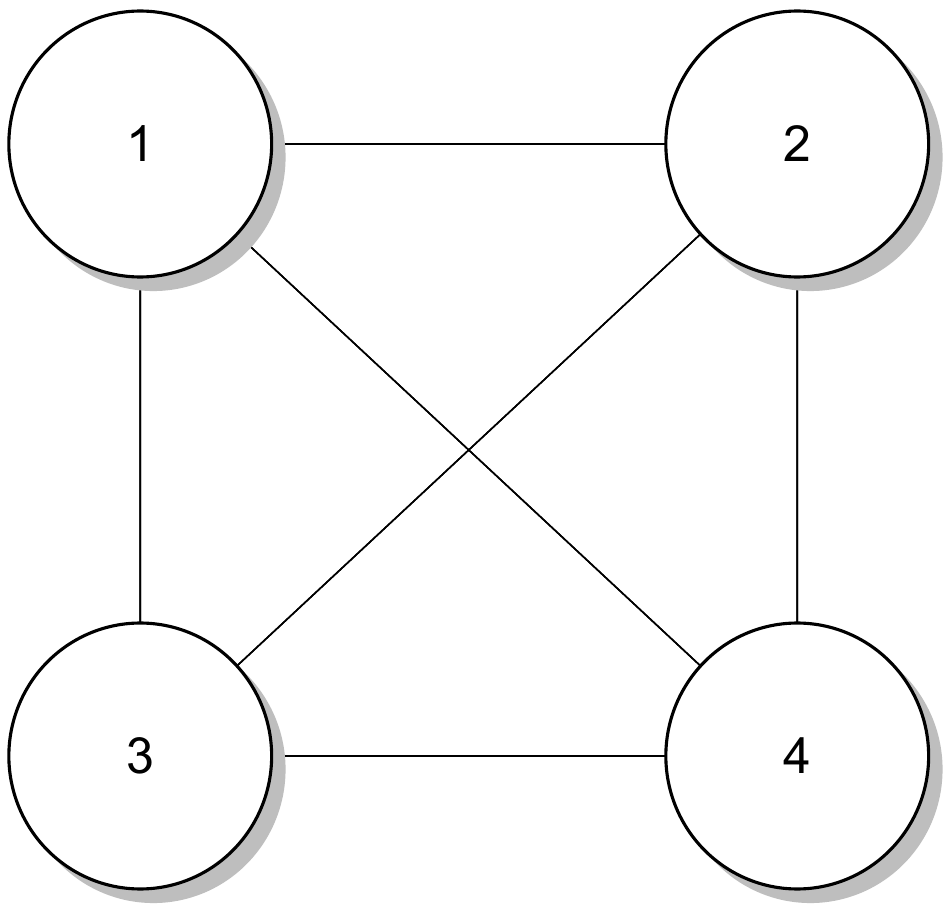}
    \end{subfigure}
    \hfill
    \begin{subfigure}[b]{0.32\textwidth}
        \includegraphics[width=\textwidth]{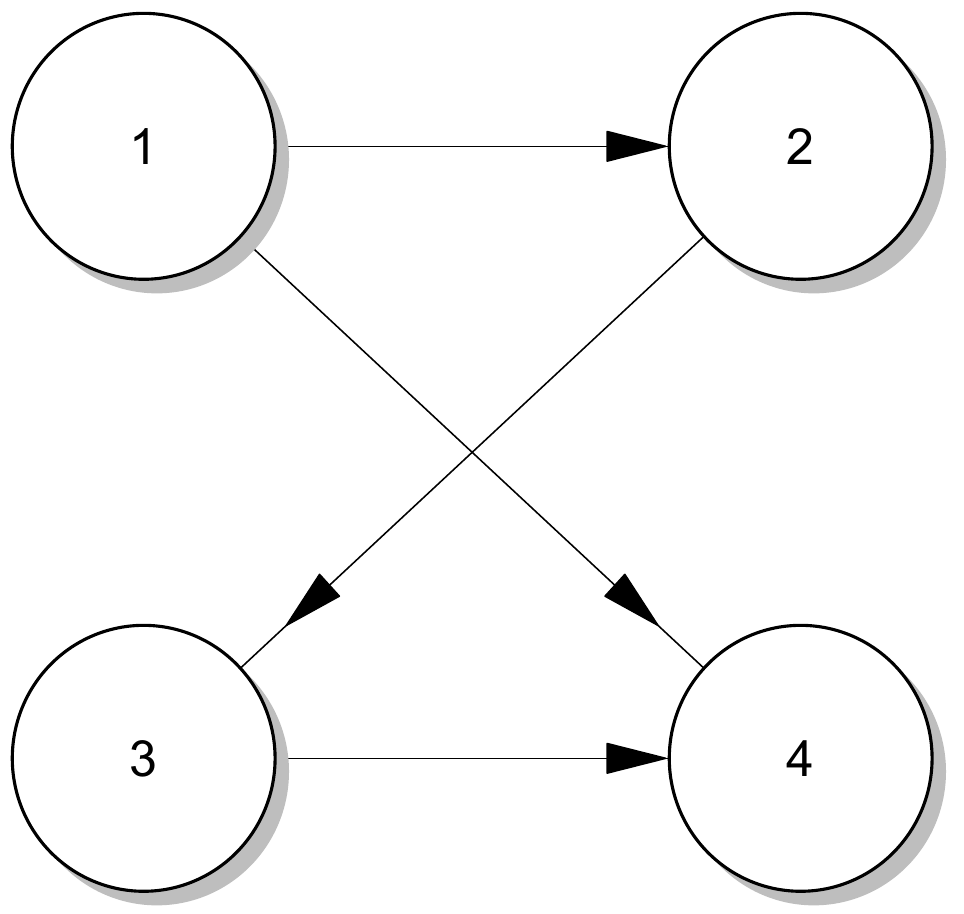}
    \end{subfigure}
    \hfill
    \begin{subfigure}[b]{0.32\textwidth}
        \includegraphics[width=\textwidth]{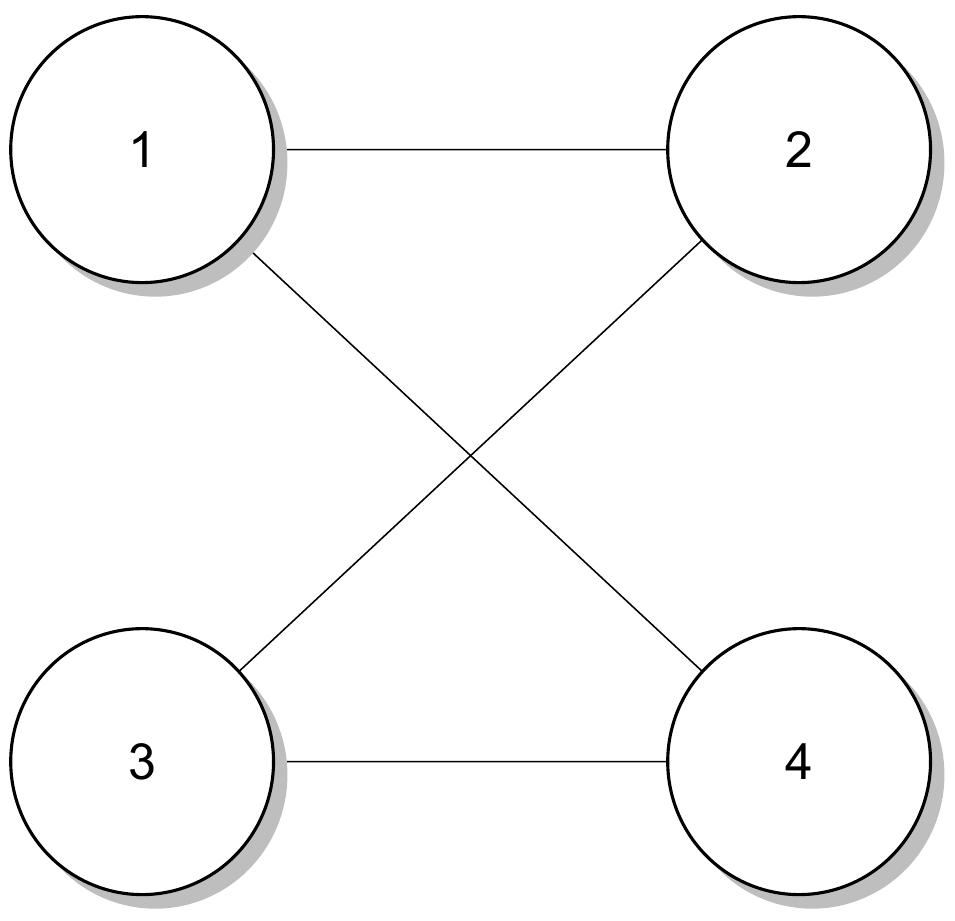}
    \end{subfigure}
    \caption{Left: unstructured graph, Center: directed acyclic graph. Right: undirected graph.}
    \label{fig:disease_graphs}
\end{figure}

We introduce dependence among outcomes using linearly transformed latent variables resulting in an unstructured graph (left panel of Figure~\ref{fig:disease_graphs}). Therefore, $\bgamma_1 = a_{11}\bbf_1$, $\bgamma_d = a_{d1}\bbf_1 + \cdots + a_{dd}\bbf_d$ for each $d=2,\dots,q$, where $a_{dh}$, $h = 1, \dots , d$, are unknown coefficients that associate spatial components for different diseases and each $\bbf_d \stackrel{ind}{\sim} \Norm(\mathbf{0},\bQ(\rho_d,\bW_{d})^{-1})$, where $\bW_{d}$ is a spatial adjacency matrix and $\rho_d$ is the spatial autocorrelation parameter for disease $d$. Therefore, each $\bgamma_d$ is a linear combination of independent DAGAR random variables. This construction makes the model independent on the disease ordering because the association among the diseases is modeled using $\bA\bA^{\T}$, where $\bA$ is the lower-triangular matrix with elements $a_{dh}$. The covariance matrix of $\bgamma$ is
\begin{equation}\label{eq:sp_cov}
    \bSigma_{\bgamma} = (\bA \otimes \bI_{N})\lsq \bigoplus_{d=1}^q \bQ(\rho_d,\bW_{d})^{-1} \rsq (\bA^\T \otimes \bI_{N})
\end{equation}
where $\otimes$ denotes the Kronecker product and $\bW_{d}$ is specified through \eqref{eq:w_binary} with a different $\bet_d$ for each disease $d$. Therefore, we are incorporating a random adjacency matrix into the model for each type of cancer. We evaluate the model's performance by accounting for the particular cancer under consideration, as well as the explanatory variables utilized in model \eqref{eq:w_binary} that informs the adjacency structure, potentially varying for each cancer type. 

We model $\bA\bA^{\T}$ using an inverse-Wishart distribution with $\bA$ identifying as the unique lower-triangular Cholesky factor. Alternatively, we introduce a graphical model for the diseases as a representation of multivariate dependencies \citep[see, e.g.,][]{cox1993graph, cox1996graph}. Such models distinguish between directed and ``undirected'' graphs (central and right panels in Figure~\ref{fig:disease_graphs}, respectively). The former assumes a topological ordering of the diseases, while the latter does not. For a directed graph, we build an inter-disease DAGAR using the ordering of the diseases and the adjacency matrix of the directed inter-disease graph. This amounts to modeling $\bA = (\bI_q - \bB)^{-1}\bLambda^{-1/2}$ in \eqref{eq:sp_cov}, where $\bB=(b_{ij})$ and $\bLambda=(\lambda_i)$ are as defined in Section~\ref{subsec: dagar_adj_modeling}, but with $\bW$ using the inter-disease graph. This inter-disease DAGAR depends upon ordering of the diseases. \cite{gao2022hierarchical} devised model choice and Bayesian model averaging over all topological orders in directed graphical models, but the approach is computationally onerous.    

Consider a fixed directed acyclic disease graph ${\cal G}_{dis} = \{{\cal V}_{dis}, {\cal E}_{dis}\}$ (the suffix $dis$ serves to distinguish this from the graph of the spatial map), where ${\cal V}_{dis}$ is a set of $q$ nodes and ${\cal E}_{dis}$ is a set of directed edges from parent nodes to children (central panel of Figure~\ref{fig:disease_graphs} with $q=4$). Let
\begin{equation}\label{eq: matrix_latent_factors}
    \bgamma_1 = \bbf_1,\quad \bgamma_d = \bA_{d,1}\bgamma_1 + \cdots + \bA_{d,d-1}\bgamma_{d-1} + \bbf_d,\; \mbox{ for }\; d=2,\dots,q\;,
\end{equation}
where each $\bbf_d \stackrel{ind}{\sim} \Norm(\mathbf{0},\bQ(\rho_d,\bW_{d})^{-1})$ and $\bA_{d,d'}$ is $n\times n$ with $\bA_{d,d'} = \bO$ whenever $d' \geq d$. The linear equations in \eqref{eq: matrix_latent_factors} encode the conditional distributions $p(\bgamma_d \given \bgamma_1,\ldots, \bgamma_{d-1})$ for $d=2,\ldots,q$. Setting $\bA_{d,d'} = \bO$ whenever node $d'$ is not a parent of $d$, i.e., there is no directed edge from $d'$ to $d$, ensures that the model conforms to the conditional independence among diseases posited in ${\cal G}_{dis}$. If $\bA = (\bA_{d,d'})$ is the $N\times N$ block matrix with the $n\times n$ matrix $\bA_{d,d'}$ occupying block $(d,d')$, then $\bA$ is strictly lower triangular, hence $\bI_N-\bA$ is nonsingular and $\bSigma_{\bgamma} = (\bI_N - \bA)^{-1}\left[\bigoplus_{d=1}^q \bQ(\rho_d,\bW_{d})^{-1}\right](\bI_N - \bA^{\T})^{-1}$ is positive definite no matter how we specify $\bA_{d,d'}$ for all $(d,d') \in {\cal E}_{dis}$, i.e, $d'$ is a parent of $d$. 

Setting $\bA_{d,d'} = \alpha_{0dd'}\bI_n + \alpha_{1dd'}\bW$, where $\bW$ is the fixed geographic map, introduces parameters $\alpha_{0dd'}$ and $\alpha_{1dd'}$ that allow the nature of dependence between diseases $d$ and $d'$ to vary across space. This flexibility is desirable because inter-disease dependencies, or lack thereof, are often manifestations of lurking shared risk factors that remain unaccounted for.

In contrast to directed acyclic graphs, ``undirected'' graphs model relationships among nodes using conditional dependencies. An edge between nodes signifies conditional dependence given all other nodes, while the absence of an edge indicates conditional independence given all other nodes (right panel in Figure~\ref{fig:disease_graphs} with $q=4$). 
Given an undirected disease graph, we adapt a multivariate conditional autoregression model \citep[MCAR,][]{mardia1988multi, carlin2003hierarchical, gelfand2003proper} by specifying the full conditional models,
\begin{equation}\label{eq:MCAR}
    \bgamma_{d} \given \bgamma_{(-d)} \sim \Norm_{n}\lrnd \sum_{h=1}^q \bA_{d,h} \bgamma_{h}, \bQ(\rho_{d},\bW_{d})^{-1}\rrnd\;,
\end{equation}
where, $\bgamma_{(-d)}$ represents the collection of all random effects except for the $d$-th disease type. The matrices $\bA_{d,h}$ represent $n \times n$ matrices that encode the relationships between various cancers, while $\bQ(\rho_{d},\bW_{d})$ refers to the univariate precision matrix of the DAGAR model applied to the $d$-th cancer, capturing the spatial correlation within each cancer case. By exploiting a multivariate extension of Brook's Lemma \citep[see, e.g.,][]{mardia1988multi, banerjee2014hierarchical}, the collection of distributions in \eqref{eq:MCAR} yield the following joint distribution:
\begin{equation}\label{eq:MCAR_joint}
    p(\bgamma) \propto \exp \lrnd -\frac{1}{2} \bgamma^\T \bSigma_{\bgamma}^{-1} \bgamma \rrnd
\end{equation}
where $\bSigma_{\bgamma}^{-1} = \bM^{-1} \lrnd \IdentityMat_{N} - \bA \rrnd$, $\bM^{-1} = \bigoplus_{d=1}^q \bQ(\rho_d,\bW_{d})$, and, analogously to the previous case, $\bA = (\bA_{d,h})$ is the $N\times N$ block matrix with the $n\times n$ matrix $\bA_{d,h}$ occupying the $(d,h)$-th block. The key consideration here is to ensure that $\bSigma^{-1}_{\bgamma}$ is positive definite. 

Let $\bLambda_{dis} = \lrnd \bD_{dis} - \rho_{dis} \bW_{dis} \rrnd$ be $q\times q$, where $\bW_{dis}$ is the $q\times q$ binary adjacency matrix of the inter-disease graph, $\bD_{dis}$ is diagonal with the number of edges incident on each node as its diagonal element, and $\rho_{dis} \in (1/\zeta_{min},\zeta_{max})$ with $\zeta_{min} < 0$ and $\zeta_{max} = 1$ denoting the smallest and largest eigenvalues of $\bD_{dis}^{-1/2}\bW_{dis}\bD_{dis}^{-1/2}$, respectively; all eigenvalues are real, $\zeta_{max}=1$ since $\bD_{dis}^{-1}\bW_{dis}$ is row-stochastic and $\zeta_{min} < 0$ since the trace of $\bD_{dis}^{-1/2}\bW_{dis}\bD_{dis}^{-1/2}$ is zero. The interval for $\rho_{dis}$ ensures that $\bLambda_{dis}$ is positive definite. Let $\bR_d^{\T}\bR_d = (1/\lambda_{dis,dd})\bQ(\rho_d, \bW_d)$ with $\sqrt{\lambda_{dis,dd}}\bR_d$ being the upper-triangular Cholesky factor of $\bQ(\rho_d, \bW_d)$. Setting $\bA_{dh} = (\lambda_{dis,dh}/\lambda_{dis,dd})\bR_d^{-1}\bR_h$ yields $\bSigma_{\bgamma}^{-1} = \left(\oplus_{d=1}^q \bR_d^{\T}\right)\left(\bLambda_{dis}\otimes \IdentityMat\right)\left(\oplus_{d=1}^q \bR_d\right)$, which is positive definite. Hence, \eqref{eq:MCAR_joint} is a proper prior distribution for the spatial effects incorporating spatial autocorrelation using DAGAR and conforming to a posited undirected conditional independence graph.  


\subsection{Model implementation}

Letting $\bet = (\bet_{1}^\T,\ldots,\bet_{q}^\T)^\T$, and $\bbeta$ defined analogously, for the unstructured disease graph we evaluate the following joint posterior distribution,
\begin{equation}\label{eq:post}
    p(\btheta,\bbeta,\tau_s,\bV,\bgamma,\brho,\bet,\bA \given \by) \propto p(\btheta,\bbeta,\tau_s,\bV,\bgamma,\brho,\bet,\bA) \times \prod_{i=1}^n \prod_{d=1}^q p(y_{id} \given \phi_{id})\;,
\end{equation}
where $\by = (\by_1^{\T},\ldots,\by_q^{\T})^{\T}$ is $N\times 1$ with $N=nq$, $n$ is the number of geographic regions, and $\by_d = (y_{1d},\ldots, y_{nd})^{\T}$ and each $y_{id}$ being modeled independently conditional on the spatial random effects using a member from the exponential family. For example, in our subsequent analysis we assume $y_{id} \given \bbeta_d, \phi_{id} \overset{ind}{\sim} Pois(E_{id}\exp{(\bx_{id}^{\T}\bbeta_d + \phi_{id})})$ and specify the joint prior distribution, $p(\btheta,\bbeta,\tau_s,\bV,\bgamma,\brho,\bet,\bA)$ proportional to
\begin{equation}\label{eq:joint_prob}
    \begin{split}
        & \prod_{d=1}^q \Norm(\bbeta_d \given \bzero, \bSigma_{\beta_d}) \times \prod_{k=1}^K \lcur \Norm(\theta_{k} \given 0, 1/\tau_s) \times Beta(V_k \given 1,\alpha) \rcur\\
        &\qquad\qquad\qquad \times IG(1/\tau_s \given a_s, b_s) \times \Norm(\bgamma \given \mathbf{0}, \bSigma_{\bgamma}) \times \mathcal{W}^{-1}(\bA\bA^\T \given \nu, \bPsi) \times \left\vert \frac{\partial \bA\bA^\T}{\partial a_{dh}}\right\vert \\ 
        &\qquad\qquad\qquad\qquad \times \prod_{d=1}^q \lcur \prod_{r=1}^{R_d} \mathcal{U}(\eta_{dr} \given 0, M_r) \times \mathcal{U}(\rho_d \given 0, 1) \rcur\;,
    \end{split} 
\end{equation}
where $\left\vert \frac{\partial \bA\bA^\T}{\partial a_{dh}}\right\vert = 2^q \prod_{d=1}^q a_{dd}^{q-d+1}$ is the Jacobian transformation for the prior on $\bA\bA^\T$ in terms of the Cholesky factor $\bA$ \citep{olkin1972jacobians}, and $R_d$ is the dimension of $\bet_{d}$. We employ Markov Chain Monte Carlo (MCMC) with Gibbs sampling and random walk Metropolis \citep{gamerman2006markov}, implemented in \texttt{R} statistical computing environment, to draw samples from the posterior distribution, as defined in \eqref{eq:post}.

The graphical models discussed above provide parametric representations for $\bA$. Therefore, the $\mathcal{W}(\bA\bA^{\T}\given \cdot, \cdot)$ in \eqref{eq:joint_prob} is replaced with priors on the parameters defining $\bA$. In the directed graphical setting, we further model $\bA_{d,h}$ in terms of parameters $\{\alpha_{0dd'}\}$ and $\alpha_{1dd'}$. These parameters are all assigned independent Gaussian prior distributions $\alpha_{0dd'} \overset{ind}{\sim} \Norm(\cdot, \cdot)$ and $\alpha_{1dd'} \overset{ind}{\sim} \Norm(\cdot, \cdot)$, respectively, resulting in a joint prior distribution for the whole $\balpha = (\alpha_{0dd'},\alpha_{1dd'})_{d,d'}$ vector parameter, i.e., $\balpha \sim \Norm(\bzero, 100\IdentityMat)$. In the undirected model, the unknown parameter is $\rho_{dis}$ which specifies $\bA$ completely so we model $\rho_{dis} \overset{ind}{\sim} \mathcal{U}(\cdot,\cdot)$. Details on the computations are presented in Appendix~\ref{sec:comp_det}.

\subsection{Difference boundaries through FDR}

We treat spatial boundary analysis as an exercise in multiple hypothesis testing, although a formal null hypothesis is not required in Bayesian inference. Instead, we pursue full stochastic quantification to derive a threshold for detecting cancer-specific spatial disparities. For each pair of adjacent regions, i.e., $i \sim j$, and each cancer $d$, we seek the posterior probability of $\phi_{id} = \phi_{jd'}$ and its complement $\phi_{id'} \neq \phi_{jd'}$ for region-disease pairs $(i,d)$ and $(j,d')$. We designate edge ($i, j$) as a difference boundary if the posterior probability of $\phi_{id} \neq \phi_{jd'}$ given $\by$ surpasses a predefined threshold $t$. 

The most general difference boundary we report is the \emph{cross-difference} boundary based upon $v_{(i,d)(j,d')} = P(\phi_{id} \neq \phi_{jd'}, \phi_{id'} \neq \phi_{jd} \given \by)$. We derive a threshold $t$ that controls the false discovery rate (FDR) below a level $\zeta = 0.05$. To this end, we define
\begin{equation}\label{eq: fdr}
    \text{FDR}_{d,d'}(t) = \frac{\sum_{i \sim j} I(\phi_{id} = \phi_{jd'}) I(v_{(i,d)(j,d')} > t)}{\sum_{i \sim j} I(v_{(i,d)(j,d')} > t)}
\end{equation}
every disease pair $d$ and $d'$. The quantity in \eqref{eq: fdr} is evaluated as
\begin{equation}\label{eq: fdr_est}
    \widehat{\text{FDR}}_{d,d'}(t) = \mathbb{E}[\text{FDR}_{d,d'} \given \by] = \frac{\sum_{i \sim j} (1-v_{(i,d)(j,d')}) I(v_{(i,d)(j,d')} > t)}{\sum_{i \sim j} I(v_{(i,d)(j,d')} > t)}\;. 
\end{equation}
Following \cite{muller2004optimal}, we define
\begin{equation}\label{eq:threshold}
    t^{*} = \sup\lcur t: \widehat{\text{FDR}}_{d,d'}(t) \leq \zeta \rcur\;,
\end{equation}
which is based upon the optimal decision that minimizes the estimated false negative rate (FNR), $\widehat{\text{FNR}}_{d,d'}(t) = \frac{\sum_{i \sim j} v_{(i,d)(j,d')} (1 - I(v_{(i,d)(j,d')} > t))}{m - \sum_{i \sim j} I(v_{(i,d)(j,d')} > t)}$, where $m$ is the total number of geographic boundaries, subject to $\widehat{\text{FDR}} \leq \zeta$ \citep[also see][who proffer a similar approach]{sun2015false}. This estimation is based on a bivariate loss function $L_{2R} = (\widehat{\text{FDR}},\widehat{\text{FNR}})$. 

Note that the case of a single disease arises by substituting $v_{i,j}^{(d)} = P(\phi_{id} \neq \phi_{jd'}, \phi_{id'} \neq \phi_{jd} \given \by)$ in the above expressions. This is analogous to the formulation in \cite{li2015bayesian} but with the added flexibility of modeling the adjacency matrices and taking into account more detailed disease dependency structures.

\section{Simulation experiments}\label{sec: simulations}

We generated count data for $q=4$ diseases using \eqref{eq:glm_link_mean} assuming $y_{id} \stackrel{\text{ind}}{\sim} \text{Pois}(\exp{(\beta_d + \phi_{id})})$ over the $N=58$ counties in California. The counts were generated after simulating spatial effects $\bphi = (\bphi_1^\T,\bphi_2^\T,\bphi_3^\T,\bphi_4^\T)^\T \sim G_N$ with parameters fixed as $K = 15$, $\alpha = 1$, $\beta_1 = -2$, $\beta_2 = 2$, $\beta_3 = 1$, $\beta_4 = -1$, and $\tau_s = 0.25$. We present simulations under three different settings corresponding to the 3 graphs in Figure~\ref{fig:disease_graphs}: (i) unstructured; (ii) directed; and (iii) undirected. These imply 3 different specifications for $\bSigma_{\bgamma}$ in $\bgamma \sim \Norm(\mathbf{0},\bSigma_{\bgamma})$. For each of these settings we generate 25 data sets by fixing all parameters required to specify the underlying distributions as described below.

\begin{figure}[t]
    \centering
    \includegraphics[width = 0.66 \textwidth]{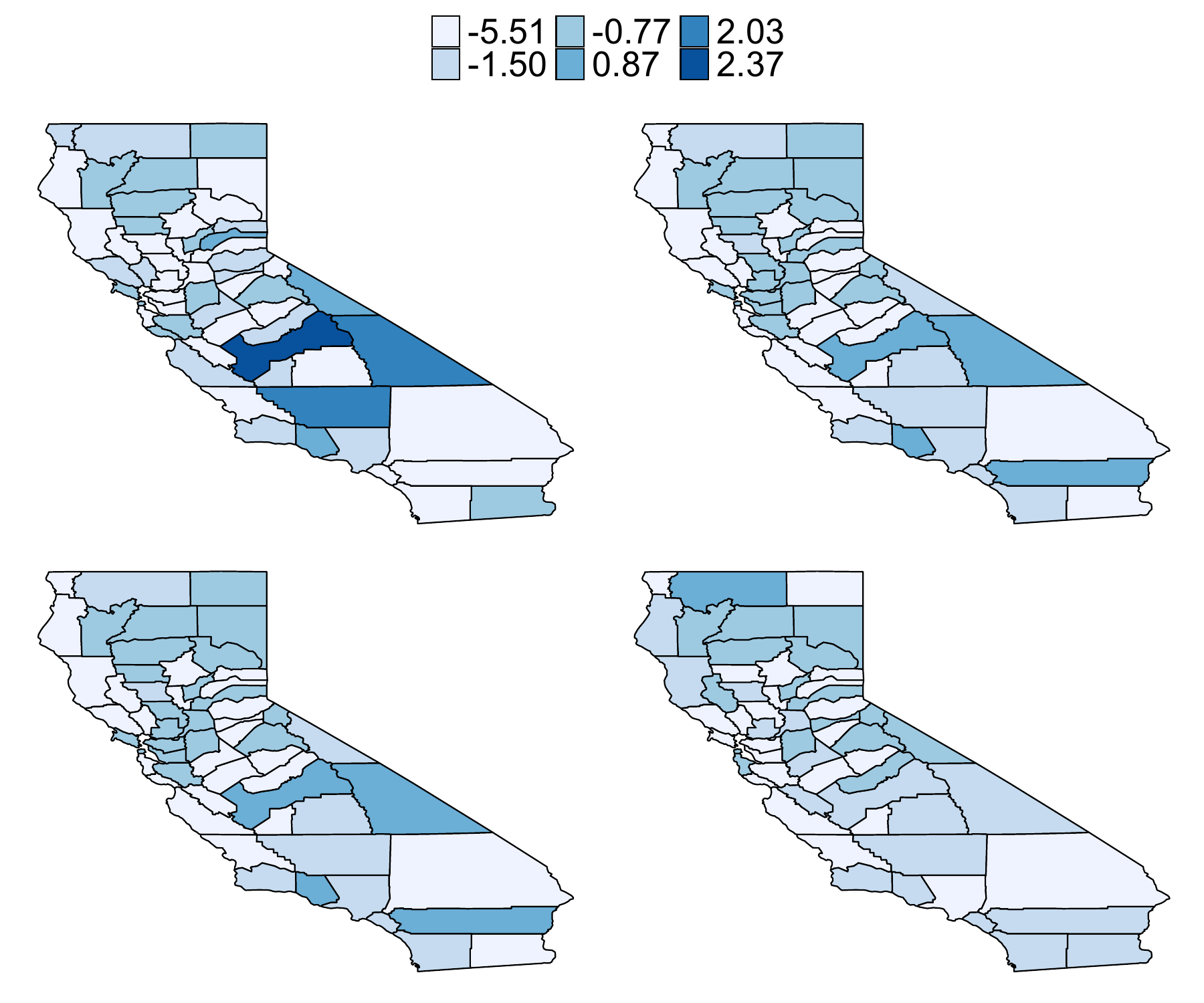}
    \caption{Simulated data maps depicting the random effects for the four cancers using a disease unstructured graph.}
    \label{fig:unstructured_map}
\end{figure}

For each of the above models, $\bSigma_{\bgamma}$ depends upon the DAGAR parameters and the inter-disease covariances. Spatial correlation parameters for the different cancers are fixed as $\brho = (0.2,0.8,0.4,0.6)^\T$. For the adjacency matrices, $\bW_{d}$, the elements $z_{ij}$ of the $\bZ$ matrix are computed as the standardized absolute difference of a generated covariate $\lvert x_{i} - x_{j}\rvert/\sigma$. We generate each $x_{i}$ from a normal distribution with mean $15$ and standard deviation $5$, while $\sigma$ represents the standard deviation of all the absolute differences between the $x_{i}$ values across all pairs of contiguous areas. The unknown parameters associated with these variables were fixed at $\bet = (0.5,0.25,0.33,0.6)^\T$. In addition to specifying the DAGAR parameters for each disease, for (i) we specify $\bA$ as lower triangular with all elements set to $1$. For (ii) we fix $(\alpha_{021},\alpha_{121})^\T = (0.3,0.5)^\T$, $(\alpha_{041},\alpha_{141})^\T = (0.5,0.4)^\T$, $(\alpha_{032},\alpha_{132})^\T = (0.4,0.4)^\T$, and $(\alpha_{043},\alpha_{143})^\T = (0.8,0.1)^\T$, and for (iii) we fix $\rho_{dis} = 0.25$. For each of these scenarios, we produce a single instance of $\bSigma_{\bgamma}$, which is used to generate a corresponding instance of $\bgamma$. This yields one set of spatial random effects which are retained in generating the counts for the 25 data sets. Figure~\ref{fig:unstructured_map} illustrates the maps of random effects for the four diseases in the generated data using the unstructured disease graph. A total of six distinct levels are observed with values ranging from $-5.51$ to $2.37$ arranged in ascending order. There are $99$ ``true difference boundaries” for disease $1$, $83$ for disease $2$, $91$ for disease $3$ and $89$ for disease $4$. The Appendix presents additional maps for data generated using the directed (Figure~\ref{fig:directed_map}) and unidrected (Figure~\ref{fig:undirected_map}) disease graphs. 

We apply our model to the 25 simulated data sets generated as described above. For the data analysis, we specify the prior hyperparameters as $\sigma^2_{\beta_{0}} = 1$, $a_s = 2$, $b_s = 1$, $\alpha = 1$, and $M = -\log(0.5)/\bZ_{0.5}$, where $\bZ_{0.5}$ represents the $0.5$th quantile of the $\bZ$ matrix. For the unstructured or unconstrained inter-disease graph (left panel in Figure~\ref{fig:disease_graphs}) we use the Wishart distribution in \eqref{eq:joint_prob} with $\nu = 2$ and $\bPsi = \text{diag}(0.1, 0.1, 0.1, 0.1)$. For the directed graph (center panel in Figure~\ref{fig:disease_graphs}), $\bA$ is determined through $\balpha \sim \Norm(\bzero, 100\bI)$, while for the undirected graph (right panel in Figure~\ref{fig:disease_graphs}) we specify $\rho_{dis} \sim {\cal U}(-1,1)$. We perform boundary detection by computing the conditional probability, $P(\phi_{id} \neq \phi_{jd'} \given \by)$, for $d$ and $d'$ ranging from 1 to 4, across neighboring regions denoted as $i$ and $j$. We use these posterior probabilities to obtain boundary detection results in terms of sensitivity and specificity for all pairs of regions and for all the disease graph models. 

For each data set, we present inference based on $2,500$ posterior samples using Markov Chain Monte Carlo (MCMC) algorithms after discarding the initial $2,500$ iterations as burn-in. Table~\ref{tab:bd_dd} presents the corresponding results. For each simulation, the values of sensitivity and specificity are determined by selecting a fixed number of edges with the highest T posterior probabilities, where $\mbox{T} \in \{80, 90, 100, 110, 120\}$. This selection accounts for false positives in all other T values for the diseases. The values in Table~\ref{tab:bd_dd} are average values of sensitivity and specificity over the 25 simulated data sets. 

\begin{table}[t]
    \centering
    \renewcommand{\arraystretch}{1.167}
    \caption{Boundary detection results in the simulation study for the three disease graph models.}
    \label{tab:bd_dd}
    \begin{adjustbox}{max width=\textwidth}
    \begin{tabular}{ccccccccccc}
        \hline
        & & \multicolumn{2}{c}{Disease 1} & \multicolumn{2}{c}{Disease 2} & \multicolumn{2}{c}{Disease 3} & \multicolumn{2}{c}{Disease 4} \\
        \cmidrule(lr){3-4} \cmidrule(lr){5-6} \cmidrule(lr){7-8} \cmidrule(lr){9-10}
        Disease graph & T & Specificity & Sensitivity & Specificity & Sensitivity & Specificity & Sensitivity & Specificity & Sensitivity \\
        \hline
        Unstructured & 80 & 0.470 & 0.596 & 0.729 & 0.790 & 0.503 & 0.602 & 0.432 & 0.582 \\
                     & 90 & 0.395 & 0.665 & 0.630 & 0.839 & 0.405 & 0.665 & 0.367 & 0.656 \\
                     & 100 & 0.302 & 0.729 & 0.519 & 0.881 & 0.319 & 0.734 & 0.293 & 0.727 \\
                     & 110 & 0.239 & 0.804 & 0.395 & 0.918 & 0.231 & 0.799 & 0.208 & 0.791 \\
                     & 120 & 0.144 & 0.867 & 0.263 & 0.949 & 0.145 & 0.866 & 0.131 & 0.861 \\
        \hline
        Directed & 80 & 0.433 & 0.619 & 0.621 & 0.821 & 0.459 & 0.600 & 0.427 & 0.578 \\
                 & 90 & 0.358 & 0.678 & 0.532 & 0.870 & 0.383 & 0.668 & 0.358 & 0.651 \\
                 & 100 & 0.298 & 0.749 & 0.435 & 0.912 & 0.304 & 0.737 & 0.287 & 0.724 \\
                 & 110 & 0.215 & 0.808 & 0.315 & 0.943 & 0.233 & 0.809 & 0.214 & 0.795 \\
                 & 120 & 0.142 & 0.875 & 0.215 & 0.973 & 0.152 & 0.874 & 0.135 & 0.863 \\
        \hline
        Undirected & 80 & 0.458 & 0.595 & 0.588 & 0.898 & 0.385 & 0.563 & 0.411 & 0.568 \\
                   & 90 & 0.377 & 0.661 & 0.497 & 0.929 & 0.316 & 0.638 & 0.340 & 0.642 \\
                   & 100 & 0.301 & 0.730 & 0.396 & 0.946 & 0.238 & 0.721 & 0.263 & 0.711 \\
                   & 110 & 0.220 & 0.799 & 0.293 & 0.958 & 0.185 & 0.795 & 0.182 & 0.777 \\
                   & 120 & 0.139 & 0.873 & 0.193 & 0.976 & 0.124 & 0.868 & 0.115 & 0.851 \\
        \hline
    \end{tabular}
    \end{adjustbox}
\end{table}

In considering the true boundaries specified above and selecting 110 boundaries from posterior computations, sensitivities of approximately $80\%$ are achieved for all diseases except the second one, which surpasses $90\%$. This may appear somewhat subdued in comparison to results from \cite{gao2022spatial}, but is not entirely unexpected. In fact, these lower values are a consequence of introducing the uncertainty in the adjacency matrix. This additional uncertainty, imposed on the adjacency structure, propagates through the model resulting in a significantly wider posterior distribution for spatial random effects. A broader posterior distribution signifies an increased occurrence of ties between random effects across various cancers and regions. Consequently, a greater number of ties produces fewer difference boundaries, which, in turn, yield lower sensitivity values. Conversely, when more boundaries are detected, the specificity is consequently reduced. Table~\ref{tab:bd_dd} also presents the results concerning the directed and undirected disease graphs, where similar considerations apply. Table~\ref{tab:cross_sens_spec} in the Appendix presents the values of specificity and sensitivity for the three cancer disease graph choices cross-cancer boundary detection. Also in this case analogous aspects and considerations to those presented here are drawn.

Regarding the adjacency model, Table~\ref{tab:adj_det} reports values of specificity and sensitivity regarding the detection of boundaries. Apparently, higher values of sensitivity are obtained for all the diseases and for all the disease graph models, while specificity values are smaller. High sensitivity and low specificity imply that our model is sensitive to detecting true positives, but may also misidentify some negative cases as positive (false positives). While false positives are undesirable, this does not represent a major issue for our application in spatial health disparities where policy is dictated by true positives.

\begin{table}[t]
    \centering
    \renewcommand{\arraystretch}{1.167}
    \caption{Adjacencies detection for the three disease graph models.}
    \label{tab:adj_det}
    \begin{adjustbox}{max width=\textwidth}
    \begin{tabular}{ccccccccc}
        \hline
         & \multicolumn{2}{c}{Disease 1} & \multicolumn{2}{c}{Disease 2} & \multicolumn{2}{c}{Disease 3} & \multicolumn{2}{c}{Disease 4} \\
        \cmidrule(lr){2-3} \cmidrule(lr){4-5} \cmidrule(lr){6-7} \cmidrule(lr){8-9}
        Disease graph & Specificity & Sensitivity & Specificity & Sensitivity & Specificity & Sensitivity & Specificity & Sensitivity \\
        \hline 
        Unstructured & 0.373 & 1.000 & 0.878 & 0.912 & 0.692 & 0.981 & 0.384 & 1.000 \\
        Directed & 0.431 & 0.999 & 0.720 & 0.926 & 0.743 & 0.976 & 0.412 & 1.000 \\
        Undirected & 0.456 & 0.995 & 0.562 & 0.952 & 0.597 & 0.962 & 0.304 & 1.000 \\
        \hline
    \end{tabular}
    \end{adjustbox}
\end{table}

\begin{figure}[t]
    \centering
    \includegraphics[width = 0.66 \textwidth]{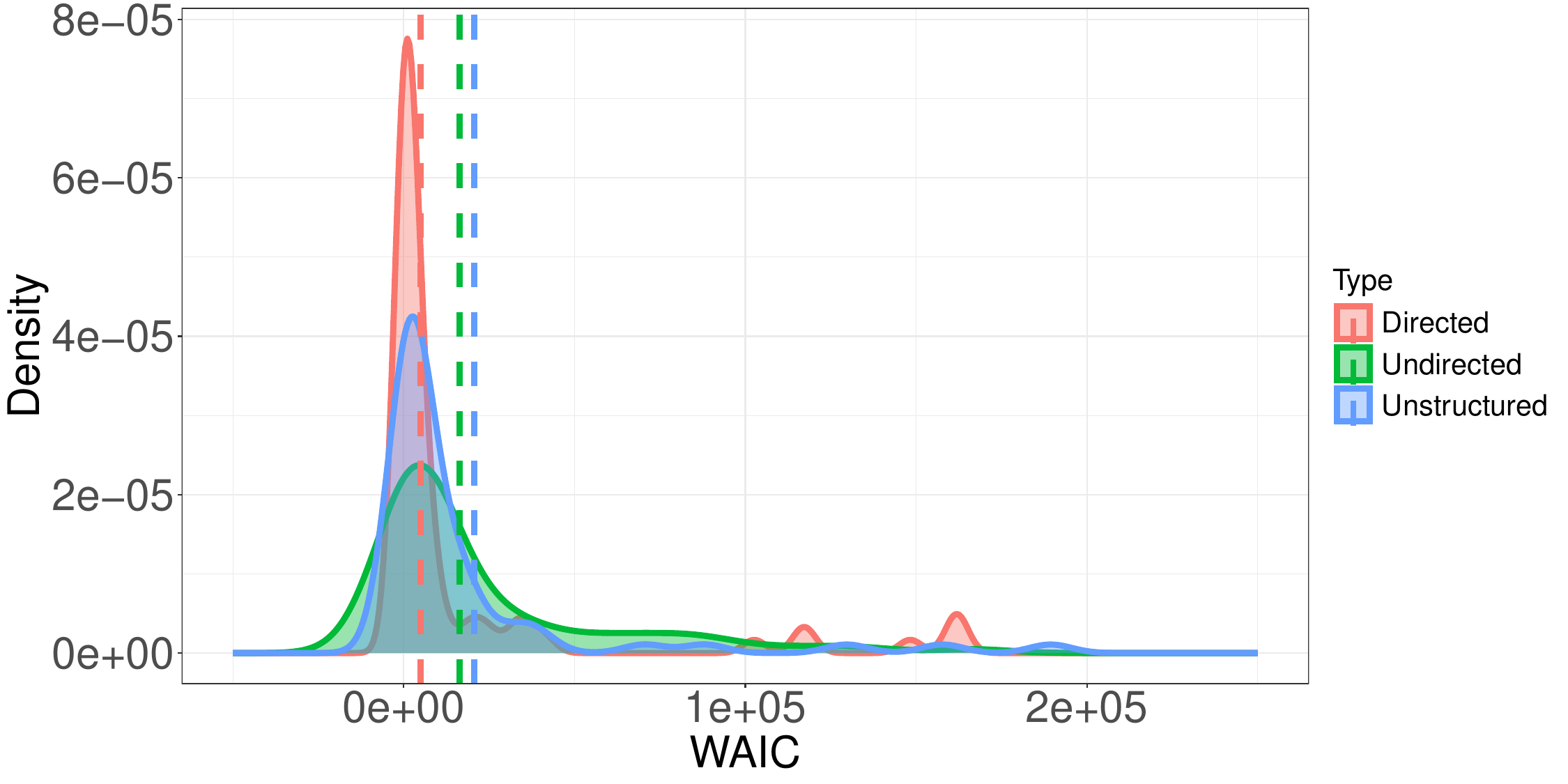}
    \caption{WAIC comparison for different choices of the disease graph.}
    \label{fig:waic_comp}
\end{figure}

Figure~\ref{fig:waic_comp} presents the distributions of WAIC values \citep{watanabe2010asymptotic}, a generalized version of the Akaike information criterion, across multiple simulations for the three disease graphs. This metric gauges the extent of information loss in a given model, the less the better, by effectively addressing the risks of both overfitting and underfitting. Notably, in terms of WAIC values, the directed graph stands out with superior performance, while the other two graphs exhibit closely aligned values. This close resemblance might stem from the fact that the unstructured and undirected graphs bear similarity in interpretation; the former is essentially analogous to an undirected graph where all nodes are interconnected.

\section{SEER cancers analysis}\label{sec: cancer_analysis}

We analyze the data set described in Section~\ref{sec: data} using a Poisson spatial regression model. Specifically, we model the observed counts of incidence, for each county and for each cancer, as $y_{id} \given \bbeta_d, \phi_{id} \overset{ind}{\sim} Pois(E_{id}\exp{(\bx_{id}^{\T}\bbeta_d + \phi_{id})})$ for $i = 1, \dots , 58$ and $d = 1, \dots , 4$. Following the prior specification outlined in the simulation study, we employ the model discussed in Section~\ref{sec: models} to represent $\bgamma$ with adjacency modeling within the MDAGAR model and unstructured disease graph. Here, we draw diagnostic inference based upon $10,000$ posterior samples using MCMC algorithms after discarding an initial $20,000$ iterations as burn-in. Throughout this Section, it is worth noting, we comment on the results by directly referring to specific counties by their names. For a visual reference of the corresponding county names, we refer the reader to the Appendix and the map there depicted in Figure~\ref{fig:named_map}.

\begin{figure}[t]
    \centering
    \includegraphics[width = 0.66\textwidth]{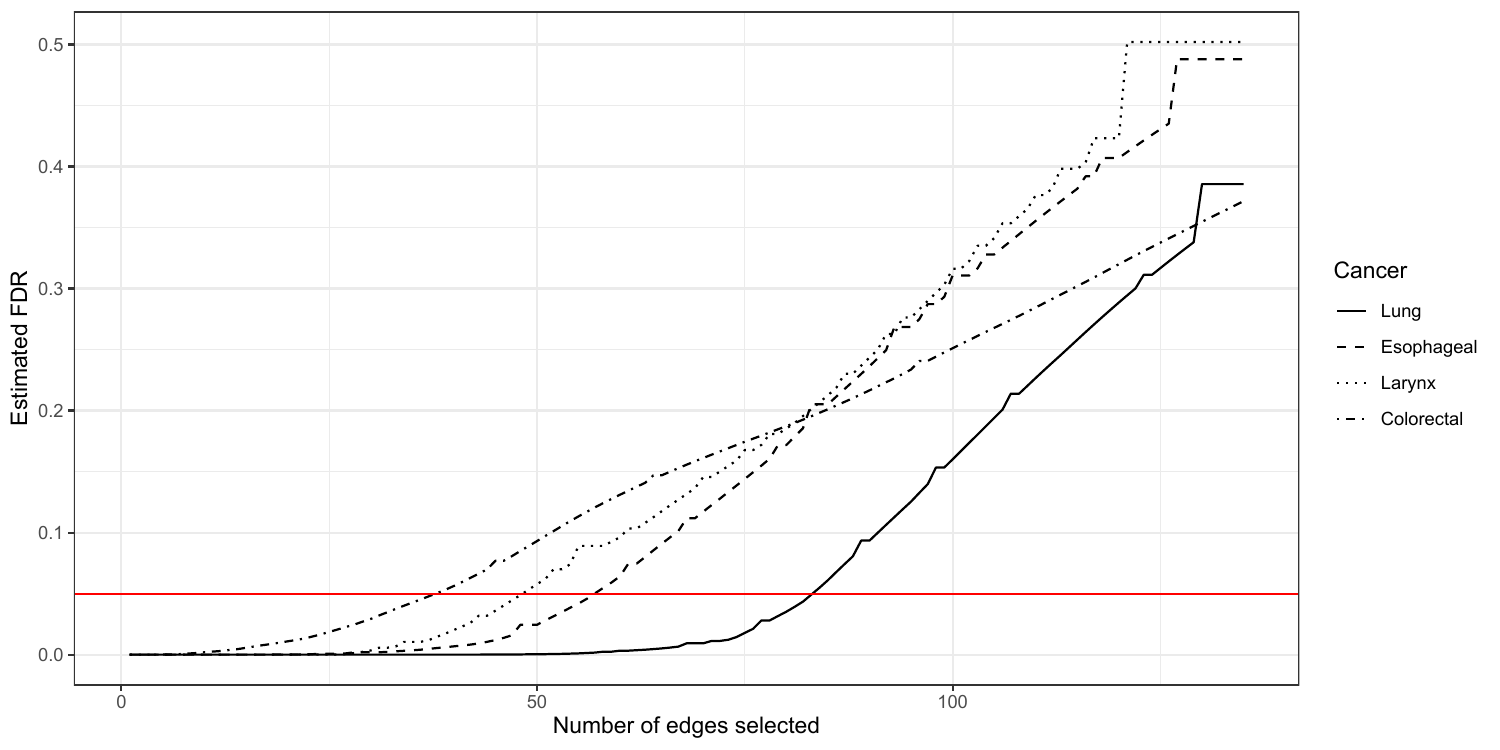}
    \caption{Estimated FDR curves plotted against the number of selected difference boundaries for four cancers}
    \label{fig:FDR}
\end{figure}

We employ a threshold to control for FDR as specified in \eqref{eq:threshold} for detecting difference boundaries for each cancer. Figure~\ref{fig:FDR} illustrates the change in estimated FDR with different numbers of selected edges as differential boundaries for the four cancers individually using adjacency modeling within MDAGAR. We observed analoguous FDR curves for lung, esophageal and larynx cancer leading us to select a similar number of boundaries for these three cancers with a threshold of $\zeta = 0.05$. Conversely, we detect a smaller number of boundaries for colorectal cancer using the same threshold.

\begin{figure}[t]
    \centering
    \includegraphics[width = 0.66 \textwidth]{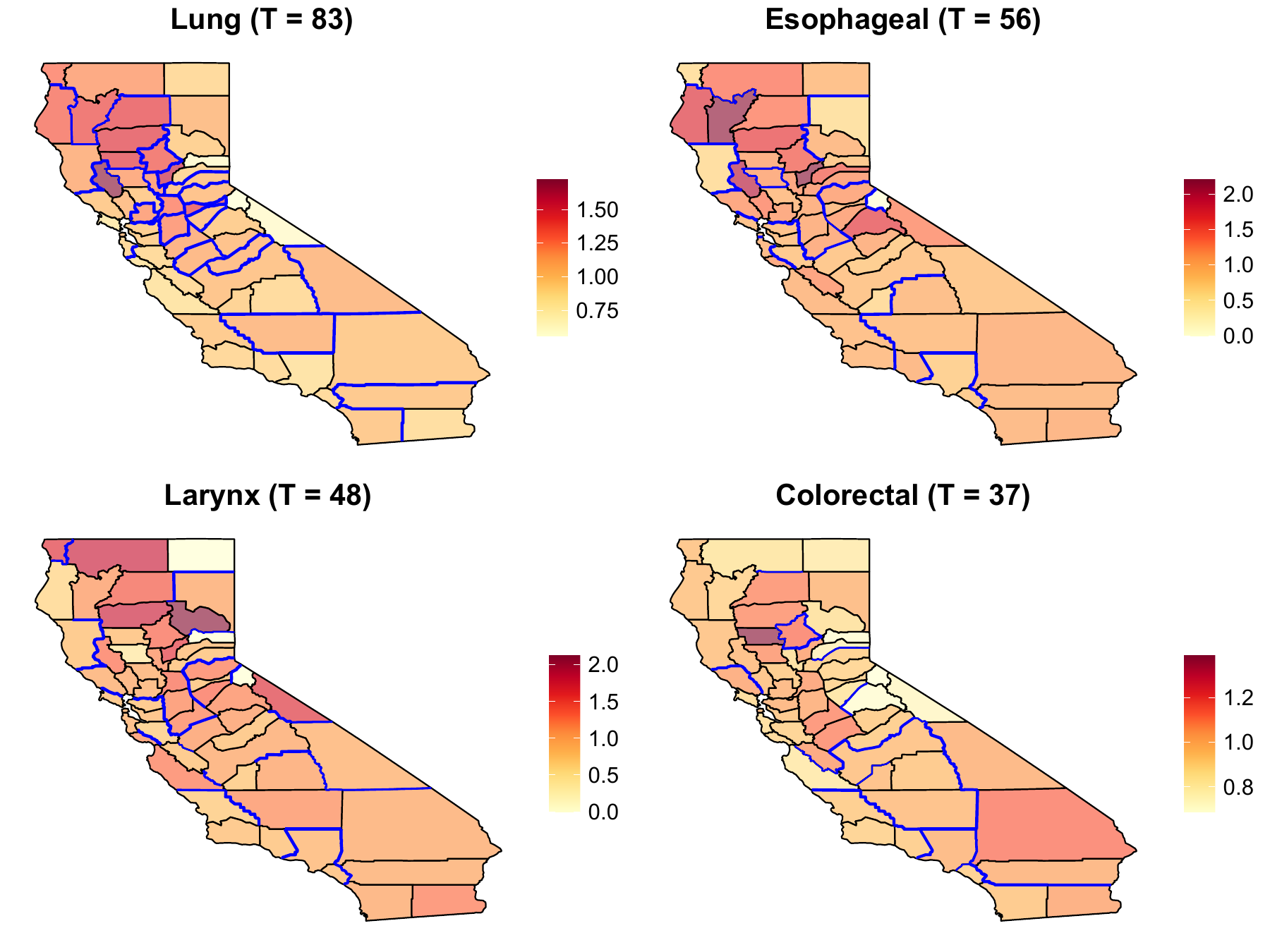}
    \caption{Difference boundaries (highlighted in red) detected by the model in SIR map for four cancers individually when $\zeta = 0.05$. The values in brackets are the number of difference boundaries detected.}
    \label{fig:bd}
\end{figure}

Setting $\zeta = 0.05$ in \eqref{eq:threshold}, Figure~\ref{fig:bd} illustrates the identified difference boundaries (highlighted in red) in the SIR maps for the four types of cancers. The width of the lines on the boundaries signify higher probabilities of detection. Notably, lung cancer exhibits the highest number of detected boundaries, i.e., $126$, with esophageal and larynx cancers demonstrating a much higher number of boundaries, i.e., $119$ and $115$, compared to colorectal cancer with $54$ boundaries. It is worth noting that the detected boundaries for the various cancers tend to form geographical clusters, which bring together regions with similar random effects. 

\begin{figure}[t]
    \centering
    \includegraphics[width = \textwidth]{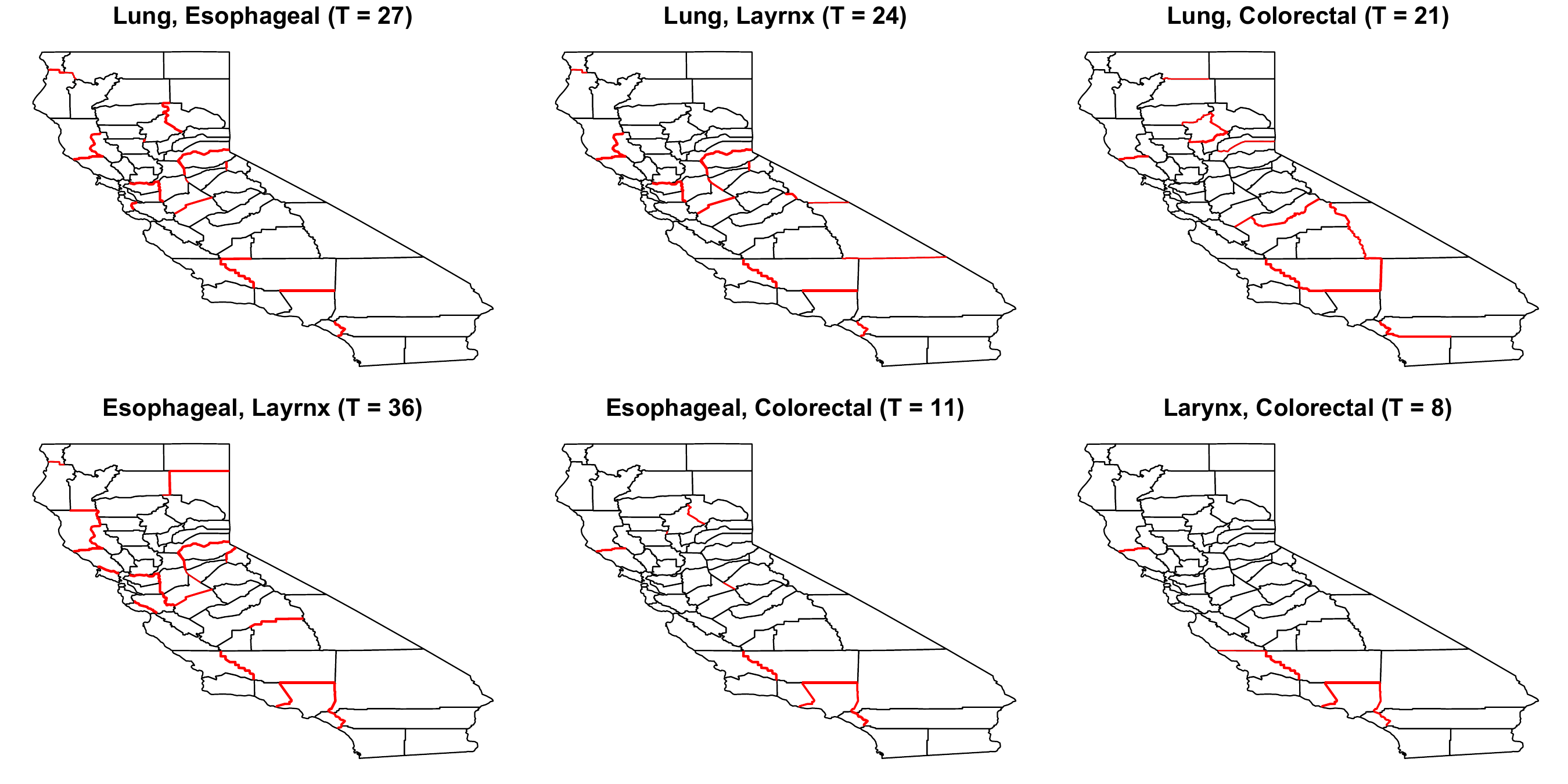}
    \caption{Shared difference boundaries (highlighted in red) detected by the model for each pair of cancers in SIR map when $\zeta = 0.05$. The values in brackets are the number of difference boundaries detected.}
    \label{fig:sbd}
\end{figure}

For instance, in northern California, we observe clustering of certain regions (Tehama, Glenn, Butte, and Yuba) with a high lung SIR value. These regions also exhibit similar patterns for esophageal and larynx cancers indicating spatial association. This is evident from the maps depicting shared difference and mutual cross-difference boundaries in Figure~\ref{fig:sbd}, where these regions display identical random effects. Such consistency is likely attributable to the strong correlation among these cancers. A similar pattern is observed in the central regions, namely Solano, Sacramento, and San Joaquin, with lung SIR values ranging from $60\%$ to $100\%$. The esophageal and larynx cancer maps also exhibit comparable dynamics. However, identifying spatial patterns for colorectal cancer is more challenging. These disparities are likely due to our accounting for smoking standardized differences in both the adjacency and mean models. It is well-known that smoking has a significant influence, particularly on the development of lung, esophageal, and larynx cancers \citep{doll2005mortality}.

For distinguishing boundaries among cancers, we examine both shared difference boundaries and mutual cross-cancer boundaries. The shared difference boundaries refer to the boundaries detected in common across different cancers. Figure~\ref{fig:sbd} illustrates the shared boundaries for each cancer pair, denoted as $P(\phi_{id} \neq \phi_{jd}, \phi_{id'} \neq \phi_{jd'} \given \by)$ where $d \neq d'$. We have observed consistent detection of specific boundaries whenever the lung is affected, including Siskiyou-Modoc, Siskiyou-Shasta, Nevada-Placer and Santa Clara-Stanislaus. Similarly, when examining all cancers, similar patterns emerge in Los Angeles county. Notably, the Los Angeles-San Bernardino border, as well as Orange-San Bernardino, Orange-Riverside and Riverside-San Diego, are consistently identified.

\begin{figure}[t]
    \centering
    \includegraphics[width = \textwidth]{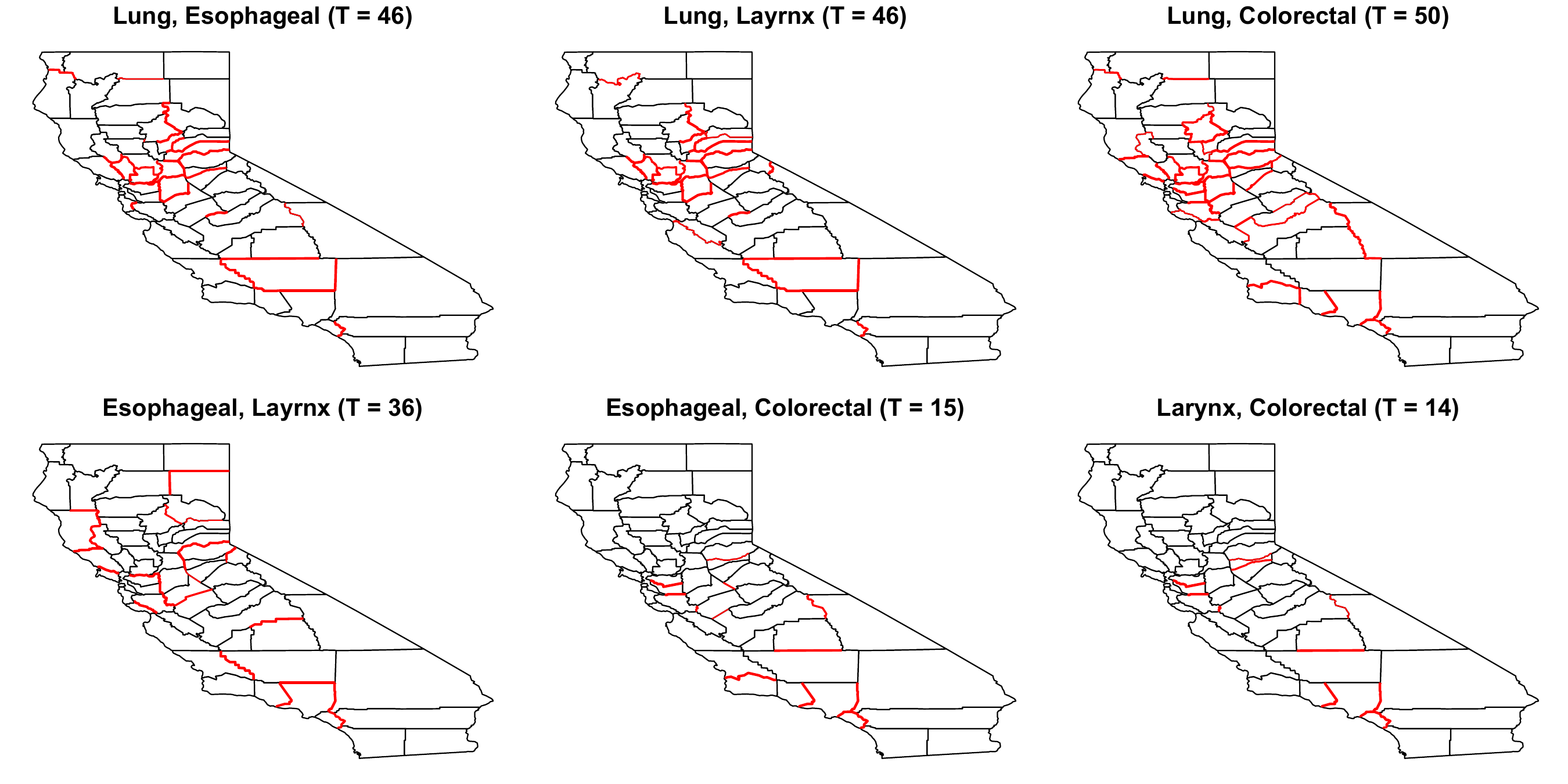}
    \caption{Mutual cross-difference boundaries (highlighted in red) detected by the model for each pair of cancers in SIR map when $\zeta = 0.05$. The values in brackets are the number of difference boundaries detected.}
    \label{fig:mbd}
\end{figure}

We introduce a mutual cross-cancer boundary denoted by $P(\phi_{id} \neq \phi_{jd'}, \phi_{id'} \neq \phi_{jd} \given \by)$, where $i \sim j$ and $i < j$, to ascertain boundaries for cross-cancer differences. This boundary effectively segregates the effects pertaining to distinct cancers across adjacent counties (see Figure~\ref{fig:mbd}). Upon analysis, we observe distinct variations between counties in terms of lung-esophageal and lung-larynx differences. It is evident that the number of detected borders is significantly higher for lung-esophageal, lung-larynx and esophageal-larynx cancers compared to other types. Moreover, when considering these three cases, Orange County demonstrates clear separation from Riverside and San Diego counties, indicating notable differences in their respective lung-related conditions.

Including explanatory variables affect difference boundary detection, leading to either larger or smaller numbers of such boundaries. For instance, a covariate intensifies the contrast in residual spatial effects between two adjacent counties, a difference boundary is more likely to be observed between them. Conversely, if a covariate reduces the disparity in residual spatial effects between neighboring counties, the presence of a difference boundary may diminish. The former occurs, for example, for lung, esophageal and larynx boundary detection individually, while the latter is probably seen for colorectal for which we detect fewer boundaries. The impact of covariates on the number of difference boundaries depends on how they affect the variation in rates across neighboring counties. It is important to note that while covariates always absorb some spatial effects, the relationships among cancers, regions, and covariates are intricate, and a definitive pattern is not always discernible.

\begin{figure}[t]
    \centering
    \includegraphics[width = 0.66 \textwidth]{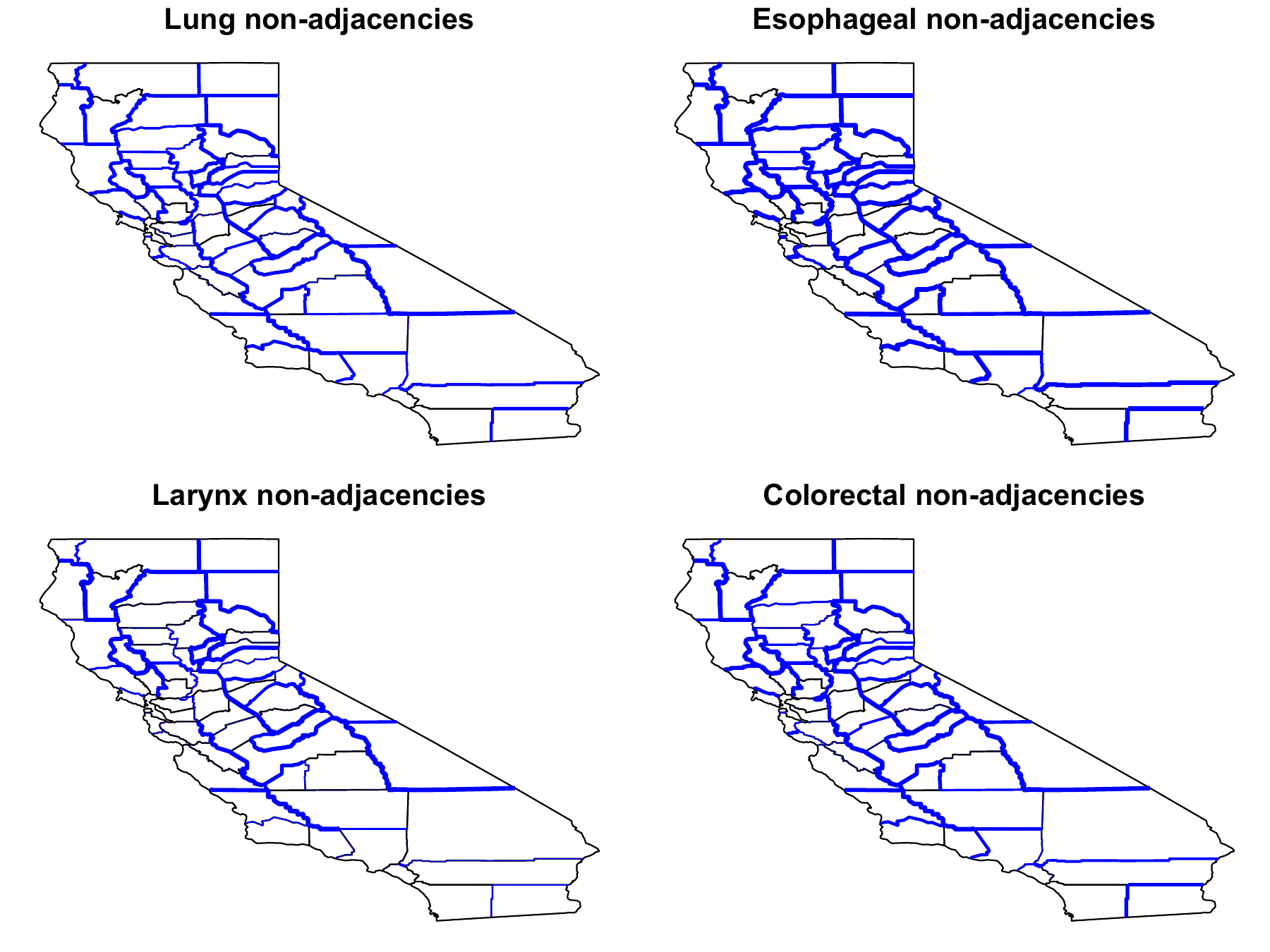}
    \caption{Non-adjacencies (shown in blue) overlaid on smoking rates. The thickness of the lines is proportional to the probability of being considered as a non-adjacency}
    \label{fig:adj_det}
\end{figure}

Turning to adjacency modeling, in the following we will say that a non-adjacency is detected based on the following probabilities:
\begin{equation}\label{eq:det_prob}
    P(w_{d,ij} = 0 \given \by, i \sim j)
\end{equation}
where $w_{d,ij}$ represents the $(i,j)$-th element of the  adjacency matrix corresponding to the $d$-th cancer. Figure~\ref{fig:adj_det} depicts a map of California with these probabilities represented by the thickness of the boundaries. As is evident in Figure~\ref{fig:adj_det}, the majority of the borders are not colored in blue. This absence of blue indicates that the probability of $w_{d,ij}$ being $0$ is null, implying that adjacency detection is performed for those geographic boundaries. Considering all the cancers under investigation, the detected non-adjacencies are $32$ for lung, $12$ for esophageal, $31$ for larynx and $29$ for colorectal. These correspond, respectively, to approximately $23\%$, $9\%$, $22\%$ and $21\%$ of the total number of borders. 

In general, it is evident that the detected non-adjacencies predominantly occur in counties with high smoking rates. Conversely, counties with lower smoking rates consistently exhibit zero detected non-adjacencies. This pattern strongly suggests that smoking serves as a suitable dissimilarity metric for identifying these boundaries, as the estimated boundaries closely align with significant variations in the smoking covariate. However, it is worth noting that a few of the identified boundaries do not show evidence of separating areas with differing smoking rates. Examples of such boundaries include Trinity-Shasta, Siskiyou-Shasta, Trinity-Tehama, Lake-Colusa, and Lake-Glenn. These cases indicate that factors other than smoking may contribute to the observed differences in cancer occurrence or that the boundaries in these particular instances are not solely determined by smoking rates. This occurrence may be attributed to the significant variations in cancer counts between the counties, as the values can differ greatly. Therefore, the presence or absence of detected non-adjacencies in these cases may be influenced by factors other than smoking rates alone. 

Through our comprehensive analysis, we discover that there exists a strong correlation among lung, esophageal, and larynx cancers. The findings of this study highlight the interconnected nature of these specific types of cancers, suggesting common risk factors or underlying mechanisms that contribute to their occurrence. A compelling aspect of this association is revealed through the visual representations of the three maps we have presented. These maps vividly illustrate remarkably similar spatial clusters for lung, esophageal, and larynx cancers. The clustering patterns observed strongly imply that the corresponding random effects associated with these cancers share identical values. This consistency in spatial distribution reinforces the notion that these cancers are closely related and potentially influenced by shared factors. The spatial clusters observed in these maps indicate that the incidence and prevalence of lung, esophageal, and larynx cancers tend to occur in close proximity to one another. This spatial pattern may be influenced by a multitude of factors, including geographical characteristics, socioeconomic factors, or specific cultural and lifestyle factors prevalent in certain regions. Understanding the spatial relationships can provide valuable insights into the complex interplay between environmental and individual risk factors.

In the Appendix two distinct settings of the model are presented for this study. The first setting involves incorporating covariates solely in the adjacency model, while the second setting integrates covariates exclusively in the mean structure while maintaining fixed adjacencies. These additional maps provide valuable insights into the impact of covariates on the overall analysis. By examining the effects of covariates within the adjacency model, we gain a better understanding of how specific factors contribute to the spatial patterns observed in the data. Similarly, the inclusion of covariates solely in the mean structure, with fixed adjacencies, allows us to investigate the influence of these variables on the overall mean values of the observed phenomena.

\section{Discussion}\label{sec: discussion}

The developments in this manuscript address statistical methods for detecting spatial health disparities based upon mapping disease rates and modeling difference boundaries. The need for formal modeling in ascertaining health disparities has been articulated in the literature \citep[see, e.g.,][and references therein]{rao2023}, but formal model based approaches for detecting geographic disparities remain relatively scarce. The inferential framework developed here achieves full probabilistic uncertainty quantification. The idea is to use a flexible (nonparametric) prior on the spatial effects that will endow discrete masses to spatial random effects to detect discrepancies among neighbors while also allowing spatial smoothing across neighbors as is customary in disease mapping. 

Our methodology extends some recent work by \cite{gao2022spatial} in multiple directions. As there, we also deal with multiple diseases, possibly associated among themselves, but now explicitly address multivariate dependencies more flexibly using posited graphical models. Second, we address a possible limitation of \cite{gao2022spatial} where factors directly influencing difference boundaries need to be evaluated using their inclusion and exclusion in models with subsequent model assessment. Instead, we introduce the possible incorporation of such factors in the adjacency modeling to evaluate its effects. This enriches earlier statistical frameworks for ``areal wombling'' that have attempted similar adjacency models with limited effectiveness since statistical significance of boundary effects is difficult to establish with such hierarchically embedded information. Our framework, on the other hand, allows flexible modeling of boundaries (as in ``wombling''), but uses a Bayesian FDR approach to ascertain spatial difference boundaries.      

Our application to the SEER database leads to data-based discoveries with regard to the impact of associations among cancers in detecting spatial disparities. The high correlation among lung, esophageal, and larynx cancers carries significant implications for public health initiatives and cancer prevention strategies. The associations among these cancers suggest that interventions targeting one of these malignancies could potentially have a positive impact on reducing the incidence and burden of the others. By recognizing the shared patterns and risk factors, healthcare professionals and policymakers can develop more targeted and effective interventions. In response to these findings, a multifaceted approach should be employed to address this correlation and reduce the overall burden of these cancers. This approach should encompass public awareness campaigns to educate individuals about the shared risk factors and promote healthy lifestyle choices.

\begin{supplement}
    Supplementary material is provided in the Appendix. It includes the sampling scheme, consideration on Monte Carlo standard errors, the map with the county names, maps regarding different model settings for the data analysis, additional plots and tables for the simulation. Computer programs to reproduce the numerical experiments and data analyses are available at \href{https://github.com/lucaaiello/graphical-models-for-boundary-detection/tree/main}{https://github.com/lucaaiello/graphical-models-for-boundary-detection/tree/main}
\end{supplement}

\bibliographystyle{imsart-nameyear}
\bibliography{biblio}

\appendix

\section{Computational details}\label{sec:comp_det}

Our models rely on posterior inference, which is achieved through MCMC posterior simulations. To accomplish this, we employ the blocked Gibbs Sampler with some adaptive and some classic Metropolis-Hastings steps, based on the work by \cite{gamerman2006markov} and \cite{miasojedow2013adaptive}. We decided not to employ an adaptive strategy for $\bgamma$ due to its high-dimensionality and for $\bA$ due to the particular sampling step involving it. This approach allows us to achieve the desired acceptance rates while updating all random parameters in our model. It is important to note that $\phi_{i}$ corresponds to $\theta_{u_{i}}$, and we update $\phi_{i}$ by updating both $\btheta$ and $u_{i}$, and, ultimately, that $\bet$ is constrained to be between $0$ and a positive constant $M$. The MCMC algorithm proceeds through the following steps, with a noteworthy observation that step 8 is further subdivided into three parts, depending on the selected model used to represent disease dependence.

\begin{noendalgorithm}
    \hrulefill
    \vspace{-0.25em}
    \caption{Posterior inference of $\btheta$, $\bbeta$, $\tau_s$, $\bV$, $\bgamma$, $\brho$, $\bet$ and $\bA$ or $\balpha$ or $\rho_{dis}$ based on our models}\label{alg:amwg}
    \vspace{-1em}
    \hrulefill
    \begin{algorithmic}[1]
        
        \State update $\bbeta \given \btheta$
        \begin{enumerate}[(a)]
            \item sample candidate $\bbeta^*$ from $\Norm(\bbeta, \xi_{\bbeta})$
            \item accept $\bbeta^*$ with probability
            \begin{equation*}
                \min\lcur 1, \frac{\exp{\lrnd \sum_{k=1}^K \sum_{i,d:\phi_{id}=\theta_{k}} y_{id}(\bx_{id}^\T \bbeta_d^* + \phi_{id}) - E_{id} \exp{\lrnd \bx_{id}^\T \bbeta_d^* + \phi_{id} \rrnd} - \frac{{\bbeta_d^*}^\T\bbeta_d^*}{2\sigmasq_{\beta}}\rrnd}}{\exp{\lrnd \sum_{k=1}^K \sum_{i,d:\phi_{id}=\theta^*_{k}} y_{id} (\bx_{id}^\T \bbeta_d + \phi_{id}) - E_{id} \exp{\lrnd \bx_{id}^\T \bbeta_d + \phi_{id} \rrnd}- \frac{\bbeta_d^\T\bbeta_d}{2\sigmasq_{\beta}} \rrnd}} \rcur
            \end{equation*}
            \item adapt $\xi_{\bbeta}$
        \end{enumerate}

        \algstore{bkbreak}
    \end{algorithmic}
\end{noendalgorithm}

\begin{notoprulealgorithm}
    \caption*{}
    \begin{algorithmic}
        \algrestore{bkbreak}

        \State update $\btheta \given \bbeta, \tau_s$
        \begin{enumerate}[(a)]
            \item sample candidate $\btheta^*$ from $\Norm(\btheta, \xi_{\btheta})$
            \item accept $\btheta^*$ with probability
            \begin{equation*}
                \min\lcur 1, \frac{\exp{\lrnd \sum_{k=1}^K \sum_{i,d:\phi_{id}=\theta^*_{k}} y_{id} (\bx_{id}^\T \bbeta_d + \phi_{id}) - E_{id} \exp{\lrnd \bx_{id}^\T \bbeta_d + \phi_{id} \rrnd} - \frac{{\theta^*_k}^2}{2}\tau_s\rrnd}}{\exp{\lrnd \sum_{k=1}^K \sum_{i,d:\phi_{id}=\theta_{k}} y_{id} (\bx_{id}^\T \bbeta_d + \phi_{id}) - E_{id} \exp{\lrnd \bx_{id}^\T \bbeta_d + \phi_{id} \rrnd} - \frac{\theta_{k}^2}{2}\tau_s\rrnd}} \rcur
            \end{equation*}
            \item adapt $\xi_{\btheta}$
        \end{enumerate}
        
        \State update $\gamma_{id} \given \bbeta, \btheta, \brho, \bet, \bA$, $i = 1,\dots,n$, $d = 1,\dots,q$
        \begin{enumerate}[(a)]
            \item sample candidate $\gamma^*_{id}$ from $\Norm(\gamma_{id}, \xi_{\bgamma})$ 
            \item compute the corresponding candidate $u^*_{id} = \sum_{k=1}^K k I\lrnd\sum_{t=1}^{k-1}p_t < F^{(id)}(\gamma^*_{id}) < \sum_{t=1}^{k}p_t\rrnd$
            \item accept $\gamma^*_{id}$ with probability
            \begin{equation*}
                \min\lcur 1, \frac{\exp{\lrnd y_{id}(\bx_{id}^\T \bbeta_d + \theta_{u^*_{id}}) - E_{i} \exp{(\bx_{id}^\T \bbeta_d + \theta_{u^*_{id}})} - \frac{\bgamma^{*T}\bSigma^{-1}_{\bgamma}\bgamma^{*}}{2} \rrnd}}{\exp{\lrnd y_{id}(\bx_{id}^\T \bbeta_d + \theta_{u_{id}}) - E_{id} \exp{(\bx_{id}^\T \bbeta_d + \theta_{u_{id}})} - \frac{\bgamma^{T}\bSigma^{-1}_{\bgamma}\bgamma}{2} \rrnd}} \rcur
            \end{equation*}    
        \end{enumerate}

        \State update $\bV \given \bbeta, \btheta$
        \begin{enumerate}[(a)]
            \item let $\widetilde{\bV} = \text{logit}(\bV)$ and sample $\widetilde{\bV}^*$ from $\Norm(\widetilde{\bV}, \xi_{\widetilde{\bV}})$ then $\bV^* = \frac{\exp{(\widetilde{\bV}^*)}}{1 + \exp{(\widetilde{\bV}^*)}}$
            \item compute the corresponding candidate $\bp^* = (p^*_1,\dots,p^*_K)$ and $\bu = \{u^*_{id}\}$
            \item accept $\bV^*$ with probability
            \begin{equation*}
                \min\lcur 1, \frac{\exp{\lrnd \sum_{i,d} y_{id} (\bx_{id}^\T \bbeta_d + \theta_{u^*_{id}}) - E_{id} \exp{(\bx_{id}^\T \bbeta_d + \theta_{u^*_{id}})} \rrnd}\prod_{k=1}^K(1-V^*_k)^{(\alpha-1)} \prod_{k=1}^K V^*_k (1-V^*_k)}{\exp{\lrnd \sum_{i,d} y_{id} (\bx_{id}^\T \bbeta_d + \theta_{u_{id}}) - E_{id} \exp{(\bx_{id}^\T \bbeta_d + \theta_{u_{id}})} \rrnd}\prod_{k=1}^K(1-V_k)^{(\alpha-1)} \prod_{k=1}^K V_k (1-V_k)} \rcur
            \end{equation*}    
            \item adapt $\xi_{\widetilde{\bV}}$
        \end{enumerate}

        \State sample $\tau_s \given \btheta, \beta_0$ from
        \begin{equation*}
            \Gamma \lrnd a_s + \frac{K}{2}, b_s + \frac{\sum_{k=1}^K \theta_k^2}{2} \rrnd
        \end{equation*}
        
        \State update $\brho \given \bgamma, \bet, \bA$
        \begin{enumerate}[(a)]
            \item let $\widetilde{\brho} = \text{logit}(\brho)$ and sample $\widetilde{\brho}^*$ from $\Norm(\widetilde{\brho}, \xi_{\widetilde{\brho}})$ then $\brho^* = \frac{\exp{(\widetilde{\brho}^*)}}{1 + \exp{(\widetilde{\brho}^*)}}$
            \item update $\bSigma^*_{\bgamma}$ with the proposed $\brho^*$
            \item accept $\widetilde{\brho}^*$ with probability 
            \begin{equation*}
                \min\lcur 1, \frac{\lvert \bSigma^*_{\bgamma} \rvert^{-\frac{N}{2}}  \exp{\lrnd - \frac{\bgamma^{T}\bSigma^{*-1}_{\bgamma}\bgamma}{2} \rrnd} \prod_{d=1}^q \rho^*_d(1-\rho^*_d)}{\lvert \bSigma_{\bgamma} \rvert^{-\frac{N}{2}}  \exp{\lrnd - \frac{\bgamma^{T}\bSigma^{-1}_{\bgamma}\bgamma}{2} \rrnd}\prod_{d=1}^q \rho_d(1-\rho_d)} \rcur
            \end{equation*}
            \item adapt $\xi_{\widetilde{\brho}}$
        \end{enumerate}

        \State update $\bet \given \bgamma, \brho, \bA$
        \begin{enumerate}[(a)]
            \item let $\widetilde{\bet} = \log \lrnd \frac{\bet}{M-\bet} \rrnd $ and sample $\widetilde{\bet}^*$ from $\Norm(\widetilde{\bet}, \xi_{\widetilde{\bet}})$ then $\bet^* = \frac{M\exp{(\widetilde{\bet}^*)}}{1 + \exp{(\widetilde{\bet}^*)}}$
            \item update $\bSigma^*_{\bgamma}$ with the proposed $\bet^*$
            \item accept $\widetilde{\bet}^*$ with probability 
            \begin{equation*}
                \min\lcur 1, \frac{\lvert \bSigma^*_{\bgamma} \rvert^{-\frac{N}{2}}  \exp{\lrnd - \frac{\bgamma^{T}\bSigma^{*-1}_{\bgamma}\bgamma}{2} \rrnd} \prod_{d=1}^q \frac{\exp{(\widetilde{\eta}^*_{d})}}{(1 + \exp{(\widetilde{\eta}^*_{d}))^2}}}{\lvert \bSigma_{\bgamma} \rvert^{-\frac{N}{2}}  \exp{\lrnd - \frac{\bgamma^{T}\bSigma^{-1}_{\bgamma}\bgamma}{2} \rrnd}\prod_{d=1}^q \frac{\exp{(\widetilde{\eta}_{d})}}{(1 + \exp{(\widetilde{\eta}_{d}))^2}}} \rcur
            \end{equation*}
            \item adapt $\xi_{\widetilde{\bet}}$
        \end{enumerate}
        
        \algstore{bkbreak}
        
    \end{algorithmic}
\end{notoprulealgorithm}

\begin{notoprulealgorithm}
    \caption*{}
    \begin{algorithmic}
        \algrestore{bkbreak}

        \State 
        \textbf{Unstructured} update $\bA \given \bgamma, \brho, \bet$
        \begin{enumerate}[(a)]
            \item let $\widetilde{a}_{dd} = \log(a_{dd})$ and sample candidates $\widetilde{a}^*_{dd}$ from  $\Norm(\widetilde{a}_{dd}, \xi_{\widetilde{\ba}})$
            \item for off-diagonal elements $a_{dh}$, $d \neq h$, $a^*_{dh}$ are sampled from $\Norm(\widetilde{a}_{dh}, \xi_{\ba})$
            \item update $\bSigma^*_{\bgamma}$ with the proposed $\bA^*$
            \item accept $\bA^*$ with probability
            \begin{equation*}
                \min\lcur 1, \frac{\lvert \bSigma^*_{\bgamma} \rvert^{-\frac{N}{2}}  \exp{\lrnd - \frac{\bgamma^{T}\bSigma^{*-1}_{\bgamma}\bgamma}{2} \rrnd} p(\bA^*)\prod_{d=1}^q a^*_{dd}}{\lvert \bSigma_{\bgamma} \rvert^{-\frac{N}{2}}  \exp{\lrnd - \frac{\bgamma^{T}\bSigma^{-1}_{\bgamma}\bgamma}{2} \rrnd}p(\bA)\prod_{d=1}^q a_{dd}} \rcur
            \end{equation*}
        \end{enumerate}
        
        \Statex
        \textbf{Directed} sample $\balpha_{d} \given \bgamma, \brho_{d}$ from
        \begin{equation*}
            \Norm(\bH_d h_d,H_d)
        \end{equation*}
        where $\bH_d = \lrnd \bdelta_d^\T Q_{d} \bdelta_d + \bSigma_{\balpha}^{-1} \rrnd^{-1}$ and $\bh_{d} = \bdelta_d^\T Q_{d} \bgamma_d + \bSigma_{\balpha}^{-1} \bmu_{\balpha}$. 
        Here $\bdelta_d = (F_h)_{h \in N_d}$, $F_h = (\bgamma_h,\bzeta_h)$, $N_d = \{h: h \sim d, h < d \}$ and $\bzeta_h = \lrnd \sum_{j\sim 1} \bgamma_{h,j}, \dots, \sum_{j \sim n} \bgamma_{h,j} \rrnd$.
        Then update $\bA_{d,h}$ matrices and consequently $\bSigma_{\bgamma}$.
        \vspace{2.5em}
        \Statex
        \textbf{Undirected} update $\rho_{dis} \given \bgamma, \brho, \bet$
        \begin{enumerate}[(a)]
            \item let $\widetilde{\rho}_{dis} = \log \lrnd \frac{\rho_{dis} + 1}{1-\rho_{dis}} \rrnd$ and sample $\widetilde{\rho}_{dis}^*$ from $\Norm(\widetilde{\rho}_{dis}, \xi_{\widetilde{\rho}_{dis}})$ then $\rho_{dis}^* = \frac{\exp{(\widetilde{\rho}_{dis}^*)}-1}{1 + \exp{(\widetilde{\rho}_{dis}^*)}}$
            \item update $\bLambda^*_{dis}$ with the proposed $\rho_{dis}^*$ and then $\bSigma^*_{\bgamma}$
            \item accept $\widetilde{\rho}_{dis}^*$ with probability 
            \begin{equation*}
                \min\lcur 1, \frac{\lvert \bSigma^*_{\bgamma} \rvert^{-\frac{N}{2}}  \exp{\lrnd - \frac{\bgamma^{T}\bSigma^{*-1}_{\bgamma}\bgamma}{2} \rrnd} \frac{\exp{(\widetilde{\rho}^*_{dis})}}{(1 + \exp{(\widetilde{\rho}^*_{dis}))^2}}}{\lvert \bSigma_{\bgamma} \rvert^{-\frac{N}{2}}  \exp{\lrnd - \frac{\bgamma^{T}\bSigma^{-1}_{\bgamma}\bgamma}{2} \rrnd} \frac{\exp{(\widetilde{\rho}_{dis})}}{(1 + \exp{(\widetilde{\rho}_{dis}))^2}}} \rcur
            \end{equation*}
            \item adapt $\xi_{\widetilde{\rho}_{dis}}$
        \end{enumerate}
        
    \end{algorithmic}
     \hrulefill
\end{notoprulealgorithm}

\newpage

\section{Markov Chain Monte Carlo Standard Error}

Monte Carlo estimates, whether for MCMC output or any other purpose, should always be accompanied by Monte Carlo standard error. This means that if the central limit theorem applies, the corresponding covariance matrices' estimates must be reported, either directly or indirectly. Due to the correlation between the obtained samples, these quantities necessitate more advanced tools compared to typical sample estimators. It's important to note that the validity of a Markov chain central limit theorem is not guaranteed in all cases, as it depends on the convergence rate of the Markov chain. Nonetheless, this assumption is also made when employing various convergence diagnostics. In this context, we present the estimates, by using \texttt{mcmcse R} package \citep{mcmcse}, of the standard error and the effective sample size, based on \cite{gong2016practical}, to demonstrate that we can maintain control over the Monte Carlo error by applying a predefined threshold \cite{vats2019multivariate}.

Table \ref{tab:mcmcse} presents the estimates of the highest level of the parameters hierarchy. Additionally, the Monte Carlo standard error is reported, revealing that the highest value is $0.059$, the only one surpassing the standard threshold of $0.05$.

\begin{table}[h]
    \centering
    \caption{Estimates and Monte Carlo standard errors for the high level parameters of the hierarchy}
    \label{tab:mcmcse}
    \begin{tabular}{ccc}
        \hline
        Variable & Estimate & Standard Error \\
        \hline
        $\beta_0$ & 0.046 & 0.001 \\
        $\tau_s$ & 6.830 & 0.012 \\
        $\theta_1$ & 0.217 & 0.001 \\
        $\theta_2$ & 0.344 & 0.002 \\
        $\theta_3$ & 0.042 & 0.001 \\
        $\theta_4$ & -0.022 & 0.008 \\
        $\theta_5$ & 0.109 & 0.002 \\
        $\theta_6$ & -0.067 & 0.001 \\
        $\theta_7$ & 0.004 & 0.006 \\
        $\theta_8$ & -0.145 & 0.018 \\
        $\theta_9$ & -0.057 & 0.005 \\
        $\theta_{10}$ & -0.062 & 0.022 \\
        $\theta_{11}$ & 0.398 & 0.042 \\
        $\theta_{12}$ & 0.268 & 0.059 \\
        $\theta_{13}$ & -0.011 & 0.012 \\
        $\theta_{14}$ & -0.169 & 0.015 \\
        $\theta_{15}$ & -0.137 & 0.002 \\
        \hline
    \end{tabular}
\end{table}

Next, we provide the multivariate effective sample size of the chain, computed as $1940.335$ using the formula $B \frac{\lvert \Lambda \rvert^{1/p}}{\lvert \Sigma \rvert^{1/p}}$. Here, $B$ is the total number of iterations post burn-in, $\Lambda$ represents the sample covariance matrix, and $\Sigma$ is an estimate of the Monte Carlo standard error. This computation accounts for the multivariate dependence structure of the process. Comparing our effective sample size to the minimum requirement of $7619$ for achieving a precision level relative to the variability in the target distribution of $0.05$, we observe that ours is significantly lower. However, it is important to note that we computed the precision level of our estimate, obtained with an estimated effective sample size of $1940.335$, which amounts to $0.099$. This precision level is higher than $0.05$ but lower than $0.1$.

\section{Additional figures and tables}

\begin{figure}[htbp!]
    \centering
    \includegraphics[width = 0.7\textwidth]{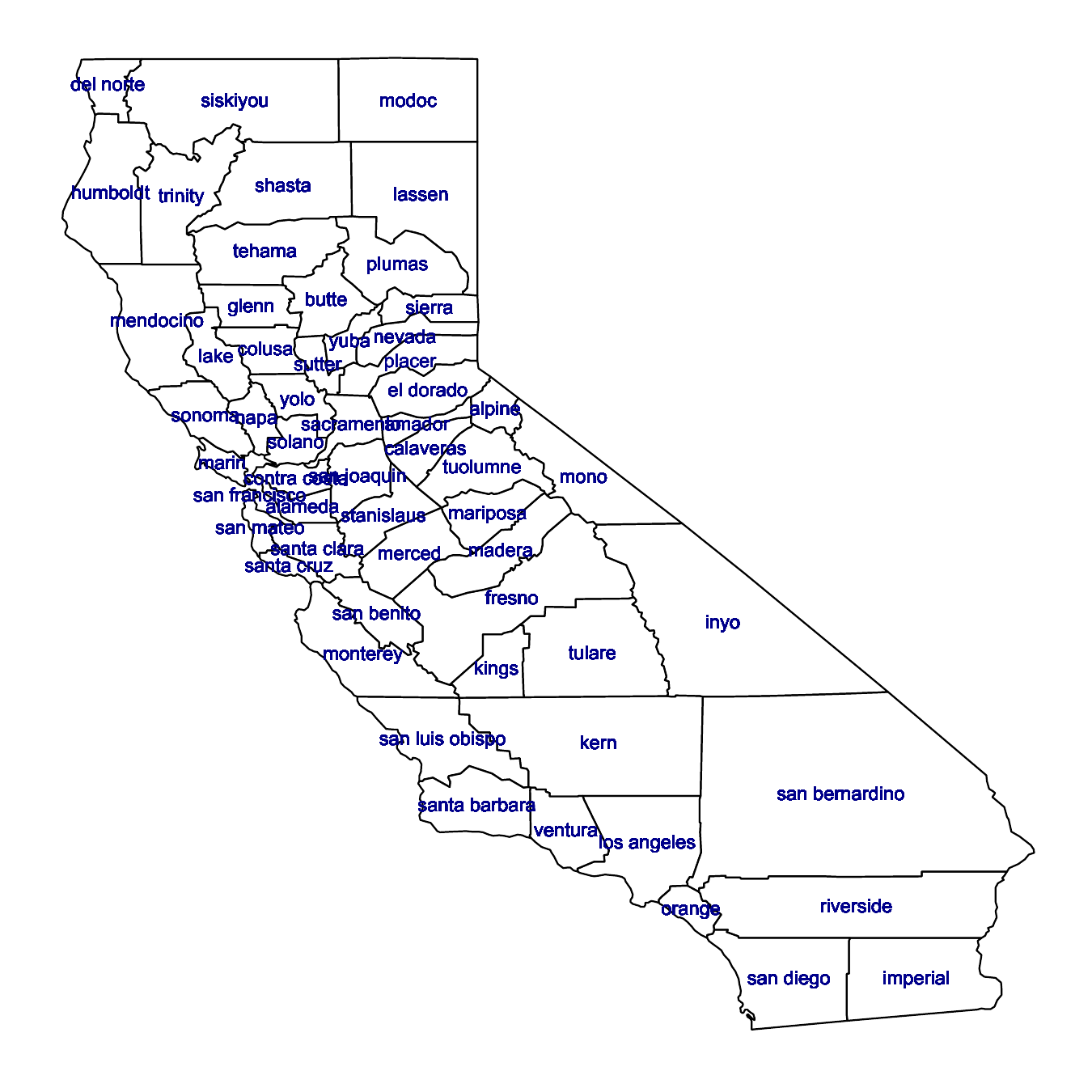}
    \caption{California map with county names labeled}
    \label{fig:named_map}
\end{figure}

\begin{figure}[htbp!]
    \centering
    \includegraphics[width = \textwidth]{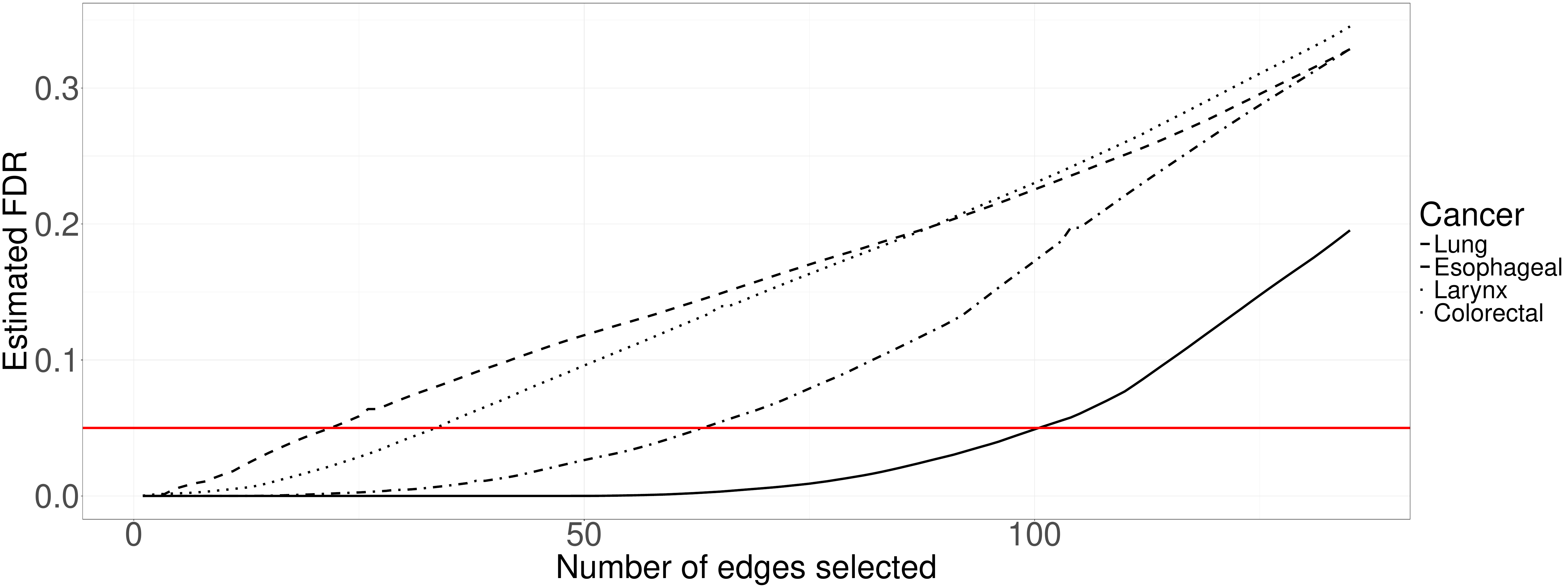}
    \caption{Estimated FDR curves plotted against the number of selected difference boundaries for four cancers in the case study with covariates only in the adjacency model}
    \label{fig:FDR_adj}
\end{figure}

\begin{figure}[htbp!]
    \centering
    \includegraphics[width = 0.75 \textwidth]{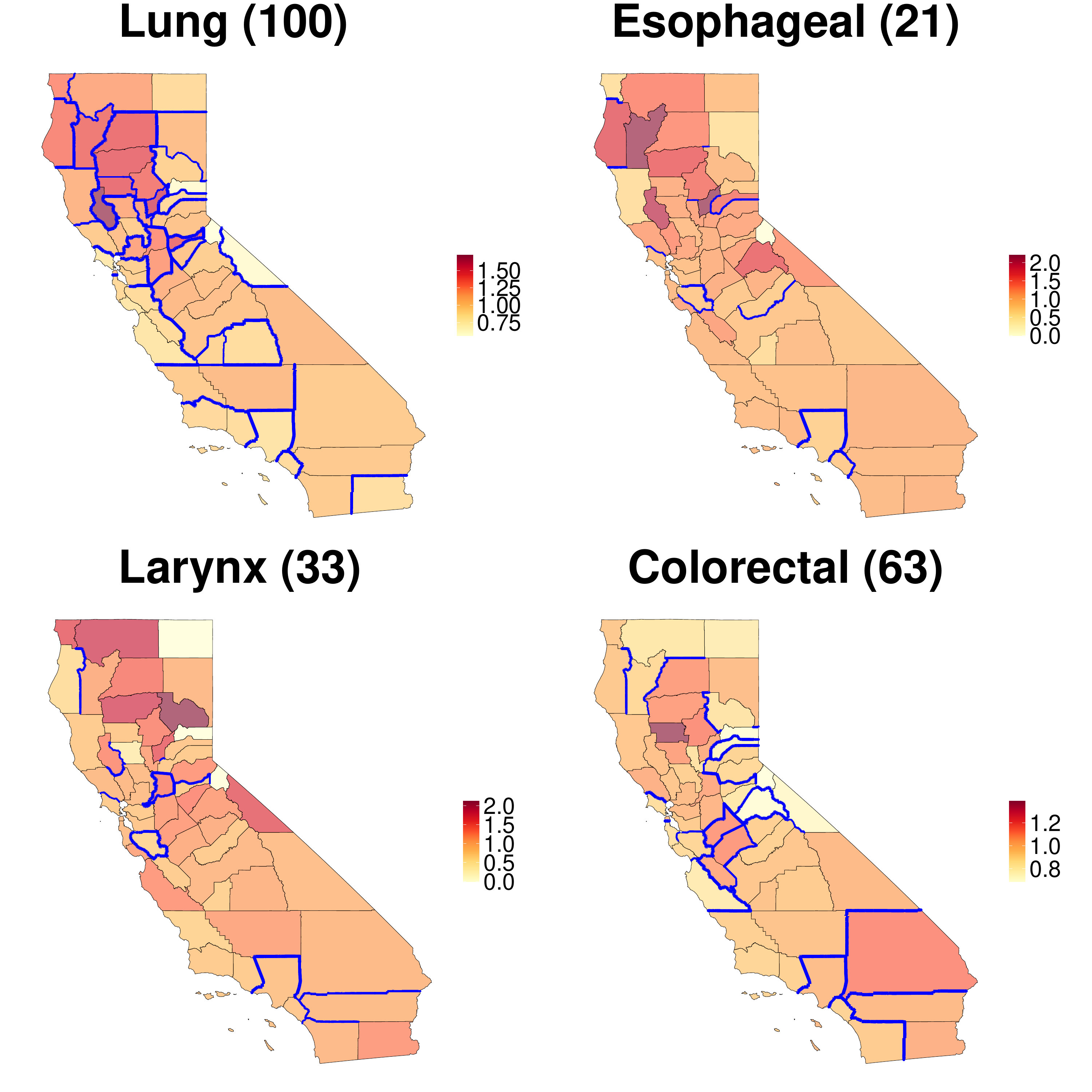}
    \caption{Difference boundaries (highlighted in red) detected by the model, in the case study with covariates only in the adjacency model, in SIR map for four cancers individually when $\zeta = 0.05$. The values in brackets are the number of difference boundaries detected.}
    \label{fig:bd_adj}
\end{figure}

\begin{figure}[htbp!]
    \centering
    \includegraphics[width = \textwidth]{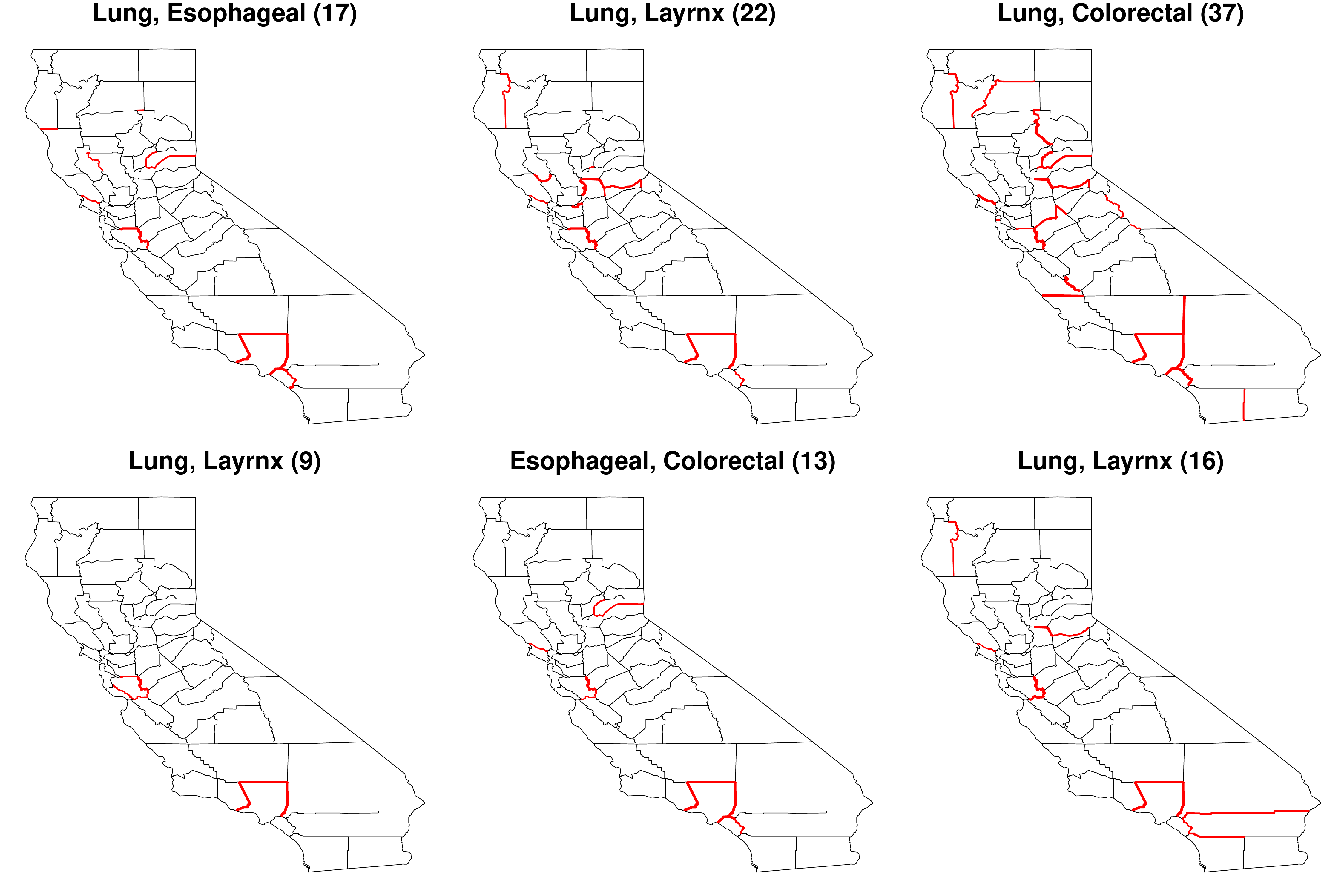}
    \caption{Shared difference boundaries (highlighted in red) detected by the model, in the case study with covariates only in the adjacency model, for each pair of cancers in SIR map when $\zeta = 0.05$. The values in brackets are the number of difference boundaries detected.}
    \label{fig:sbd_adj}
\end{figure}

\begin{figure}[htbp!]
    \centering
    \includegraphics[width = \textwidth]{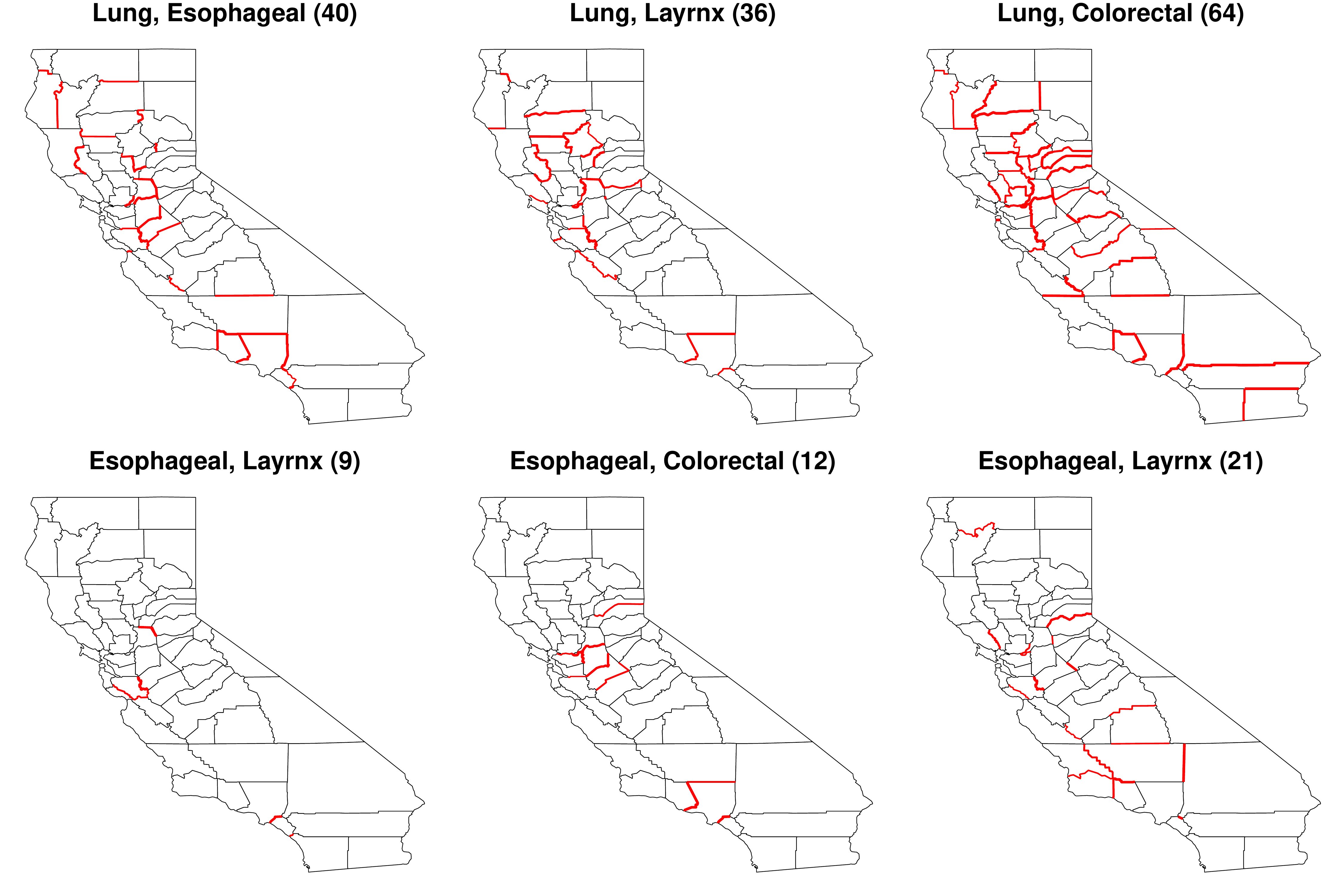}
    \caption{Mutual cross-difference boundaries (highlighted in red) detected by the model, in the case study with covariates only in the adjacency model, for each pair of cancers in SIR map when $\zeta = 0.05$. The values in brackets are the number of difference boundaries detected.}
    \label{fig:mbd_adj}
\end{figure}

\begin{figure}[htbp!]
    \centering
    \includegraphics[width = 0.75 \textwidth]{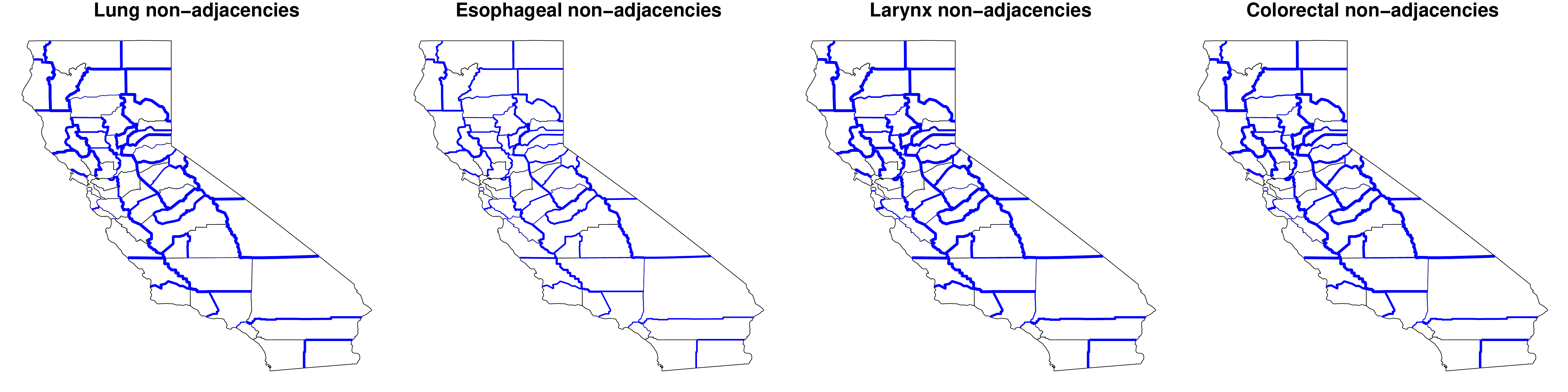}
    \caption{Non-adjacencies (shown in blue), in the case study with covariates only in the adjacency model, overlaid on smoking rates. The thickness of the lines is proportional to the probability of being considered as a non-adjacency}
    \label{fig:adj_adj}
\end{figure}

\begin{figure}[htbp!]
    \centering
    \includegraphics[width = \textwidth]{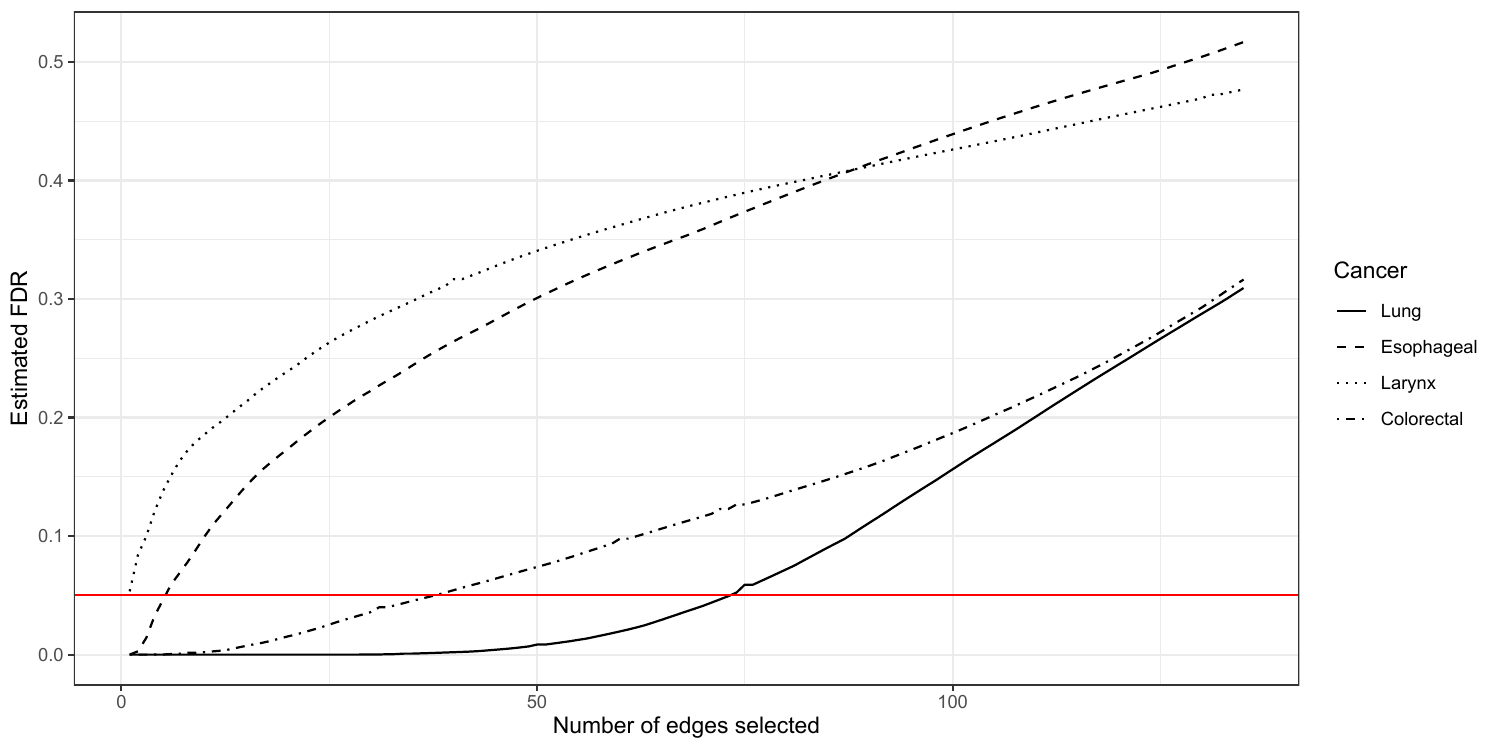}
    \caption{Estimated FDR curves plotted against the number of selected difference boundaries for four cancers in the case study with covariates only in the mean with fixed adjacencies}
    \label{fig:FDR_mean}
\end{figure}

\begin{figure}[htbp!]
    \centering
    \includegraphics[width = 0.75 \textwidth]{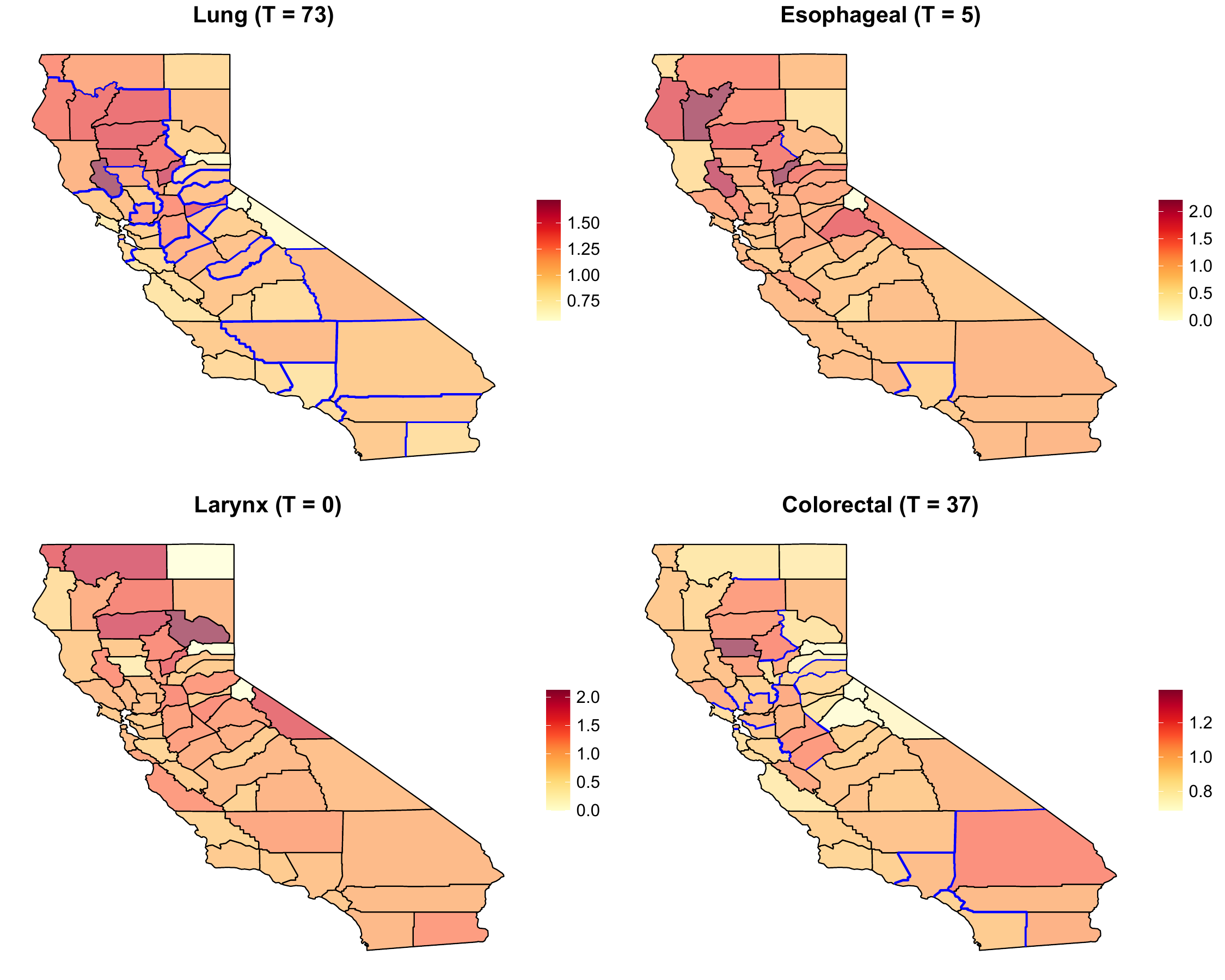}
    \caption{Difference boundaries (highlighted in red) detected by the model, in the case study with covariates only in the mean with fixed adjacencies, in SIR map for four cancers individually when $\zeta = 0.05$. The values in brackets are the number of difference boundaries detected.}
    \label{fig:bd_mean}
\end{figure}

\begin{figure}[htbp!]
    \centering
    \includegraphics[width = \textwidth]{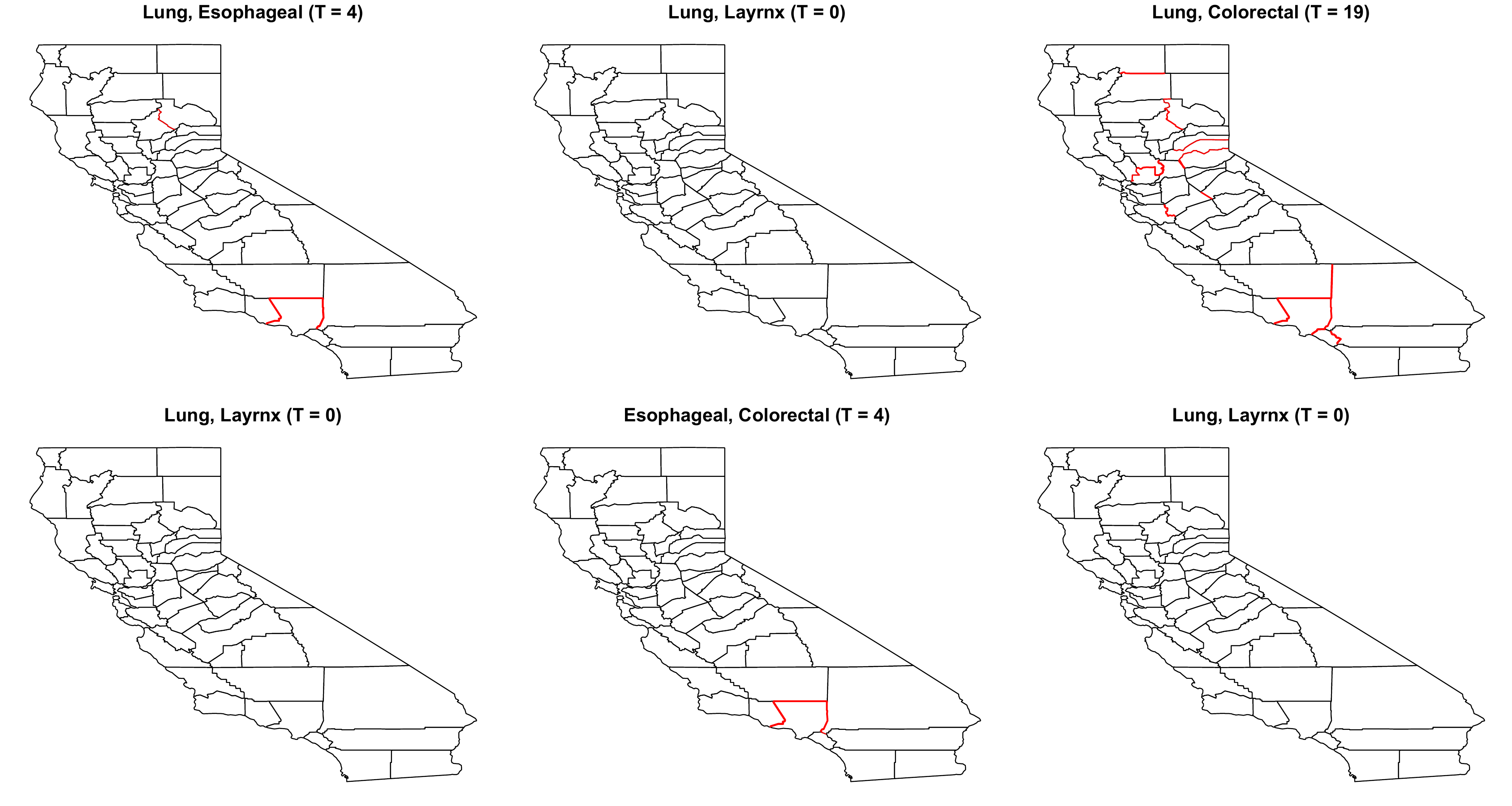}
    \caption{Shared difference boundaries (highlighted in red) detected by the model, in the case study with covariates only in the mean with fixed adjacencies, for each pair of cancers in SIR map when $\zeta = 0.05$. The values in brackets are the number of difference boundaries detected.}
    \label{fig:sbd_mean}
\end{figure}

\begin{figure}[htbp!]
    \centering
    \includegraphics[width = \textwidth]{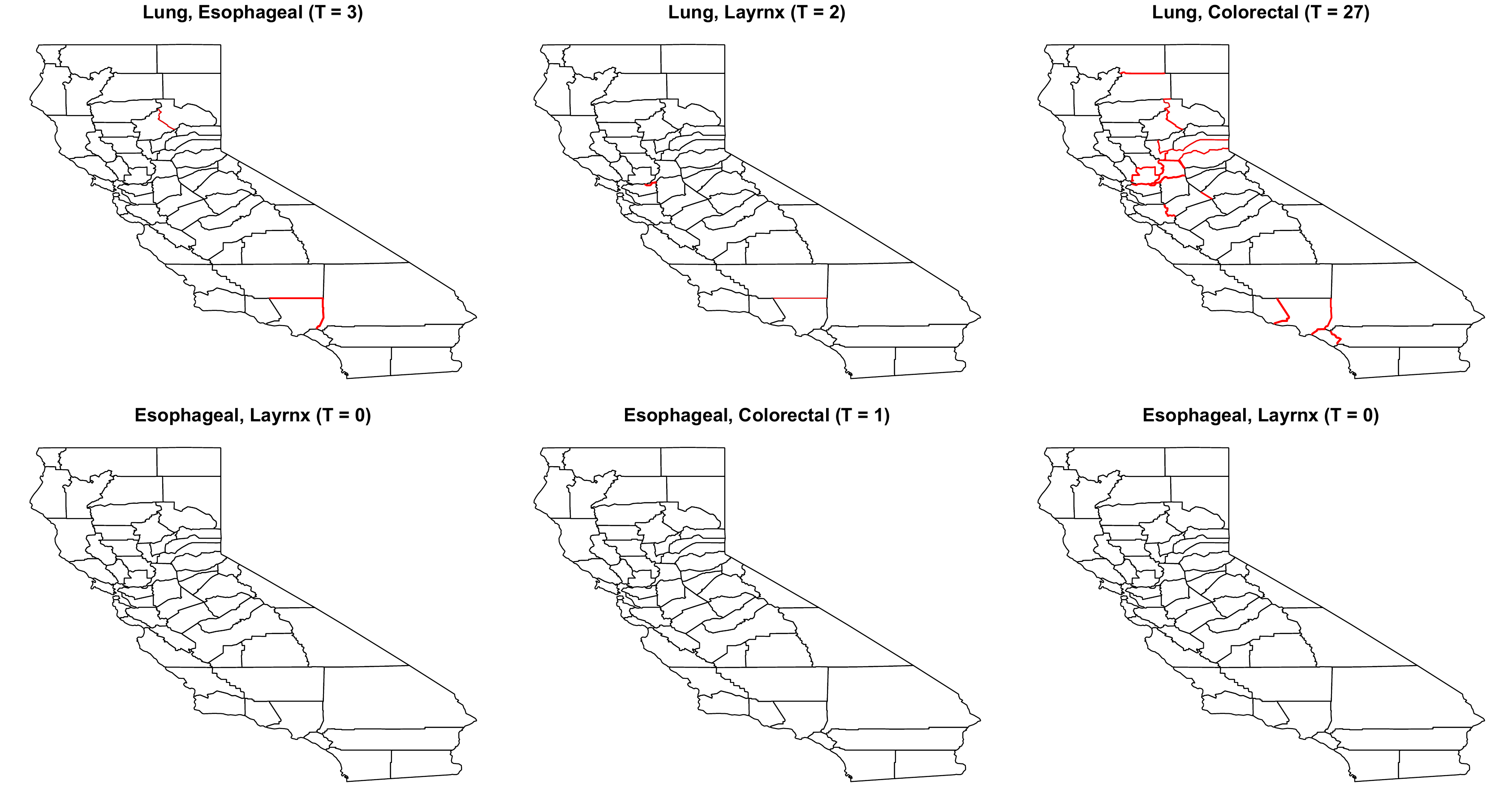}
    \caption{Mutual cross-difference boundaries (highlighted in red) detected by the model, in the case study with covariates only in the mean with fixed adjacencies, for each pair of cancers in SIR map when $\zeta = 0.05$. The values in brackets are the number of difference boundaries detected.}
    \label{fig:mbd_mean}
\end{figure}

\begin{figure}[htbp!]
    \centering
    \includegraphics[width = 0.75 \textwidth]{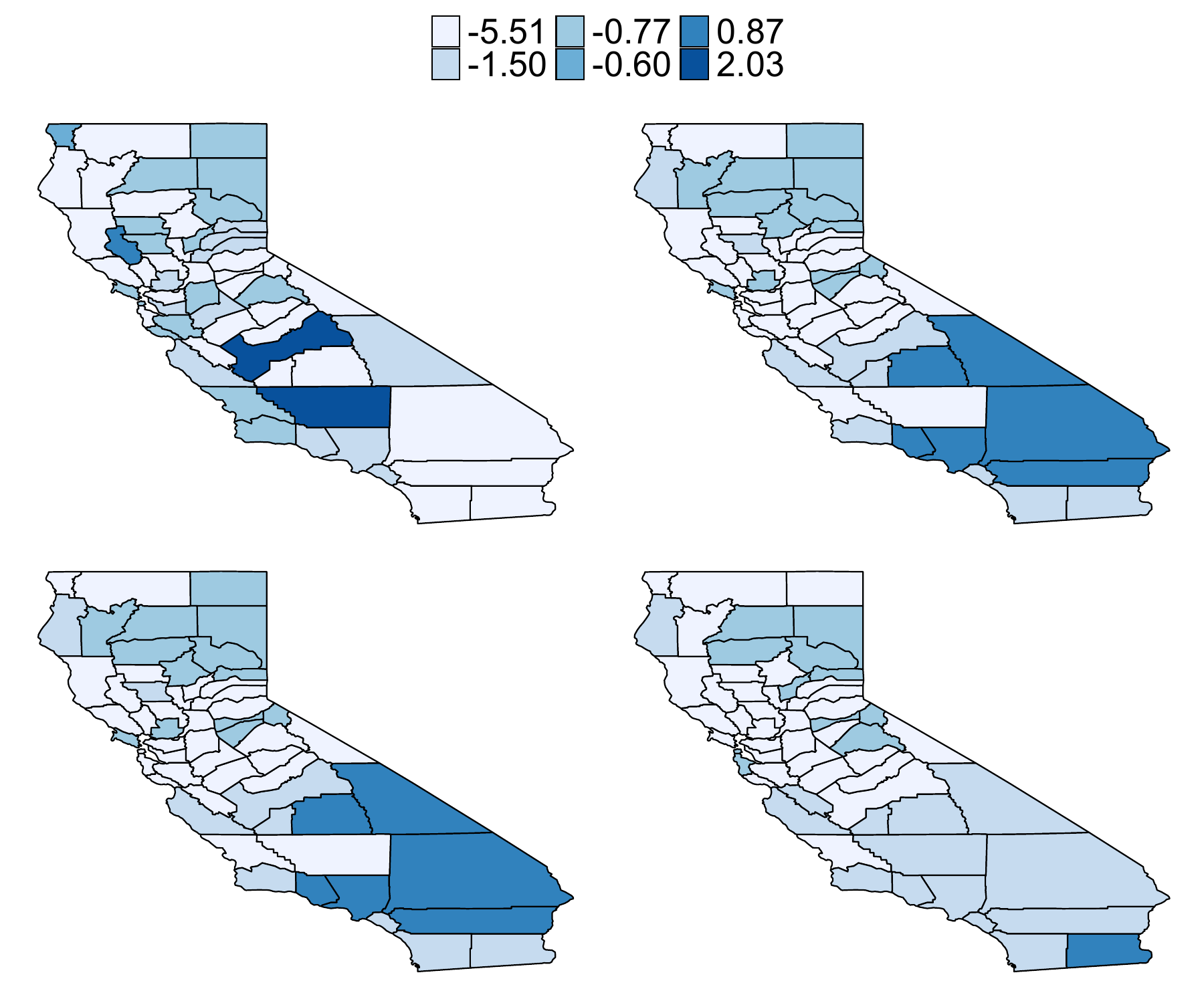}
    \caption{Simulated data maps depicting the random effects for the four cancers using a disease directed graph.}
    \label{fig:directed_map}
\end{figure}

\begin{figure}[t!]
    \centering
    \includegraphics[width = 0.75 \textwidth]{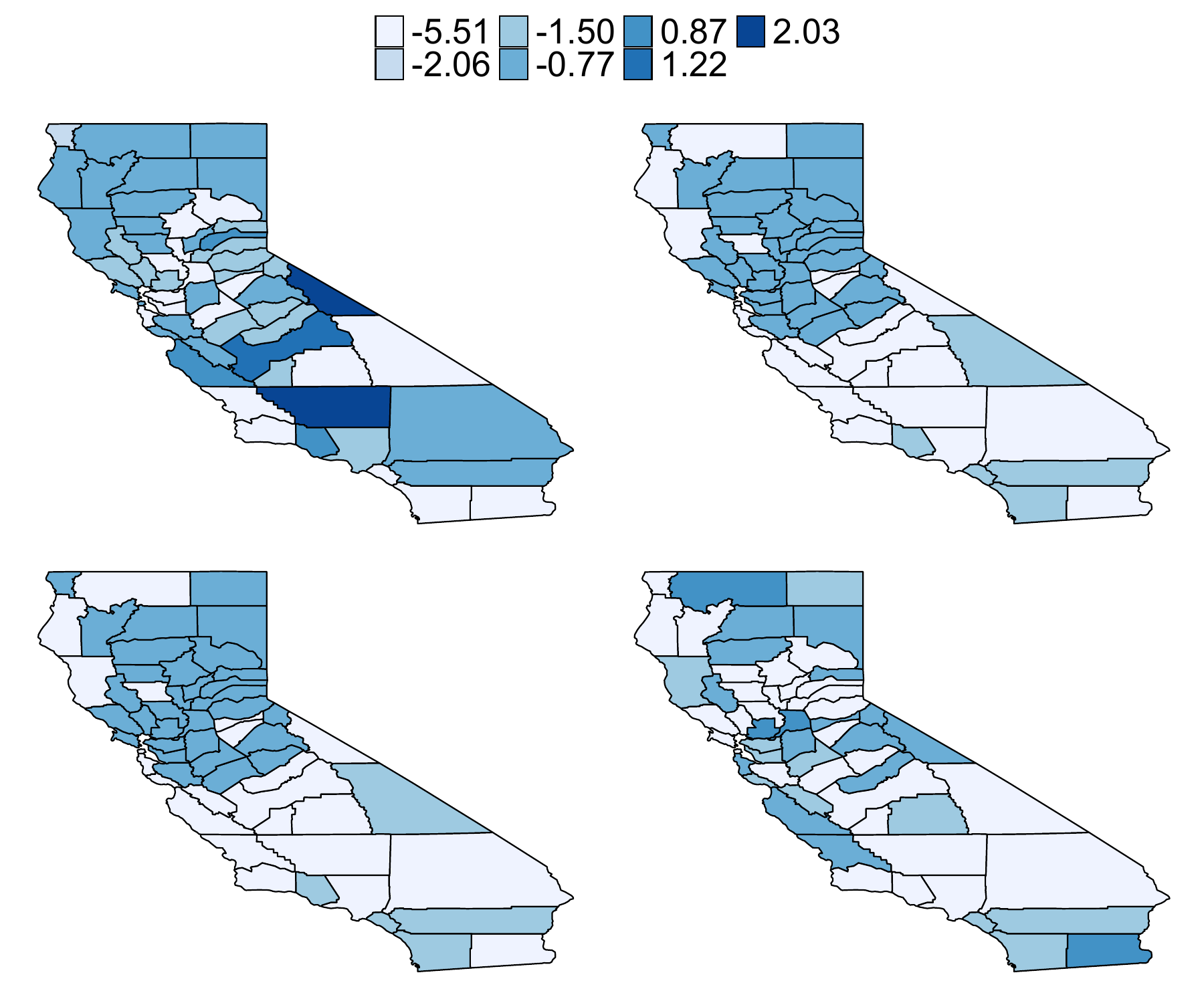}
    \caption{Simulated data maps depicting the random effects for the four cancers using a disease undirected graph.}
    \label{fig:undirected_map}
\end{figure}

\clearpage

\begin{sidewaystable}
\centering
\renewcommand{\arraystretch}{1.5}
\caption{Sensitivities and specificities for boundary detection across diseases}
\label{tab:cross_sens_spec}
\begin{adjustbox}{max width=\textheight}
\begin{tabular}{cccccccccccccccccccccccccc}
\hline
 & & \multicolumn{2}{c}{1 vs 2} & \multicolumn{2}{c}{2 vs 1} & \multicolumn{2}{c}{1 vs 3} & \multicolumn{2}{c}{3 vs 1} & \multicolumn{2}{c}{1 vs 4} & \multicolumn{2}{c}{4 vs 1} & \multicolumn{2}{c}{2 vs 3} & \multicolumn{2}{c}{3 vs 2} & \multicolumn{2}{c}{2 vs 4} & \multicolumn{2}{c}{4 vs 2} & \multicolumn{2}{c}{3 vs 4} & \multicolumn{2}{c}{4 vs 3} \\
\cmidrule(lr){3-4} \cmidrule(lr){5-6} \cmidrule(lr){7-8} \cmidrule(lr){9-10} \cmidrule(lr){11-12} \cmidrule(lr){13-14} \cmidrule(lr){15-16} \cmidrule(lr){17-18} \cmidrule(lr){19-20} \cmidrule(lr){21-22} \cmidrule(lr){23-24} \cmidrule(lr){25-26} 
Disease graph & T & Spec & Sens & Spec & Sens & Spec & Sens & Spec & Sens & Spec & Sens & Spec & Sens & Spec & Sens & Spec & Sens & Spec & Sens & Spec & Sens & Spec & Sens & Spec & Sens \\
\hline
Unstructured & 80 & 0.514 & 0.611 & 0.176 & 0.885 & 0.517 & 0.618 & 0.466 & 0.596 & 0.478 & 0.604 & 0.410 & 0.572 & 0.565 & 0.664 & 0.602 & 0.655 & 0.457 & 0.601 & 0.443 & 0.589 & 0.443 & 0.591 & 0.422 & 0.577 \\
& 90 & 0.445 & 0.681 & 0.176 & 0.885 & 0.433 & 0.685 & 0.399 & 0.672 & 0.412 & 0.677 & 0.348 & 0.646 & 0.477 & 0.720 & 0.513 & 0.720 & 0.391 & 0.673 & 0.370 & 0.659 & 0.378 & 0.663 & 0.350 & 0.647 \\
& 100 & 0.358 & 0.748 & 0.176 & 0.885 & 0.357 & 0.755 & 0.321 & 0.740 & 0.334 & 0.746 & 0.284 & 0.724 & 0.390 & 0.784 & 0.411 & 0.779 & 0.310 & 0.740 & 0.303 & 0.734 & 0.300 & 0.732 & 0.285 & 0.723 \\
& 110 & 0.263 & 0.811 & 0.176 & 0.885 & 0.270 & 0.819 & 0.238 & 0.807 & 0.243 & 0.811 & 0.219 & 0.798 & 0.298 & 0.844 & 0.310 & 0.838 & 0.234 & 0.808 & 0.229 & 0.806 & 0.215 & 0.796 & 0.213 & 0.796 \\
& 120 & 0.165 & 0.873 & 0.176 & 0.885 & 0.185 & 0.885 & 0.150 & 0.876 & 0.157 & 0.874 & 0.139 & 0.865 & 0.199 & 0.900 & 0.205 & 0.895 & 0.158 & 0.876 & 0.155 & 0.875 & 0.139 & 0.865 & 0.140 & 0.867 \\

\hline
Directed & 80 & 0.451 & 0.606 & 0.114 & 0.860 & 0.436 & 0.633 & 0.436 & 0.617 & 0.442 & 0.607 & 0.428 & 0.584 & 0.526 & 0.746 & 0.573 & 0.783 & 0.434 & 0.604 & 0.443 & 0.612 & 0.419 & 0.587 & 0.418 & 0.572 \\
& 90 & 0.376 & 0.664 & 0.114 & 0.860 & 0.364 & 0.685 & 0.379 & 0.683 & 0.374 & 0.669 & 0.352 & 0.648 & 0.444 & 0.802 & 0.483 & 0.830 & 0.369 & 0.669 & 0.371 & 0.680 & 0.352 & 0.648 & 0.350 & 0.647 \\
& 100 & 0.294 & 0.733 & 0.114 & 0.860 & 0.300 & 0.755 & 0.309 & 0.748 & 0.297 & 0.737 & 0.276 & 0.717 & 0.361 & 0.856 & 0.397 & 0.883 & 0.288 & 0.733 & 0.298 & 0.746 & 0.274 & 0.715 & 0.279 & 0.716 \\
& 110 & 0.221 & 0.801 & 0.114 & 0.860 & 0.223 & 0.814 & 0.227 & 0.811 & 0.220 & 0.803 & 0.213 & 0.796 & 0.273 & 0.899 & 0.299 & 0.921 & 0.217 & 0.804 & 0.224 & 0.816 & 0.199 & 0.779 & 0.202 & 0.776 \\
& 120 & 0.144 & 0.871 & 0.114 & 0.860 & 0.145 & 0.875 & 0.144 & 0.871 & 0.144 & 0.871 & 0.135 & 0.863 & 0.181 & 0.938 & 0.201 & 0.954 & 0.143 & 0.876 & 0.144 & 0.876 & 0.129 & 0.857 & 0.131 & 0.852 \\
\hline
Undirected & 80 & 0.406 & 0.582 & 0.155 & 0.876 & 0.469 & 0.596 & 0.425 & 0.577 & 0.430 & 0.580 & 0.434 & 0.582 & 0.518 & 0.651 & 0.570 & 0.647 & 0.394 & 0.573 & 0.399 & 0.575 & 0.444 & 0.597 & 0.426 & 0.592 \\
& 90 & 0.338 & 0.650 & 0.155 & 0.876 & 0.382 & 0.661 & 0.344 & 0.648 & 0.343 & 0.647 & 0.360 & 0.654 & 0.429 & 0.705 & 0.508 & 0.712 & 0.328 & 0.642 & 0.326 & 0.640 & 0.371 & 0.663 & 0.349 & 0.655 \\
& 100 & 0.274 & 0.722 & 0.155 & 0.876 & 0.308 & 0.732 & 0.264 & 0.718 & 0.270 & 0.718 & 0.288 & 0.724 & 0.340 & 0.760 & 0.417 & 0.773 & 0.254 & 0.711 & 0.266 & 0.715 & 0.306 & 0.736 & 0.287 & 0.727 \\
& 110 & 0.202 & 0.792 & 0.155 & 0.876 & 0.217 & 0.798 & 0.197 & 0.789 & 0.203 & 0.792 & 0.221 & 0.798 & 0.257 & 0.821 & 0.331 & 0.834 & 0.189 & 0.783 & 0.208 & 0.794 & 0.230 & 0.803 & 0.209 & 0.795 \\
& 120 & 0.140 & 0.867 & 0.155 & 0.876 & 0.133 & 0.864 & 0.135 & 0.864 & 0.126 & 0.862 & 0.146 & 0.871 & 0.173 & 0.885 & 0.234 & 0.897 & 0.120 & 0.857 & 0.137 & 0.866 & 0.163 & 0.877 & 0.145 & 0.869 \\
\hline
\end{tabular}
\end{adjustbox}
\end{sidewaystable}

\end{document}


\maketitle

\begin{appendix}

\renewcommand{\thesection}{S\arabic{section}}
\renewcommand{\thefigure}{S\arabic{figure}}
\renewcommand{\thetable}{S\arabic{table}}

\color{orange}

\section{MDAGAR details}

In this section, we outline the details of the MDAGAR model. First, we explicitly show the DAGAR structure in the univariate case. Then, we explain the computational advantages of the proposed approaches using simplified algebraic expressions.

\subsection{DAGAR details}\label{sec: DAGAR_details}
The DAGAR \citep{datta2019spatial} precision matrix has the following structure:
\begin{align*}
    \bQ(\rho) &= (\IdentityMat - \bB)^\T \bLambda (\IdentityMat - \bB) \\
        &=  \begin{bmatrix}
                1 & -b_{12} & \cdots & -b_{1n} \\
                0 & 1 & \cdots & -b_{2n} \\
                \vdots & \vdots & \ddots & \vdots \\
                0 & 0 & \cdots & 1
            \end{bmatrix}
            \begin{bmatrix}
                \lambda_1 & 0 & \cdots & 0 \\
                0 & \lambda_2 & \cdots & 0 \\
                \vdots & \vdots & \ddots & \vdots \\
                0 & 0 & \cdots & \lambda_n
            \end{bmatrix}
            \begin{bmatrix}
                1 & 0 & \cdots & 0 \\
                -b_{12} & 1 & \cdots & 0 \\
                \vdots & \vdots & \ddots & \vdots \\
                -b_{1n} & -b_{2n} & \cdots & 1
            \end{bmatrix} \\
        &= \begin{bmatrix}
                \lambda_1 + \sum_{i=2}^n \lambda_i b_{1i}^2 & -\lambda_2 b_{12} + \sum_{i=3}^n \lambda_i b_{1i} b_{2i} & \cdots & -\lambda_n b_{1n} \\
                -\lambda_2 b_{12} + \sum_{i=3}^n \lambda_i b_{1i} b_{2i} & \lambda_2 + \sum_{i=3}^n \lambda_i b_{2i}^2 & \cdots & -\lambda_n b_{2n} \\
                \vdots & \vdots & \ddots & \vdots \\
                -\lambda_n b_{1n} & -\lambda_n b_{2n} & \cdots & \lambda_n
            \end{bmatrix}
\end{align*}
One of the main advantages of the DAGAR formulation is the computation of the determinant and the quadratic form in the multivariate normal probability density function. Specifically:
\begin{align*}
    \lvert \bQ(\rho) \rvert &= \lvert \IdentityMat - \bB \rvert \lvert \bLambda \rvert \lvert \IdentityMat - \bB \rvert = \prod_{i=1}^n \lambda_{i} \\
    \bgamma^\T \bQ(\rho) \bgamma &= \bgamma^\T \lrnd \IdentityMat - \bB \rrnd^\T \bLambda \lrnd \IdentityMat - \bB \rrnd \bgamma = \lambda_1 \gamma_1^2 + \sum_{i=1}^n \lambda_i (\gamma_i - \sum_{j<i} \gamma_j b_{ij})^2 
\end{align*}
which are way less heavy to compute from a computational perspective. In particular the determinant computation reduces to computing a product of $n$ terms while the quadratic form is lighter since it takes into account the sparsity of the structure and the strictly lower triangularity of $\bB$.

\subsection{Computational advantages}\label{sec: advantages}

Taking into account the multiple disease framework, our model allow for the following multivariate gaussian log-density evaluation:
\begin{align*}
    \log{\lrnd p(\bgamma \given \bSigma_{\bgamma})\rrnd} &= \log{\lrnd 2 \pi \rrnd^{-N/2} \lrnd\lvert \bSigma_{\bgamma} \rvert^{-1/2} \exp{\lrnd - \frac{1}{2} \bgamma^\T \bSigma_{\bgamma}^{-1} \bgamma \rrnd}\rrnd} \\
    &= -\frac{nq}{2} \log \lrnd 2 \pi \rrnd + \frac{1}{2} \log{\lvert \bSigma_{\bgamma}^{-1} \rvert - \frac{1}{2} \bgamma^\T \bSigma_{\bgamma}^{-1} \bgamma} 
\end{align*}

\paragraph{Unstructured graph}
We first need to compute $\lvert \bSigma_{\bgamma}^{-1} \rvert$:
\begin{align*}
    \lvert \bSigma_{\bgamma}^{-1} \rvert = \lvert \bA^\T \otimes \IdentityMat \rvert^{-1} \prod_{d=1}^q \lvert \bQ_{d} \rvert \lvert \bA \otimes \IdentityMat \rvert^{-1} = \lvert \bA \rvert^{-2n} \prod_{d=1}^q \lvert \bQ_{d} \rvert = \lrnd \prod_{d=1}^q \bA_{dd} \rrnd^{-2n} \prod_{d=1}^q \prod_{i=1}^n \lrnd \bLambda_{d} \rrnd_{ii}
\end{align*}
where $\lvert \bQ_{d} \rvert = \lvert \IdentityMat - \bB_{d} \rvert^2 \lvert \bLambda_{d} \rvert = \lvert \bLambda_{d} \rvert$ is motivated by the strictly lower-triangularity of $\bB_d$ and $\lvert \bA \rvert = \prod_{d=1}^q \bA_{dd}$ by the lower-triangularity of $\bA$. Then we need to compute $\bgamma^\T \bSigma_{\bgamma}^{-1} \bgamma$:
\begin{align*}
    \bgamma^\T \bSigma_{\bgamma}^{-1} \bgamma &= \bgamma^\T \lrnd \bA^{-1} \otimes \IdentityMat \rrnd^\T \lsq \bigoplus_{d}^q \bQ_{d} \rsq \lrnd \bA^{-1} \otimes \IdentityMat \rrnd \bgamma \\
    &=\lsq \lrnd \bA^{-1} \otimes \IdentityMat \rrnd \bgamma \rsq^\T \lsq \bigoplus_{d}^q \bQ_{d} \rsq \lsq \lrnd \bA^{-1} \otimes \IdentityMat \rrnd \bgamma \rsq
\end{align*}
Here we need some algebra to simplify the expression. Let's start by defining $\bGamma = \lsq \bgamma_{1} \mid \dots \mid \bgamma_{q} \rsq$ and by exploiting some properties of the Kronecker product:
\begin{align*}
    \lrnd \bA^{-1} \otimes \IdentityMat \rrnd \bgamma = \lrnd \bA^{-1} \otimes \IdentityMat \rrnd \text{vec}\lrnd\bGamma\rrnd = \text{vec}\lrnd \IdentityMat \bGamma \bA^{-\T} \rrnd = \text{vec}\lrnd \bGamma \bA^{-\T} \rrnd = \text{vec} \lrnd \bC \rrnd
\end{align*}
Here, matrix $\bC$ is computed as:
\begin{align*}
    \bC &= \lsq \bgamma_{1} \mid \bgamma_{2} \mid \dots \mid \bgamma_{q} \rsq 
    \begin{bmatrix} 
        \tilde{a}_{11} & \tilde{a}_{21} & \cdots & \tilde{a}_{q1} \\
        0 & \tilde{a}_{22} & \cdots & \tilde{a}_{q2} \\
        \vdots & \vdots & \ddots & \vdots \\
        0 & 0 & \cdots & \tilde{a}_{qq}
    \end{bmatrix} \\
    &= \lsq \tilde{a}_{11} \bgamma_{1} \mid \tilde{a}_{21} \bgamma_{1} + \tilde{a}_{22} \bgamma_{2} \mid \dots \mid \tilde{a}_{q1} \bgamma_{1} + \dots + \tilde{a}_{qq} \bgamma_{q} \rsq \\
    &= \lsq \bc_{1} \mid \bc_{2} \mid \dots \mid \bc_{q} \rsq
\end{align*}
where $\tilde{a}_{dh}$ are the elements of the lower triangular matrix $\bA^{-1}$. Hence,
\begin{align*}
    \bgamma^\T \bSigma_{\bgamma}^{-1} \bgamma &= \lsq \bc_{1}^\T, \bc_{2}^\T, \dots, \bc_{q}^\T \rsq
    \begin{bmatrix} 
        \bQ_1 & \bO & \cdots & \bO \\
        \bO & \bQ_2 & \cdots & \bO \\
        \vdots & \vdots & \ddots & \vdots \\
        \bO & \bO & \cdots & \bQ_q
    \end{bmatrix}
    \begin{bmatrix}
        \bc_{1} \\
        \bc_{2} \\ 
        \vdots \\
        \bc_{q}
    \end{bmatrix} = \sum_{d=1}^q \bc_{d}^\T \bQ_d \bc_{d}
\end{align*}
The log-density of $\bgamma$ finally results in:
\begin{equation}\label{eq:unstr_ll}
    \log{\lrnd p(\bgamma \given \bSigma_{\bgamma})\rrnd} = -\frac{nq}{2} \log \lrnd 2 \pi \rrnd - n \sum_{d=1}^q \log\lrnd \bA_{dd} \rrnd + \frac{1}{2} \sum_{d=1}^q \sum_{i=1}^n \log\lrnd \lrnd \bLambda_d \rrnd_{ii} \rrnd - \frac{1}{2} \sum_{d=1}^q \bc_{d}^\T \bQ_d \bc_{d} 
\end{equation}
This log-density evaluation is computationally efficient because it requires only one matrix inversion, which pertains to $\bA$. In applications similar to this work, the dimensions of $\bA$ are typically small ($q \leq 10$).

\paragraph{Directed graph}

In this case the determinant of $\bSigma_{\bgamma}^{-1}$ is very simple:
\begin{align*}
    \lvert \bSigma_{\bgamma}^{-1} \rvert = \lvert \IdentityMat_{N} - \bA \rvert \prod_{d=1}^q \lvert \bQ_d \rvert \lvert \IdentityMat_{N} - \bA \rvert = \prod_{d=1}^q \lvert \bQ_d \rvert = \prod_{d=1}^q \prod_{i=1}^n \lrnd \bLambda_{d} \rrnd_{ii}
\end{align*}
where we have exploited the strictly lower triangularity of the $N \times N$ block matrix $\bA$. Now we need to compute $\bgamma^\T \bSigma_{\bgamma}^{-1} \bgamma$:
\begin{align*}
    \bgamma^\T \bSigma_{\bgamma}^{-1} \bgamma &= \bgamma^\T \lrnd \IdentityMat_N - \bA \rrnd^\T \lsq \bigoplus_{d=1}^q \bQ_d \rsq \lrnd \IdentityMat_N - \bA \rrnd \bgamma 
\end{align*}
where
\begin{align*}
    \lrnd \IdentityMat_N - \bA \rrnd \bgamma &= 
    \begin{bmatrix} 
        \IdentityMat_n & \bO & \cdots & \bO \\
        -\bA_{21} & \IdentityMat_n & \cdots & \bO \\
        \vdots & \vdots & \ddots & \vdots \\
        -\bA_{q1} & -\bA_{q2} & \cdots & \IdentityMat_n
    \end{bmatrix}
    \begin{bmatrix} 
        \bgamma_1 \\
        \bgamma_2 \\
        \vdots \\
        \bgamma_q 
    \end{bmatrix} =
    \begin{bmatrix} 
        \bgamma_1 \\
        -\bA_{21}\bgamma_1 + \bgamma_2 \\
        \vdots \\
        -\bA_{q1}\bgamma_1 - \dots - \bA_{qq-1} \bgamma_{q-1} + \bgamma_{q} 
    \end{bmatrix}
\end{align*}
Hence
\begin{align*}
    \scriptsize
    \bgamma^\T \bSigma_{\bgamma}^{-1} \bgamma &= 
    \scriptsize
    \begin{bmatrix} 
        \bgamma_1 \\
        -\bA_{21}\bgamma_1 + \bgamma_2 \\
        \vdots \\
        -\bA_{q1}\bgamma_1 - \dots - \bA_{qq-1} \bgamma_{q-1} + \bgamma_{q} 
    \end{bmatrix}^\T
    \begin{bmatrix} 
        \bQ_1 & \bO & \cdots & \bO \\
        \bO & \bQ_2 & \cdots & \bO \\
        \vdots & \vdots & \ddots & \vdots \\
        \bO & \bO & \cdots & \bQ_q
    \end{bmatrix}
    \begin{bmatrix} 
        \bgamma_1 \\
        -\bA_{21}\bgamma_1 + \bgamma_2 \\
        \vdots \\
        -\bA_{q1}\bgamma_1 - \dots - \bA_{qq-1} \bgamma_{q-1} + \bgamma_{q} 
    \end{bmatrix} \\
    &= \sum_{d=1}^q \lsq \lrnd \bgamma_d - \sum_{h=1}^{d-1} \mathbf{A}_{dh} \bgamma_h \rrnd^\T \bQ_d \lrnd \bgamma_d - \sum_{h=1}^{d-1} \mathbf{A}_{dh} \bgamma_h \rrnd \rsq
\end{align*}
In this case the log-density is defined as  
\begin{equation}\label{eq:dir_ll}
    \begin{aligned}
        \log{\lrnd p(\bgamma \given \bSigma_{\bgamma})\rrnd} = &-\frac{nq}{2} \log \lrnd 2 \pi \rrnd + \frac{1}{2} \sum_{d=1}^q \sum_{i=1}^n \log\lrnd \lrnd \bLambda_d \rrnd_{ii} \rrnd \\
        &- \frac{1}{2} \sum_{d=1}^q \lsq \lrnd \bgamma_d - \sum_{h=1}^{d-1} \mathbf{A}_{dh} \bgamma_h \rrnd^\T \bQ_d \lrnd \bgamma_d - \sum_{h=1}^{d-1} \mathbf{A}_{dh} \bgamma_h \rrnd \rsq
    \end{aligned}
\end{equation}
where, it is clear that there is no need to invert any matrices.

\paragraph{Undirected graph}

Again, the first thing we need to compute is $\lvert \bSigma_{\bgamma}^{-1} \rvert$:
\begin{align*}
    \lvert \bSigma_{\bgamma}^{-1} \rvert = \left | \bigoplus_{d=1}^q \bU_d \right | \lvert \bLambda_{dis} \otimes \IdentityMat_{n} \rvert \left | \bigoplus_{d=1}^q \bU_d^{\T} \right | = \lvert \bLambda_{dis} \rvert^{n} \lrnd \prod_{d=1}^q \lvert \bU_d \rvert \rrnd^2 = \lvert \bLambda_{dis} \rvert^{n} \prod_{d=1}^q \prod_{i=1}^n \lrnd \bLambda_d \rrnd_{ii} 
\end{align*}
where for the computation of $\lvert \bU_d \rvert$ we have exploited $\bU_d = \lrnd \IdentityMat - \bB_d \rrnd^\T \bLambda_{d}^{1/2}$ and consequently $\lvert \bU_d \rvert = \lvert \IdentityMat - \bB_d \rvert \lvert \bLambda_{d}^{1/2}\rvert = \prod_{i=1}^n \sqrt{\lrnd \bLambda_d \rrnd_{ii}}$. Then, it is the turn of $\bgamma^\T \bSigma_{\bgamma}^{-1} \bgamma$:
\begin{align*}
    \bgamma^\T \bSigma_{\bgamma}^{-1} \bgamma &= \bgamma^\T \lsq \bigoplus_{d=1}^q \bU_d \rsq \lrnd \bLambda_{dis} \otimes \IdentityMat_{n} \rrnd \lsq \bigoplus_{d=1}^q \bU_d^{\T} \rsq \bgamma \\
    &= \lrnd \lsq \bigoplus_{d=1}^q \bU_d^{\T} \rsq \bgamma \rrnd^{\T} \lrnd \bLambda_{dis} \otimes \IdentityMat_{n} \rrnd \lrnd \lsq \bigoplus_{d=1}^q \bU_d^{\T} \rsq \bgamma \rrnd
\end{align*}
In this case, to simplify the expression we need to develop the computations regarding
\begin{align*}
    \lsq \bigoplus_{d=1}^q \bU_d^{\T} \rsq \bgamma = \begin{bmatrix} 
        \bU_1^{\T} & \bO & \cdots & \bO \\
        \bO & \bU_2^{\T} & \cdots & \bO \\
        \vdots & \vdots & \ddots & \vdots \\
        \bO & \bO & \cdots & \bU_q^{\T}
    \end{bmatrix} 
    \begin{bmatrix}
        \bgamma_{1} \\
        \bgamma_{2} \\ 
        \vdots \\
        \bgamma_{q}
    \end{bmatrix} =
    \begin{bmatrix}
        \bU_1^{\T} \bgamma_{1} \\
        \bU_2^{\T} \bgamma_{2} \\ 
        \vdots \\
        \bU_q^{\T} \bgamma_{q}
    \end{bmatrix}
\end{align*}
Hence,
\begin{align*}
    \bgamma^\T \bSigma_{\bgamma}^{-1} \bgamma &= \lsq \bgamma_{1}^\T \bU_1, \bgamma_{2}^\T \bU_2, \dots, \bgamma_{q}^\T \bU_q \rsq 
    \begin{bmatrix} 
        \lambda_{dis,11} \IdentityMat & \lambda_{dis,12} \IdentityMat & \cdots & \lambda_{dis,1q} \IdentityMat \\
        \lambda_{dis,21} \IdentityMat & \lambda_{dis,22} \IdentityMat & \cdots & \lambda_{dis,2q} \IdentityMat \\
        \vdots & \vdots & \ddots & \vdots \\
        \lambda_{dis,q1} \IdentityMat & \lambda_{dis,q2} \IdentityMat & \cdots & \lambda_{dis,qq} \IdentityMat
    \end{bmatrix}
    \begin{bmatrix}
        \bU_1^{\T} \bgamma_{1} \\
        \bU_2^{\T} \bgamma_{2} \\ 
        \vdots \\
        \bU_q^{\T} \bgamma_{q}
    \end{bmatrix} \\ 
    &= \sum_{d=1}^q \sum_{h=1}^q \lambda_{dis,dh} \bgamma_d^\T \bU_d \bU_h^\T \bgamma_h
\end{align*}
Finally the log-density in this case results in:
\begin{equation}\label{eq:undir_ll}
    \begin{aligned}
        \log{\lrnd p(\bgamma \given \bSigma_{\bgamma})\rrnd} = &-\frac{nq}{2} \log \lrnd 2 \pi \rrnd + \frac{1}{2}\sum_{d=1}^q \sum_{i=1}^n \log{\lrnd \lrnd \bLambda_d \rrnd_{ii} \rrnd} + \frac{n}{2} \log{\lvert \bLambda_{dis} \rvert} \\
        &- \frac{1}{2} \sum_{d=1}^q \sum_{h=1}^q \lambda_{dis,dh} \bgamma_d^\T \bU_d \bU_h^\T \bgamma_h
    \end{aligned}
\end{equation}
Also in this case no matrix inversion is required, since we exploit the DAGAR construction of each spatial precision matrix.

\paragraph{Advantages with respect to the MCAR specification}

In all the aforementioned log-density specifications \eqref{eq:unstr_ll}, \eqref{eq:dir_ll}, and \eqref{eq:undir_ll}, the log-determinant of the spatial precision matrix, $\lvert \bQ_{d} \rvert$, was simplified to the summation of the diagonal elements of the diagonal matrix $\lvert \bLambda_{d} \rvert$, resulting, for all the $d$, in a computational complexity of $\mathcal{O}(qn)$, due to the DAGAR construction. However, this simplification does not apply to an MCAR specification, i.e., $\bQ_{d} = \lrnd \bD_d - \rho_d \bW_d \rrnd$, where $\bD_d$ is a diagonal matrix with elements equal to the number of neighbors for each region. For MCAR, the precision matrix determinant cannot be simplified without computing the eigenvalues $\omega_{di}$ of $\bD_d^{-1/2}\bW_{d}\bD_d^{-1/2}$, from which $\lvert \bQ_{d} \rvert \propto \prod_{i=1}^n \lrnd 1 - \rho_d \omega_{di} \rrnd$.

In a framework with a fixed adjacency structure, this is not problematic because the eigenvalues can be precomputed. At each iteration of the MCMC, only $\rho_d$ would need to be updated. However, if the adjacency structure changes at every iteration, the eigendecomposition would need to be recomputed each time, negating the computational advantage of precomputation. In this case, for all the $d$, the computational complexity is $\mathcal{O}(qn^3)$.

Only in the case of the undirected graph, the computational complexity associated with the quadratic form differs. Specifically, for the MDAGAR, we can leverage the construction method of each $\bQ_{d}$, leading to a computational complexity of $\mathcal{O}(q^2 n^2)$. However, for the MCAR variant, Cholesky factorization of each $\bQ_{d}$ is necessary, resulting in a computational complexity of $\mathcal{O}(q n^3 + q^2 n^2)$, where the first term represents the additional complexity introduced by the Cholesky decomposition. In all the other cases also the computational complexity associated to the MCAR construction of $\bQ_d$ is $\mathcal{O}(q^2 n^2)$.

\color{black}
\section{Computational details}\label{sec:comp_det}

Our models rely on posterior inference, which is achieved through MCMC posterior simulations. To accomplish this, we employ the blocked Gibbs Sampler with some adaptive and some classic Metropolis-Hastings steps, based on the work by \cite{gamerman2006markov} and \cite{miasojedow2013adaptive}. This approach allows us to achieve the desired acceptance rates while updating all random parameters in our model. It is important to note that $\phi_{i}$ corresponds to $\theta_{u_{i}}$, and we update $\phi_{i}$ by updating both $\btheta$ and $u_{i}$, and, ultimately, that $\bet$ is constrained to be between $0$ and a positive constant $M$. The MCMC algorithm proceeds through the following steps, with a noteworthy observation that step 8 is further subdivided into three parts, depending on the selected model used to represent disease dependence. Here, we present the algorithm used for the data analysis described in Section 5. In the simulation setting, there would be an additional step to calculate the error variances, along with the closed-form full conditionals for the error variances, as well as for $\bbeta$ and $\btheta$.
\begin{noendalgorithm}
    \hrulefill
    \vspace{-0.75em}
    \caption{Posterior inference of $\btheta$, $\bbeta$, $\tau_s$, $\bV$, $\bgamma$, $\brho$, $\bet$ and $\bA$ or $\balpha$ or $\rho_{dis}$ based on our models}\label{alg:amwg}
    \vspace{-0.75em}
    \hrulefill
    \begin{algorithmic}[1]
        \scriptsize

        \State update $\bbeta \given \btheta$
        \begin{enumerate}[(a)]
            \item sample candidate $\bbeta^*$ from $\Norm(\bbeta, \xi_{\bbeta})$
            \item accept $\bbeta^*$ with probability
            \begin{equation*}
                \min\lcur 1, \frac{\exp{\lrnd \sum_{k=1}^K \sum_{i,d:\phi_{id}=\theta_{k}} y_{id}(\bx_{id}^\T \bbeta_d^* + \phi_{id}) - E_{id} \exp{\lrnd \bx_{id}^\T \bbeta_d^* + \phi_{id} \rrnd} - \frac{{\bbeta_d^*}^\T\bbeta_d^*}{2\sigmasq_{\beta}}\rrnd}}{\exp{\lrnd \sum_{k=1}^K \sum_{i,d:\phi_{id}=\theta^*_{k}} y_{id} (\bx_{id}^\T \bbeta_d + \phi_{id}) - E_{id} \exp{\lrnd \bx_{id}^\T \bbeta_d + \phi_{id} \rrnd}- \frac{\bbeta_d^\T\bbeta_d}{2\sigmasq_{\beta}} \rrnd}} \rcur
            \end{equation*}
            \item adapt $\xi_{\bbeta}$
        \end{enumerate}

        \algstore{bkbreak}
    \end{algorithmic}
\end{noendalgorithm}
\begin{notoprulealgorithm}
    \caption*{}
    \begin{algorithmic}
        \algrestore{bkbreak}
        \scriptsize

        \State update $\btheta \given \bbeta, \tau_s$
        \begin{enumerate}[(a)]
            \item sample candidate $\btheta^*$ from $\Norm(\btheta, \xi_{\btheta})$
            \item accept $\btheta^*$ with probability
            \begin{equation*}
                \min\lcur 1, \frac{\exp{\lrnd \sum_{k=1}^K \sum_{i,d:\phi_{id}=\theta^*_{k}} y_{id} (\bx_{id}^\T \bbeta_d + \phi_{id}) - E_{id} \exp{\lrnd \bx_{id}^\T \bbeta_d + \phi_{id} \rrnd} - \frac{{\theta^*_k}^2}{2}\tau_s\rrnd}}{\exp{\lrnd \sum_{k=1}^K \sum_{i,d:\phi_{id}=\theta_{k}} y_{id} (\bx_{id}^\T \bbeta_d + \phi_{id}) - E_{id} \exp{\lrnd \bx_{id}^\T \bbeta_d + \phi_{id} \rrnd} - \frac{\theta_{k}^2}{2}\tau_s\rrnd}} \rcur
            \end{equation*}
            \item adapt $\xi_{\btheta}$
        \end{enumerate}

        \State update $\gamma_{id} \given \bbeta, \btheta, \brho, \bet, \bA$, $i = 1,\dots,n$, $d = 1,\dots,q$
        \begin{enumerate}[(a)]
            \item sample candidate $\gamma^*_{id}$ from $\Norm(\gamma_{id}, \xi_{\gamma_{id}})$ 
            \item compute the corresponding candidate $u^*_{id} = \sum_{k=1}^K k I\lrnd\sum_{t=1}^{k-1}p_t < F^{(id)}(\gamma^*_{id}) < \sum_{t=1}^{k}p_t\rrnd$
            \item accept $\gamma^*_{id}$ with probability
            \begin{equation*}
                \min\lcur 1, \frac{\exp{\lrnd y_{id}(\bx_{id}^\T \bbeta_d + \theta_{u^*_{id}}) - E_{i} \exp{(\bx_{id}^\T \bbeta_d + \theta_{u^*_{id}})} - \frac{\bgamma^{*T}\bSigma^{-1}_{\bgamma}\bgamma^{*}}{2} \rrnd}}{\exp{\lrnd y_{id}(\bx_{id}^\T \bbeta_d + \theta_{u_{id}}) - E_{id} \exp{(\bx_{id}^\T \bbeta_d + \theta_{u_{id}})} - \frac{\bgamma^{T}\bSigma^{-1}_{\bgamma}\bgamma}{2} \rrnd}} \rcur
            \end{equation*}    
            \item adapt $\xi_{\gamma_{id}}$
        \end{enumerate}
        
        \State update $\bV \given \bbeta, \btheta$
        \begin{enumerate}[(a)]
            \item let $\widetilde{\bV} = \text{logit}(\bV)$ and sample $\widetilde{\bV}^*$ from $\Norm(\widetilde{\bV}, \xi_{\widetilde{\bV}})$ then $\bV^* = \frac{\exp{(\widetilde{\bV}^*)}}{1 + \exp{(\widetilde{\bV}^*)}}$
            \item compute the corresponding candidate $\bp^* = (p^*_1,\dots,p^*_K)$ and $\bu = \{u^*_{id}\}$
            \item accept $\bV^*$ with probability
            \begin{equation*}
                \min\lcur 1, \frac{\exp{\lrnd \sum_{i,d} y_{id} (\bx_{id}^\T \bbeta_d + \theta_{u^*_{id}}) - E_{id} \exp{(\bx_{id}^\T \bbeta_d + \theta_{u^*_{id}})} \rrnd}\prod_{k=1}^K(1-V^*_k)^{(\alpha-1)} \prod_{k=1}^K V^*_k (1-V^*_k)}{\exp{\lrnd \sum_{i,d} y_{id} (\bx_{id}^\T \bbeta_d + \theta_{u_{id}}) - E_{id} \exp{(\bx_{id}^\T \bbeta_d + \theta_{u_{id}})} \rrnd}\prod_{k=1}^K(1-V_k)^{(\alpha-1)} \prod_{k=1}^K V_k (1-V_k)} \rcur
            \end{equation*}    
            \item adapt $\xi_{\widetilde{\bV}}$
        \end{enumerate}
        
        \State sample $\tau_s \given \btheta, \beta_0$ from
        \begin{equation*}
            \Gamma \lrnd a_s + \frac{K}{2}, b_s + \frac{\sum_{k=1}^K \theta_k^2}{2} \rrnd
        \end{equation*}

        \State update $\brho \given \bgamma, \bet, \bA$
        \begin{enumerate}[(a)]
            \item let $\widetilde{\brho} = \text{logit}(\brho)$ and sample $\widetilde{\brho}^*$ from $\Norm(\widetilde{\brho}, \xi_{\widetilde{\brho}})$ then $\brho^* = \frac{\exp{(\widetilde{\brho}^*)}}{1 + \exp{(\widetilde{\brho}^*)}}$
            \item update $\bSigma^*_{\bgamma}$ with the proposed $\brho^*$
            \item accept $\widetilde{\brho}^*$ with probability 
            \begin{equation*}
                \min\lcur 1, \frac{\lvert \bSigma^*_{\bgamma} \rvert^{-\frac{N}{2}}  \exp{\lrnd - \frac{\bgamma^{T}\bSigma^{*-1}_{\bgamma}\bgamma}{2} \rrnd} \prod_{d=1}^q \rho^*_d(1-\rho^*_d)}{\lvert \bSigma_{\bgamma} \rvert^{-\frac{N}{2}}  \exp{\lrnd - \frac{\bgamma^{T}\bSigma^{-1}_{\bgamma}\bgamma}{2} \rrnd}\prod_{d=1}^q \rho_d(1-\rho_d)} \rcur
            \end{equation*}
            \item adapt $\xi_{\widetilde{\brho}}$
        \end{enumerate}

        \algstore{bkbreak}
        
    \end{algorithmic}
\end{notoprulealgorithm}
\begin{notoprulealgorithm}
    \caption*{}
    \begin{algorithmic}
        \algrestore{bkbreak}
        \scriptsize

        \State update $\bet \given \bgamma, \brho, \bA$
        \begin{enumerate}[(a)]
            \item let $\widetilde{\bet} = \log \lrnd \frac{\bet}{M-\bet} \rrnd $ and sample $\widetilde{\bet}^*$ from $\Norm(\widetilde{\bet}, \xi_{\widetilde{\bet}})$ then $\bet^* = \frac{M\exp{(\widetilde{\bet}^*)}}{1 + \exp{(\widetilde{\bet}^*)}}$
            \item update $\bSigma^*_{\bgamma}$ with the proposed $\bet^*$
            \item accept $\widetilde{\bet}^*$ with probability 
            \begin{equation*}
                \min\lcur 1, \frac{\lvert \bSigma^*_{\bgamma} \rvert^{-\frac{N}{2}}  \exp{\lrnd - \frac{\bgamma^{T}\bSigma^{*-1}_{\bgamma}\bgamma}{2} \rrnd} \prod_{d=1}^q \frac{\exp{(\widetilde{\eta}^*_{d})}}{(1 + \exp{(\widetilde{\eta}^*_{d}))^2}}}{\lvert \bSigma_{\bgamma} \rvert^{-\frac{N}{2}}  \exp{\lrnd - \frac{\bgamma^{T}\bSigma^{-1}_{\bgamma}\bgamma}{2} \rrnd}\prod_{d=1}^q \frac{\exp{(\widetilde{\eta}_{d})}}{(1 + \exp{(\widetilde{\eta}_{d}))^2}}} \rcur
            \end{equation*}
            \item adapt $\xi_{\widetilde{\bet}}$
        \end{enumerate}
        
        \State 
        \textbf{Unstructured} update $\bA \given \bgamma, \brho, \bet$
        \begin{enumerate}[(a)]
            \item let diagonal elements $\widetilde{a}_{dd} = \log{(a_{dd})}$ and off-diagonal elements $\widetilde{a}_{dh} = a_{dh}$ for $h < d$. Then sample candidates $\widetilde{\ba}^*$ (lower triangular elements of $\bA$) from  $\Norm(\widetilde{\ba}, \xi_{\widetilde{\ba}})$ and set $a^*_{dd} = \exp{(\widetilde{a}^*_{dd})}$ and for $h < d$, $a^*_{dh} = \widetilde{a}^*_{dh}$
            \item construct the proposed $\bA^*$ and update $\bSigma^*_{\bgamma}$ accordingly
            \item accept $\bA^*$ with probability
            \begin{equation*}
                \min\lcur 1, \frac{\lvert \bSigma^*_{\bgamma} \rvert^{-\frac{N}{2}}  \exp{\lrnd - \frac{\bgamma^{T}\bSigma^{*-1}_{\bgamma}\bgamma}{2} \rrnd} p(\bA^*)\prod_{d=1}^q a^*_{dd}}{\lvert \bSigma_{\bgamma} \rvert^{-\frac{N}{2}}  \exp{\lrnd - \frac{\bgamma^{T}\bSigma^{-1}_{\bgamma}\bgamma}{2} \rrnd}p(\bA)\prod_{d=1}^q a_{dd}} \rcur
            \end{equation*}
            \item adapt $\xi_{\widetilde{\ba}}$
        \end{enumerate}
        \Statex
        \textbf{Directed} sample $\balpha_{d} \given \bgamma, \brho_{d}$ from
        \begin{equation*}
            \Norm(\bH_d h_d,H_d)
        \end{equation*}
        where $\bH_d = \lrnd \bdelta_d^\T Q_{d} \bdelta_d + \bSigma_{\balpha}^{-1} \rrnd^{-1}$ and $\bh_{d} = \bdelta_d^\T Q_{d} \bgamma_d + \bSigma_{\balpha}^{-1} \bmu_{\balpha}$. 
        Here $\bdelta_d = (F_h)_{h \in N_d}$, $F_h = (\bgamma_h,\bzeta_h)$, $N_d = \{h: h \sim d, h < d \}$ and $\bzeta_h = \lrnd \sum_{j\sim 1} \bgamma_{h,j}, \dots, \sum_{j \sim n} \bgamma_{h,j} \rrnd$.
        Then update $\bA_{d,h}$ matrices and consequently $\bSigma_{\bgamma}$.
        \vspace{2.5em}
        \Statex
        \textbf{Undirected} update $\rho_{dis} \given \bgamma, \brho, \bet$
        \begin{enumerate}[(a)]
            \item let $\widetilde{\rho}_{dis} = \log \lrnd \frac{\rho_{dis} + 1}{1-\rho_{dis}} \rrnd$ and sample $\widetilde{\rho}_{dis}^*$ from $\Norm(\widetilde{\rho}_{dis}, \xi_{\widetilde{\rho}_{dis}})$ then $\rho_{dis}^* = \frac{\exp{(\widetilde{\rho}_{dis}^*)}-1}{1 + \exp{(\widetilde{\rho}_{dis}^*)}}$
            \item update $\bLambda^*_{dis}$ with the proposed $\rho_{dis}^*$ and then $\bSigma^*_{\bgamma}$
            \item accept $\widetilde{\rho}_{dis}^*$ with probability 
            \begin{equation*}
                \min\lcur 1, \frac{\lvert \bSigma^*_{\bgamma} \rvert^{-\frac{N}{2}}  \exp{\lrnd - \frac{\bgamma^{T}\bSigma^{*-1}_{\bgamma}\bgamma}{2} \rrnd} \frac{\exp{(\widetilde{\rho}^*_{dis})}}{(1 + \exp{(\widetilde{\rho}^*_{dis}))^2}}}{\lvert \bSigma_{\bgamma} \rvert^{-\frac{N}{2}}  \exp{\lrnd - \frac{\bgamma^{T}\bSigma^{-1}_{\bgamma}\bgamma}{2} \rrnd} \frac{\exp{(\widetilde{\rho}_{dis})}}{(1 + \exp{(\widetilde{\rho}_{dis}))^2}}} \rcur
            \end{equation*}
            \item adapt $\xi_{\widetilde{\rho}_{dis}}$
        \end{enumerate}
        
    \end{algorithmic}
     \hrulefill
\end{notoprulealgorithm}

\clearpage

\section{Simulations}\label{sec: simul}
\color{orange}
\begin{table}[htbp!]
    \centering
    \renewcommand{\arraystretch}{1.167}
    \caption{\color{orange} Root median squared error for all the model parameters under all the simulation settings.}
    \label{tab:rmse}
    \begin{adjustbox}{max width=\textwidth}
    \begin{tabular}{cccccccccc}
        \hline
        True graph & \multicolumn{3}{c}{Unstructured} & \multicolumn{3}{c}{Directed} & \multicolumn{3}{c}{Undirected} \\
        \cmidrule(lr){1-1} \cmidrule(lr){2-4} \cmidrule(lr){5-7} \cmidrule(lr){8-10} 
        Fitted graph & Unstructured & Directed & Undirected & Unstructured & Directed & Undirected & Unstructured & Directed & Undirected  \\
        \hline
         $\bbeta$      & 1.585      & 1.475      & 1.107      & 0.993      & 0.869      & 0.793      & 0.642      & 0.548      & 0.466      \\
$\bphi_1$     & 7.364      & 7.286      & 6.541      & 15.345     & 6.862      & 15.915     & 16.559     & 16.478     & 6.052      \\
$\bphi_2$     & 7.937      & 7.088      & 7.033      & 14.943     & 6.491      & 16.339     & 16.226     & 16.431     & 6.447      \\
$\bphi_3$     & 7.971      & 7.126      & 6.832      & 15.386     & 6.875      & 15.549     & 16.440     & 15.226     & 6.265      \\
$\bphi_4$     & 7.908      & 6.924      & 6.923      & 15.254     & 6.805      & 15.488     & 16.566     & 15.638     & 6.345      \\
$\bgamma_1$   & 8.160      & 8.215      & 7.995      & 8.727      & 8.623      & 8.342      & 7.995      & 8.004      & 7.487      \\
$\bgamma_2$   & 12.448     & 14.472     & 4.419      & 14.008     & 16.880     & 5.140      & 10.869     & 13.610     & 4.031      \\
$\bgamma_3$   & 14.770     & 29.796     & 7.681      & 16.414     & 33.943     & 7.792      & 13.457     & 29.285     & 6.770      \\
$\bgamma_4$   & 18.784     & 45.308     & 9.233      & 23.094     & 72.170     & 9.337      & 15.280     & 44.172     & 6.264      \\
$\btheta$     & 7.783      & 7.579      & 7.892      & 8.675      & 8.196      & 9.208      & 8.782      & 8.418      & 8.479      \\
$\btau_{dis}$     & 7.869      & 7.864      & 8.931      & 15.854     & 12.490     & 21.795     & 21.822     & 17.967     & 26.407     \\
$\bV$          & 1.739      & 1.716      & 1.762      & 1.888      & 1.830      & 1.882      & 1.851      & 1.865      & 1.876      \\
$\tau$       & 0.311      & 0.348      & 0.253      & 0.305      & 0.404      & 0.261      & 0.387      & 0.512      & 0.351      \\
$\brho$       & 0.763      & 0.784      & 0.798      & 0.817      & 0.750      & 0.707      & 0.726      & 0.695      & 0.666      \\
$\bet$       & 0.332      & 0.359      & 0.360      & 0.368      & 0.329      & 0.344      & 0.332      & 0.340      & 0.301      \\
$\bA$ & 7.625      & -      & -      & -      & -     & -      & -      & -      & -      \\
$\balpha$ & -      & -      & -      & -      & 7.789      & -      & -      & -      & -      \\
$\rho_{dis}$ & -      & -      & -      & -      & -      & -      & -      & -      & 0.250      \\
        \hline
    \end{tabular}
    \end{adjustbox}
\end{table}

\begin{table}[htbp!]
    \centering
    \renewcommand{\arraystretch}{1.167}
    \caption{\color{orange} Boundary detection results in the simulation study with DAGAR spatial dependence for the three disease graph models with data generated under the corresponding true model.}
    \label{tab:bd_dd}
    \begin{adjustbox}{max width=\textwidth}
    \begin{tabular}{ccccccccccc}
        \hline
        & & \multicolumn{2}{c}{Disease 1} & \multicolumn{2}{c}{Disease 2} & \multicolumn{2}{c}{Disease 3} & \multicolumn{2}{c}{Disease 4} \\
        \cmidrule(lr){3-4} \cmidrule(lr){5-6} \cmidrule(lr){7-8} \cmidrule(lr){9-10}
        Disease graph & T & Specificity & Sensitivity & Specificity & Sensitivity & Specificity & Sensitivity & Specificity & Sensitivity \\
        \hline
        Unstructured & 85 & 0.553 & 0.868 & 0.513 & 0.891 & 0.538 & 0.885 & 0.544 & 0.880 \\ 
                     & 90 & 0.503 & 0.882 & 0.454 & 0.906 & 0.482 & 0.900 & 0.490 & 0.893 \\ 
                     & 95 & 0.433 & 0.906 & 0.402 & 0.921 & 0.428 & 0.913 & 0.434 & 0.904 \\ 
                     & 100 & 0.383 & 0.921 & 0.342 & 0.932 & 0.359 & 0.930 & 0.379 & 0.918 \\ 
                     & 105 & 0.329 & 0.931 & 0.297 & 0.940 & 0.296 & 0.942 & 0.330 & 0.929 \\ 
        \hline
        Directed & 85 & 0.563 & 0.822 & 0.527 & 0.867 & 0.491 & 0.877 & 0.480 & 0.872 \\
                 & 90 & 0.528 & 0.838 & 0.482 & 0.881 & 0.450 & 0.893 & 0.437 & 0.881 \\
                 & 95 & 0.487 & 0.854 & 0.436 & 0.896 & 0.405 & 0.903 & 0.395 & 0.892 \\
                 & 100 & 0.441 & 0.871 & 0.386 & 0.906 & 0.359 & 0.915 & 0.351 & 0.903 \\
                 & 105 & 0.390 & 0.890 & 0.333 & 0.921 & 0.310 & 0.924 & 0.305 & 0.914 \\
        \hline
        Undirected & 85 & 0.645 & 0.838 & 0.516 & 0.864 & 0.589 & 0.861 & 0.596 & 0.857 \\ 
                   & 90 & 0.593 & 0.856 & 0.470 & 0.881 & 0.539 & 0.875 & 0.548 & 0.876 \\ 
                   & 95 & 0.534 & 0.871 & 0.424 & 0.892 & 0.487 & 0.888 & 0.494 & 0.889 \\ 
                   & 100 & 0.476 & 0.887 & 0.377 & 0.906 & 0.437 & 0.904 & 0.434 & 0.902 \\ 
                   & 105 & 0.419 & 0.905 & 0.330 & 0.917 & 0.379 & 0.918 & 0.381 & 0.916 \\ 
        \hline
    \end{tabular}
    \end{adjustbox}
\end{table}

\begin{table}[htbp!]
    \centering
    \renewcommand{\arraystretch}{1.167}
    \caption{\color{orange} Boundary detection results in the simulation study with CAR spatial dependence for the three disease graph models with data generated under the corresponding true model.}
    \label{tab:bd_dd_MCAR}
    \begin{adjustbox}{max width=\textwidth}
    \begin{tabular}{ccccccccccc}
        \hline
        & & \multicolumn{2}{c}{Disease 1} & \multicolumn{2}{c}{Disease 2} & \multicolumn{2}{c}{Disease 3} & \multicolumn{2}{c}{Disease 4} \\
        \cmidrule(lr){3-4} \cmidrule(lr){5-6} \cmidrule(lr){7-8} \cmidrule(lr){9-10}
        Disease graph & T & Specificity & Sensitivity & Specificity & Sensitivity & Specificity & Sensitivity & Specificity & Sensitivity \\
        \hline
        Unstructured & 85 & 0.652 & 0.859 & 0.630 & 0.869 & 0.650 & 0.873 & 0.655 & 0.871 \\
                     & 90 & 0.601 & 0.875 & 0.581 & 0.883 & 0.599 & 0.886 & 0.604 & 0.888 \\
                     & 95 & 0.544 & 0.890 & 0.527 & 0.897 & 0.542 & 0.903 & 0.545 & 0.900 \\
                     & 100 & 0.480 & 0.904 & 0.463 & 0.913 & 0.485 & 0.918 & 0.487 & 0.913 \\
                     & 105 & 0.418 & 0.920 & 0.408 & 0.926 & 0.425 & 0.932 & 0.426 & 0.926 \\
        \hline
        Directed & 85 & 0.669 & 0.860 & 0.637 & 0.895 & 0.551 & 0.896 & 0.537 & 0.896 \\
                 & 90 & 0.618 & 0.877 & 0.584 & 0.909 & 0.501 & 0.907 & 0.485 & 0.907 \\
                 & 95 & 0.561 & 0.893 & 0.527 & 0.919 & 0.452 & 0.919 & 0.437 & 0.918 \\
                 & 100 & 0.503 & 0.908 & 0.469 & 0.929 & 0.400 & 0.930 & 0.387 & 0.928 \\
                 & 105 & 0.445 & 0.924 & 0.410 & 0.938 & 0.350 & 0.941 & 0.323 & 0.941 \\
        \hline
        Undirected & 85 & 0.666 & 0.871 & 0.618 & 0.885 & 0.643 & 0.871 & 0.648 & 0.879 \\ 
                   & 90 & 0.611 & 0.887 & 0.567 & 0.900 & 0.589 & 0.883 & 0.596 & 0.894 \\ 
                   & 95 & 0.553 & 0.899 & 0.514 & 0.915 & 0.535 & 0.898 & 0.538 & 0.907 \\ 
                   & 100 & 0.498 & 0.915 & 0.459 & 0.926 & 0.480 & 0.911 & 0.480 & 0.921 \\ 
                   & 105 & 0.440 & 0.929 & 0.405 & 0.939 & 0.423 & 0.924 & 0.422 & 0.933 \\ 
        \hline
    \end{tabular}
    \end{adjustbox}
\end{table}

\begin{table}[htbp!]
    \centering
    \renewcommand{\arraystretch}{1.167}
    \caption{\color{orange} Adjacencies detection with DAGAR spatial dependence for the three disease graph models with data generated under the corresponding true model.}
    \label{tab:adj_det}
    \begin{adjustbox}{max width=\textwidth}
    \begin{tabular}{ccccccccc}
        \hline
         & \multicolumn{2}{c}{Disease 1} & \multicolumn{2}{c}{Disease 2} & \multicolumn{2}{c}{Disease 3} & \multicolumn{2}{c}{Disease 4} \\
        \cmidrule(lr){2-3} \cmidrule(lr){4-5} \cmidrule(lr){6-7} \cmidrule(lr){8-9}
        Disease graph & Specificity & Sensitivity & Specificity & Sensitivity & Specificity & Sensitivity & Specificity & Sensitivity \\
        \hline 
        Unstructured & 0.386 & 0.986 & 0.598 & 0.928 & 0.600 & 0.959 & 0.261 & 1.000 \\
        Directed & 0.406 & 0.985 & 0.786 & 0.939 & 0.668 & 0.954 & 0.213 & 1.000 \\
        Undirected & 0.415 & 0.981 & 0.753 & 0.927 & 0.571 & 0.969 & 0.225 & 0.990 \\
        \hline
    \end{tabular}
    \end{adjustbox}
\end{table}

\begin{table}[htbp!]
    \centering
    \renewcommand{\arraystretch}{1.167}
    \caption{\color{orange} Adjacencies detection with CAR spatial dependence for the three disease graph models with data generated under the corresponding true model.}
    \label{tab:adj_det_MCAR}
    \begin{adjustbox}{max width=\textwidth}
    \begin{tabular}{ccccccccc}
        \hline
         & \multicolumn{2}{c}{Disease 1} & \multicolumn{2}{c}{Disease 2} & \multicolumn{2}{c}{Disease 3} & \multicolumn{2}{c}{Disease 4} \\
        \cmidrule(lr){2-3} \cmidrule(lr){4-5} \cmidrule(lr){6-7} \cmidrule(lr){8-9}
        Disease graph & Specificity & Sensitivity & Specificity & Sensitivity & Specificity & Sensitivity & Specificity & Sensitivity \\
        \hline 
        Unstructured & 0.084 & 1.000 & 0.305 & 0.998 & 0.142 & 1.000 & 0.045 & 1.000 \\
        Directed & 0.160 & 0.995 & 0.445 & 0.963 & 0.306 & 0.973 & 0.211 & 1.000 \\
        Undirected & 0.846 & 0.776 & 0.834 & 0.632 & 0.888 & 0.672 & 0.848 & 0.947 \\
        \hline
    \end{tabular}
    \end{adjustbox}
\end{table}

\begin{figure}[htbp!]
    \centering
    \includegraphics[width = 0.66 \textwidth]{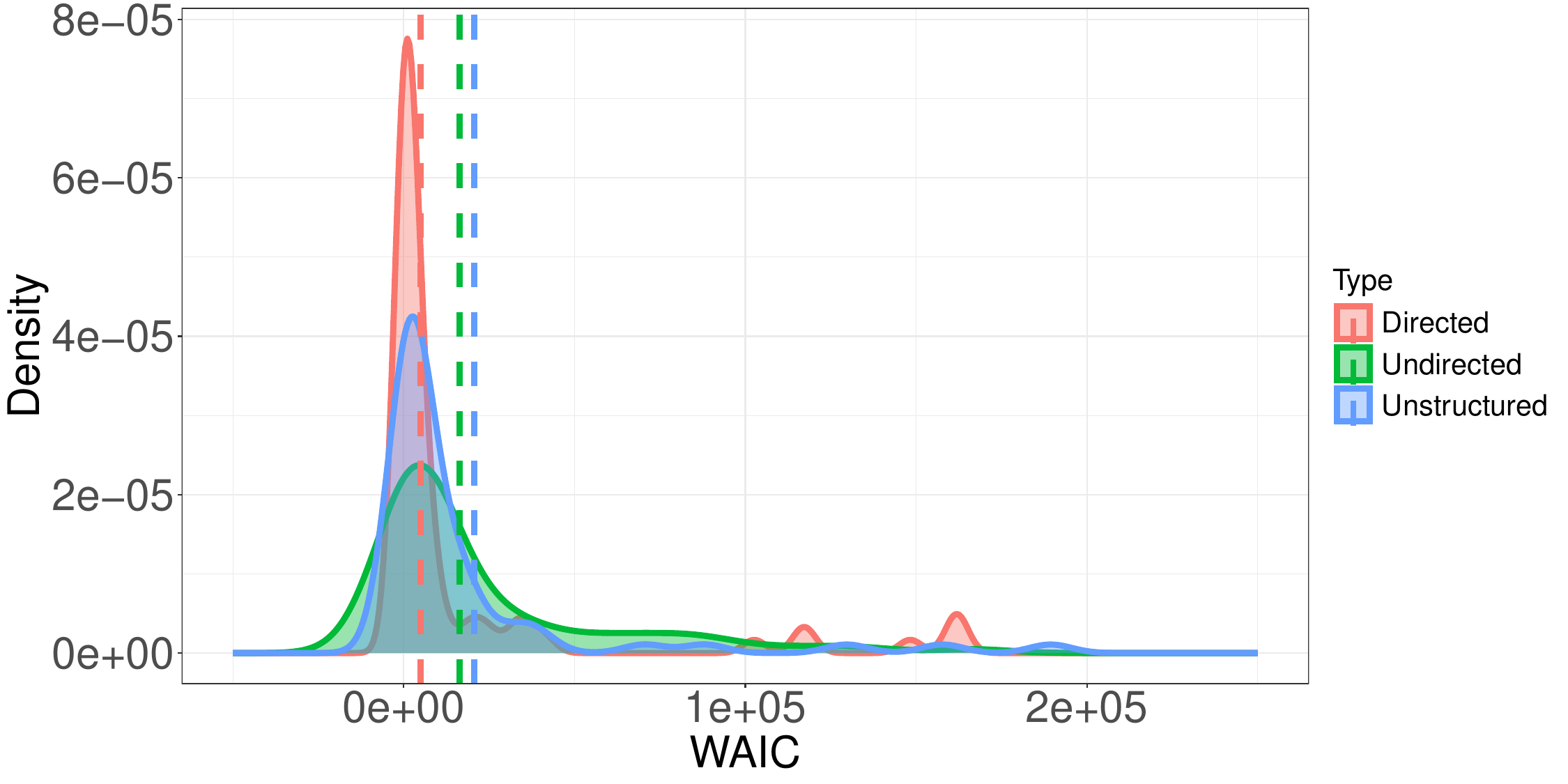}
    \caption{\color{orange} WAIC densities along with the median values (dotted lines) in the DAGAR spatial dependence case for different choices of the disease graph with data generated under the corresponding true model.}
    \label{fig:waic_comp}
\end{figure}

\begin{figure}[htbp!]
    \centering
    \includegraphics[width = 0.66 \textwidth]{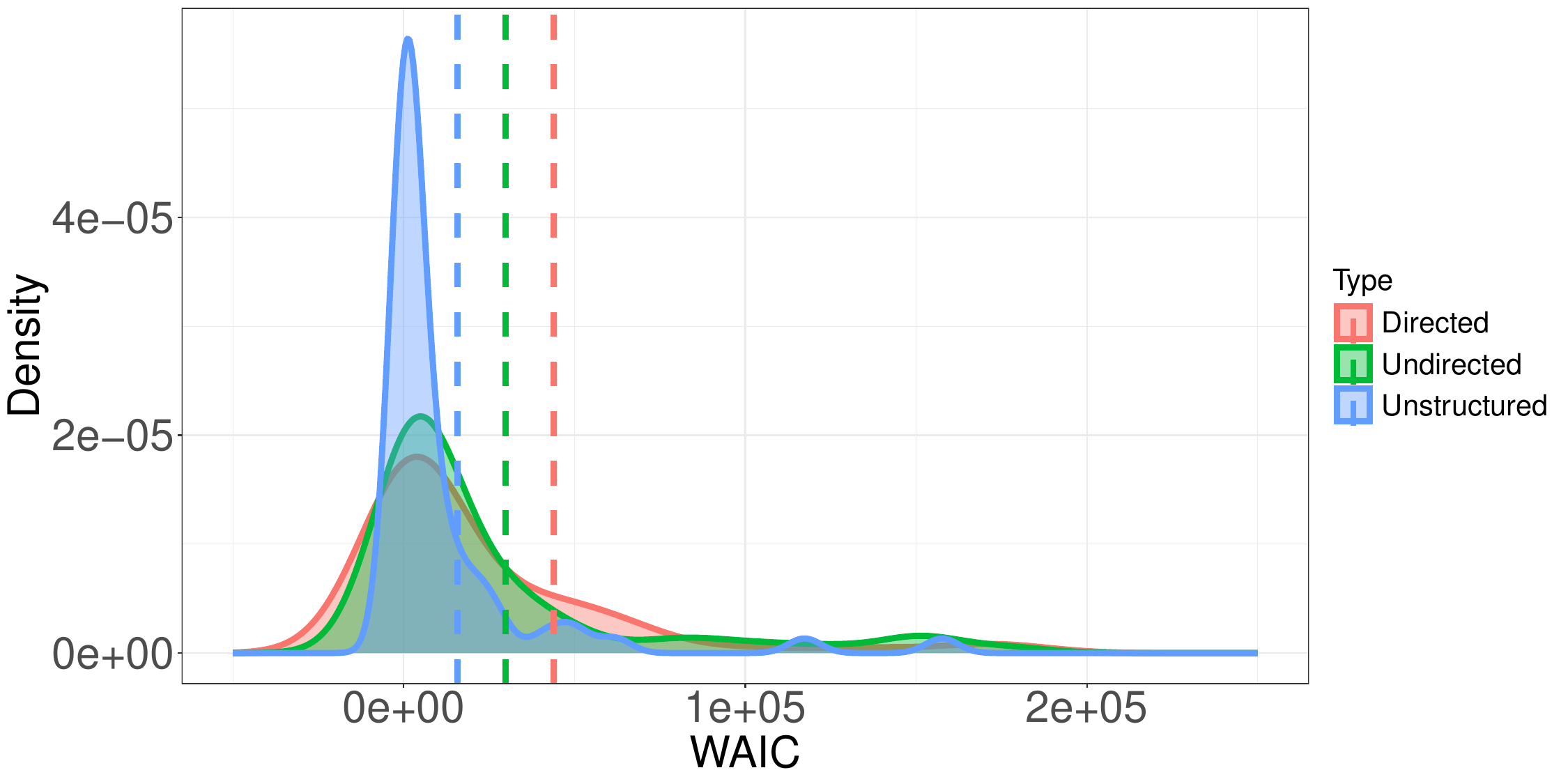}
    \caption{\color{orange} WAIC densities along with the median values (dotted lines) in the CAR spatial dependence case for different choices of the disease graph with data generated under the corresponding true model.}
    \label{fig:waic_comp_MCAR}
\end{figure}

\newpage

\color{orange}
\section{Posterior summaries for SEER data analysis}\label{sec: post_inf}

\begin{figure}[htbp!]
    \centering
    \includegraphics[width=0.49\linewidth]{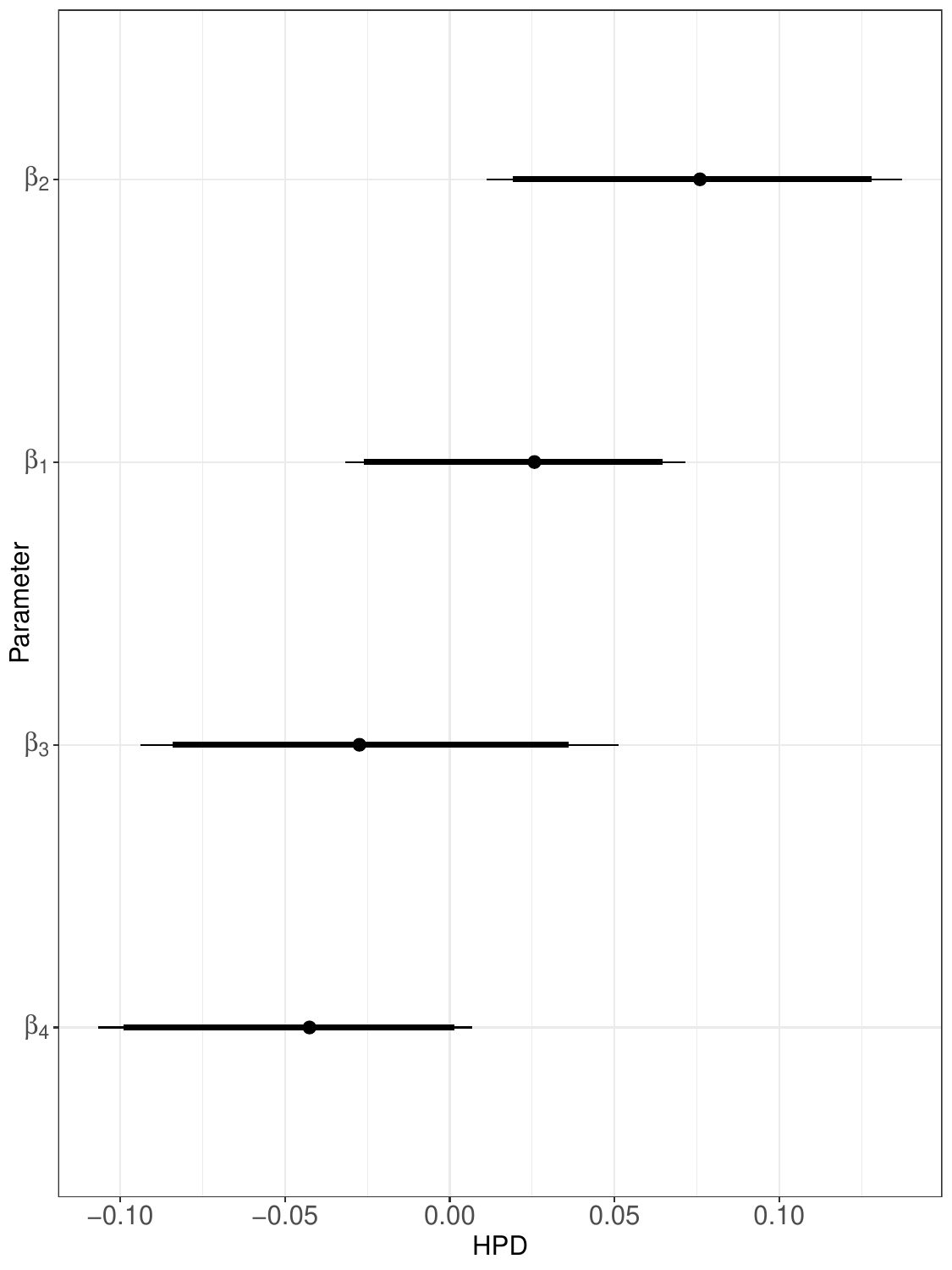}
    \includegraphics[width=0.49\linewidth]{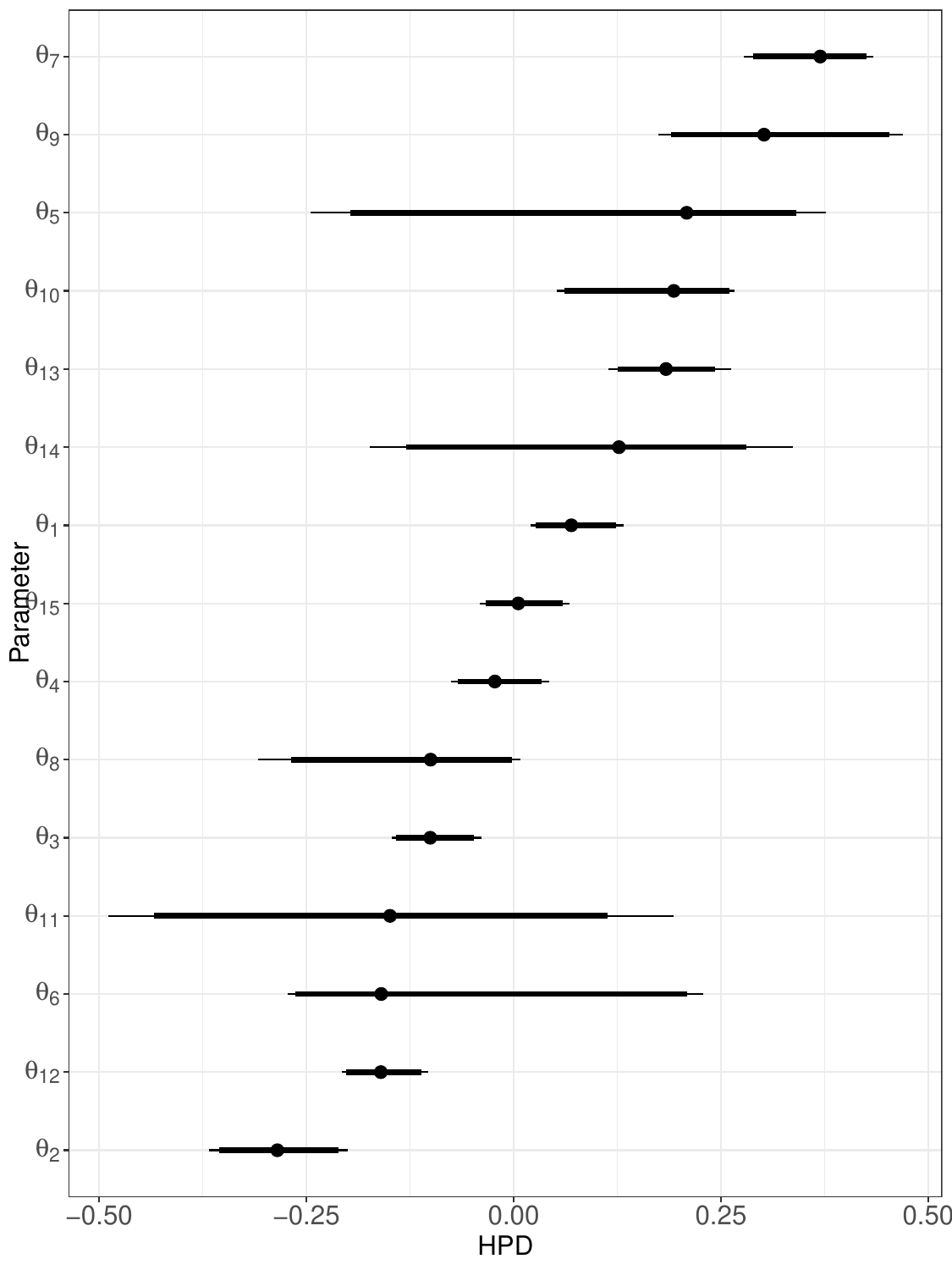}
    \caption{\color{orange} Highest posterior density (HPD) intervals at 90\% (thick lines) and 95\% (thin lines) levels for $\bbeta$ (left panel) and $\btheta$ (right panel). These are ordered according to their posterior medians. As with the posterior means in Table~\ref{tab:post_inf}, we see $\beta_1$ (lung) and $\beta_2$ (esophageal) having positive posterior medians with substantial mass greater than $0$, while $\beta_3$ (larynx) and $\beta_4$ (colorectal) having negative posterior medians with substantial mass less than $0$. The $90\%$ interval for colorectal cancer is almost entirely to the left of $0$. Turning to the right panel, we see about 4 atoms ($\theta_2$, $\theta_{12}$, $\theta_3$ and $\theta_8$) to have their HPD intervals almost entirely to the left of $0$, while 5 atoms ($\theta_1$, $\theta_{13}$, $\theta_{10}$, $\theta_9$ and $\theta_7$) have intervals situated entirely to the right of $0$. The remaining 6 atoms (out of a total of $K=15$) are not significantly different from $0$. The variation in the posterior distribution of these atoms reflect substantial posterior learning and information from the data that propagates to the spatial random effects.}
    \label{fig:post_beta_theta}
\end{figure}

\begin{figure}[htbp!]
    \centering
    \includegraphics[width=0.49\linewidth]{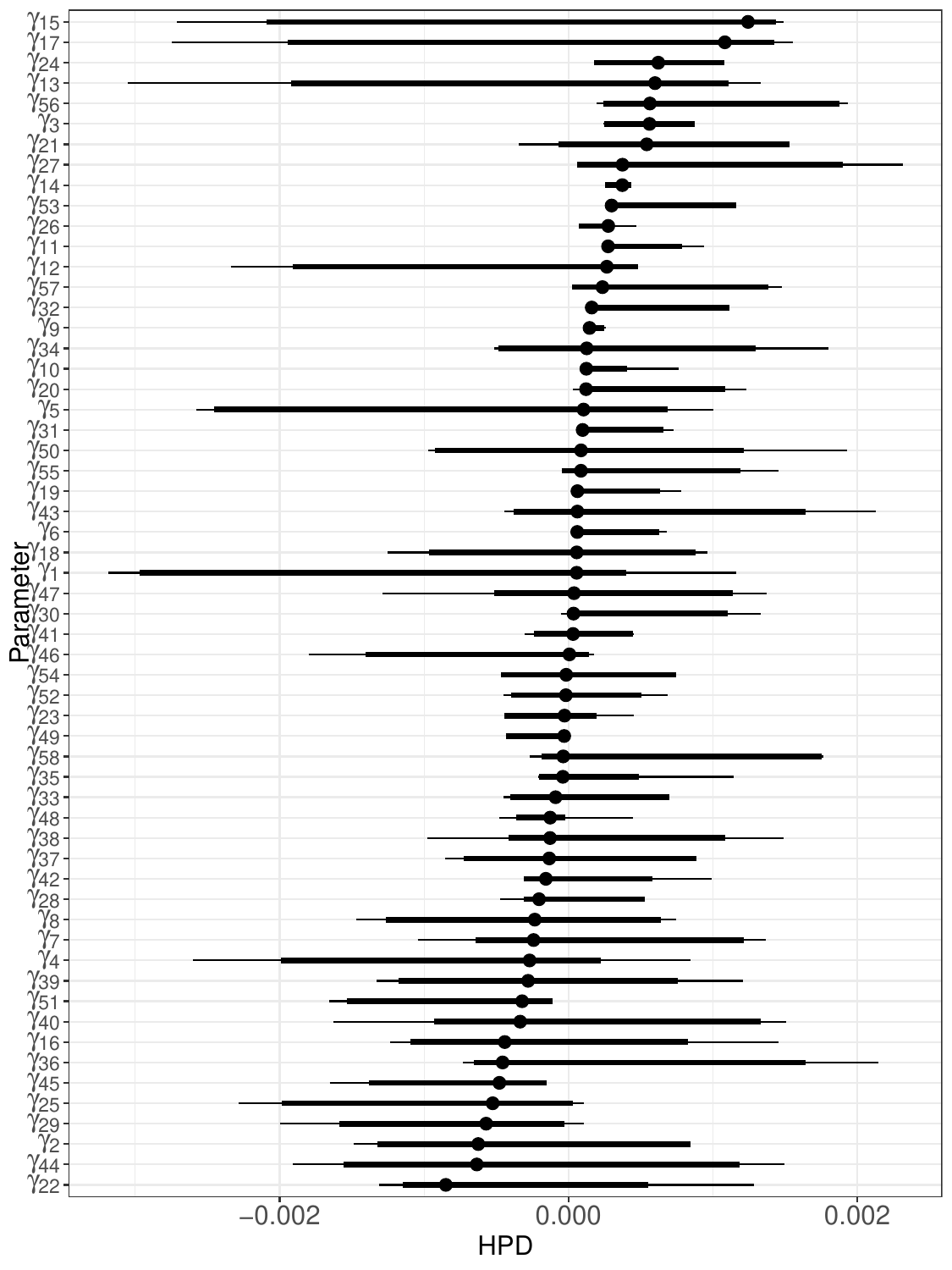}
    \includegraphics[width=0.49\linewidth]{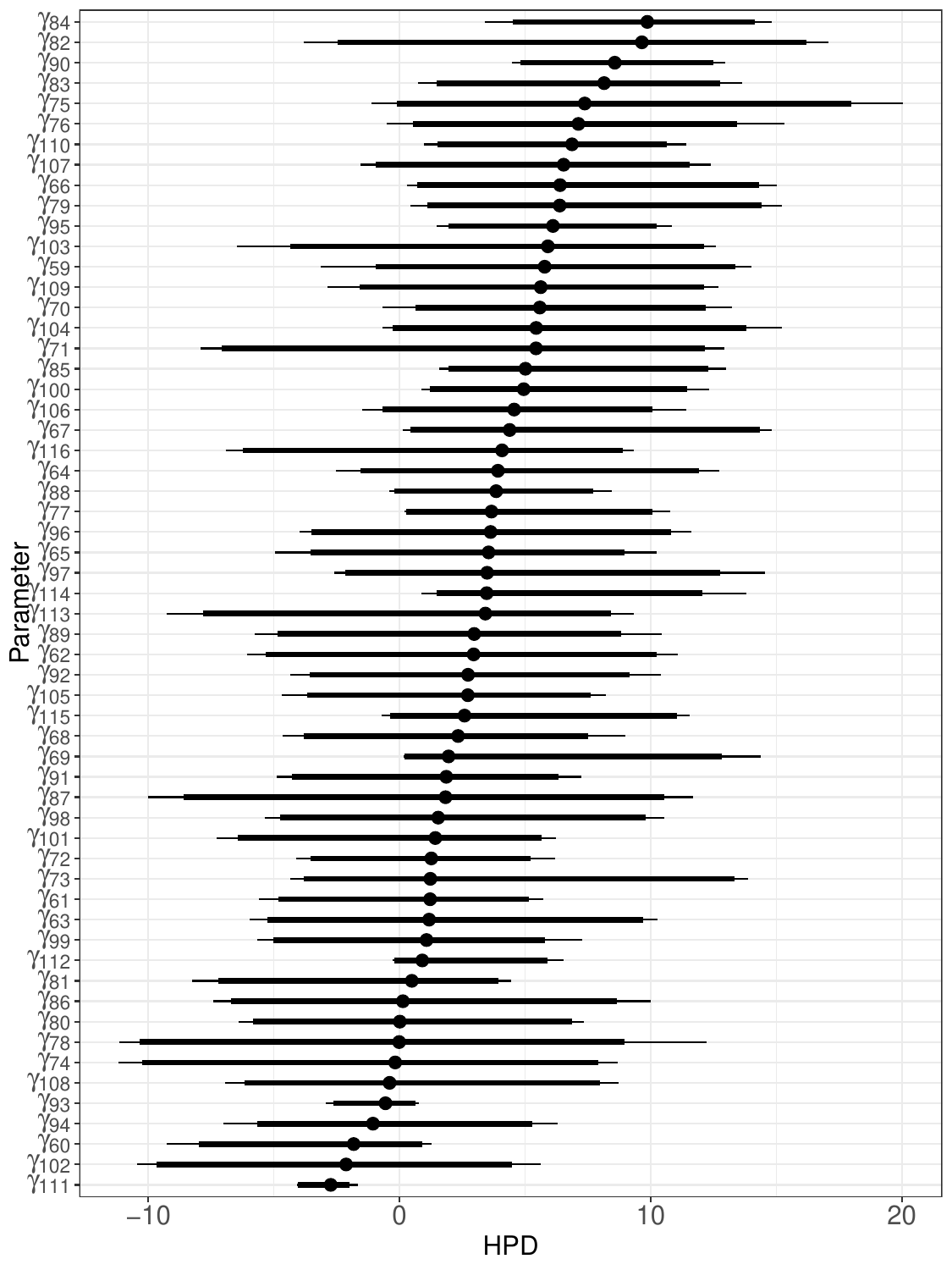}\\
    \includegraphics[width=0.49\linewidth]{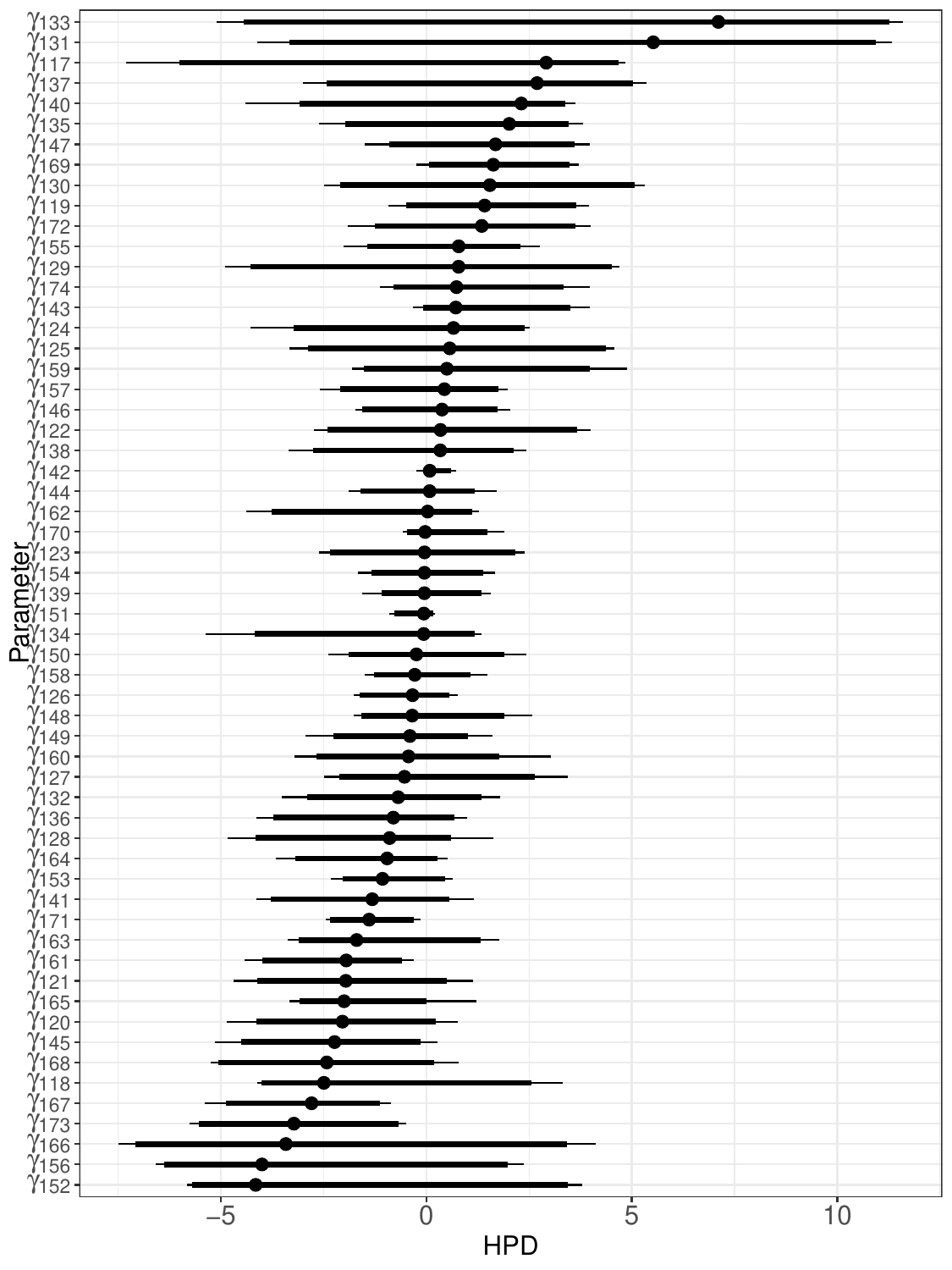}
    \includegraphics[width=0.49\linewidth]{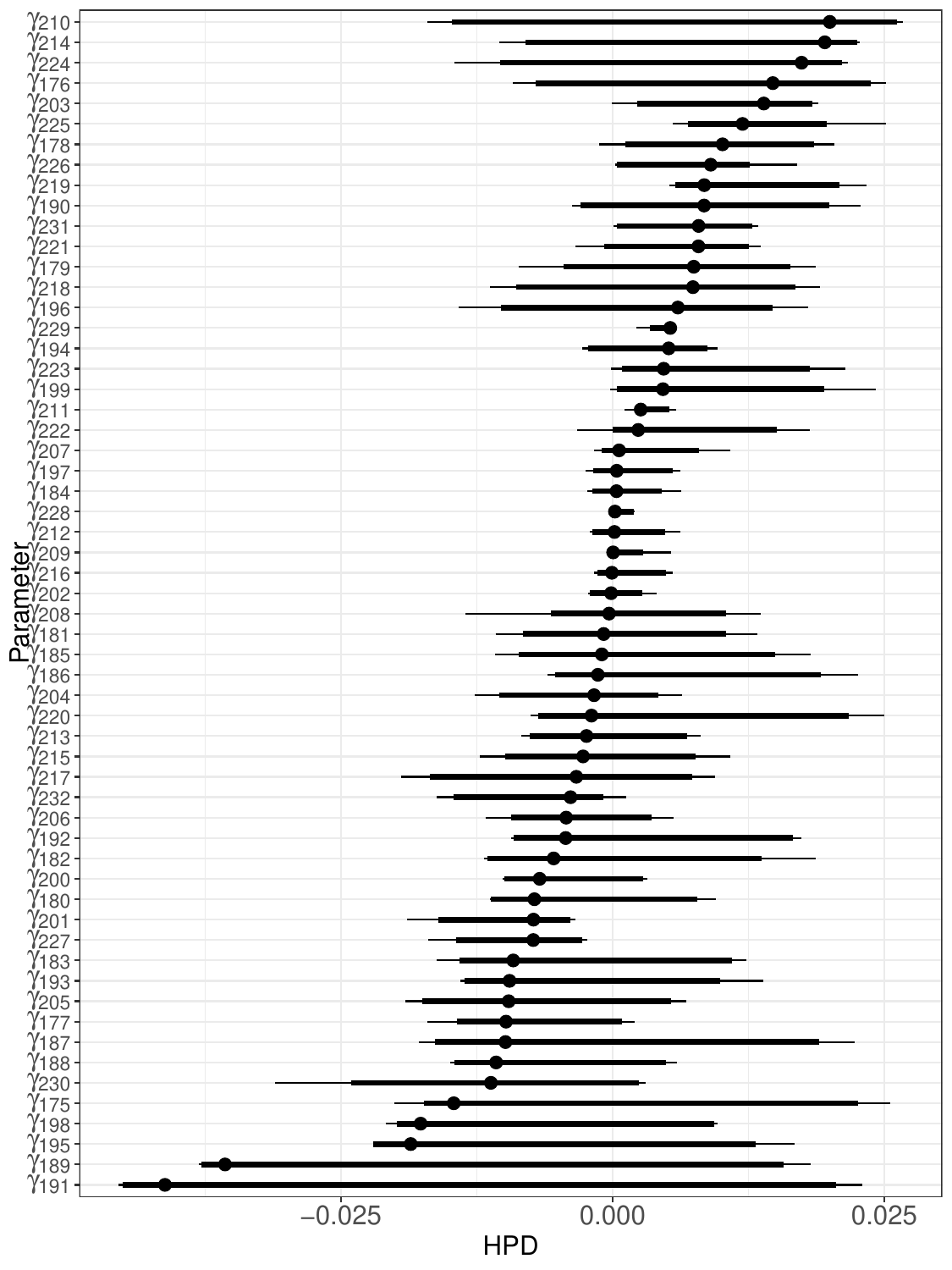}
    \caption{\color{orange} Highest posterior density (HPD) intervals at 90\% (thick lines) and 95\% (thin lines) levels for the elements of the disease-specific spatial effects in $\bgamma$ using the MDAGAR specification in Section~\ref{sec: unstructured_graph} and the precision matrix given in Eq.~\ref{eq:sp_cov}. The four panels correspond to the 4 cancers: lung (top left), esophageal (top right), larynx (bottom left) and colorectal (bottom right). The intervals are arranged in increasing order of value for posterior medians. It is expected that the intervals for these random effects will mostly not be significantly different from $0$, but they reveal substantial variation indicating enough information in the data to discern spatial patterns. It is worth pointing out that the dispersion varies substantially among the cancers with the intervals for lung and colorectal spanning a range much smaller than esophageal and larynx.}
    \label{fig:post_gamma}
\end{figure}

\begin{figure}[htbp!]
    \centering
    \includegraphics[width=0.49\linewidth]{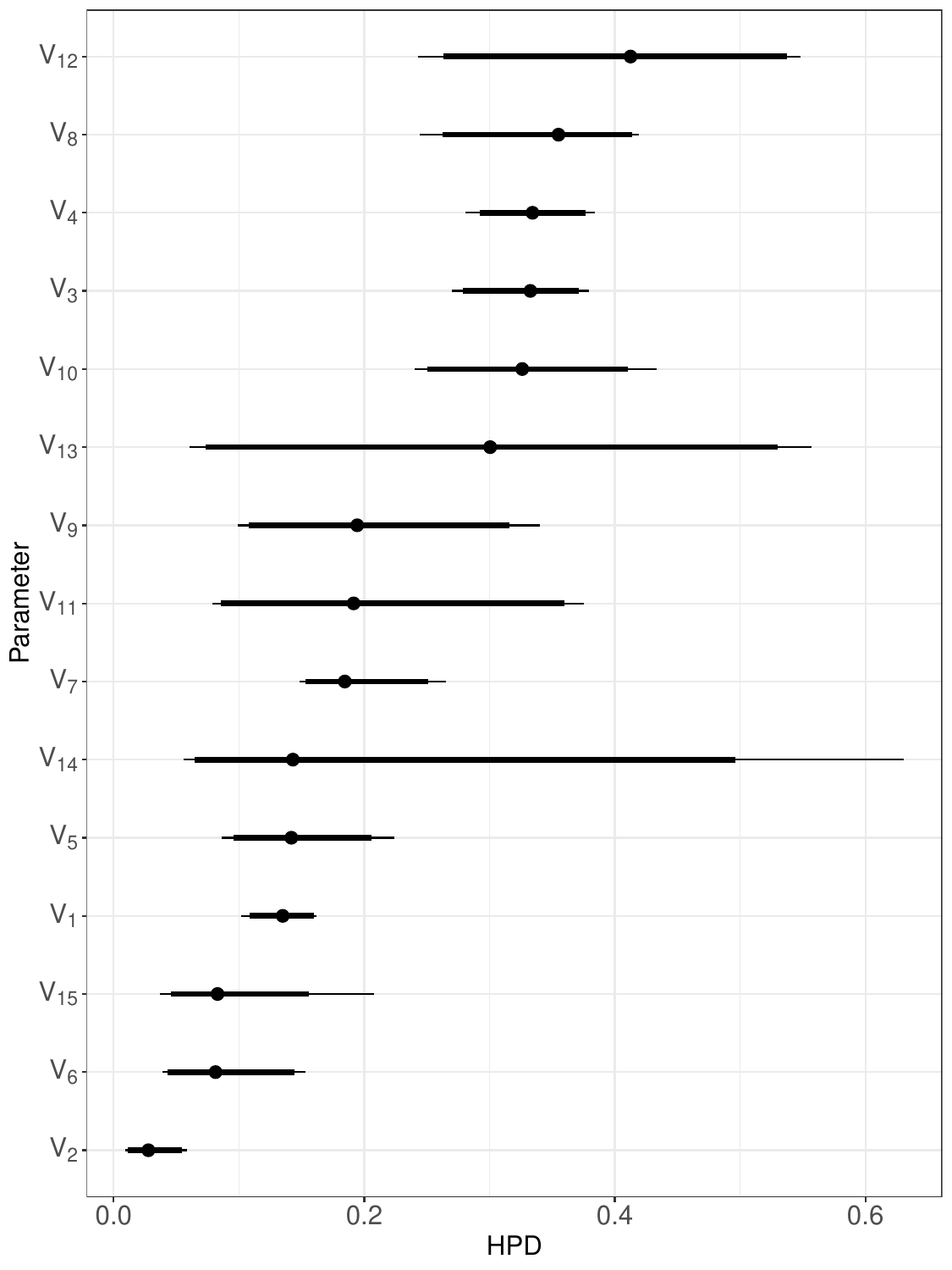}
    \includegraphics[width=0.49\linewidth]{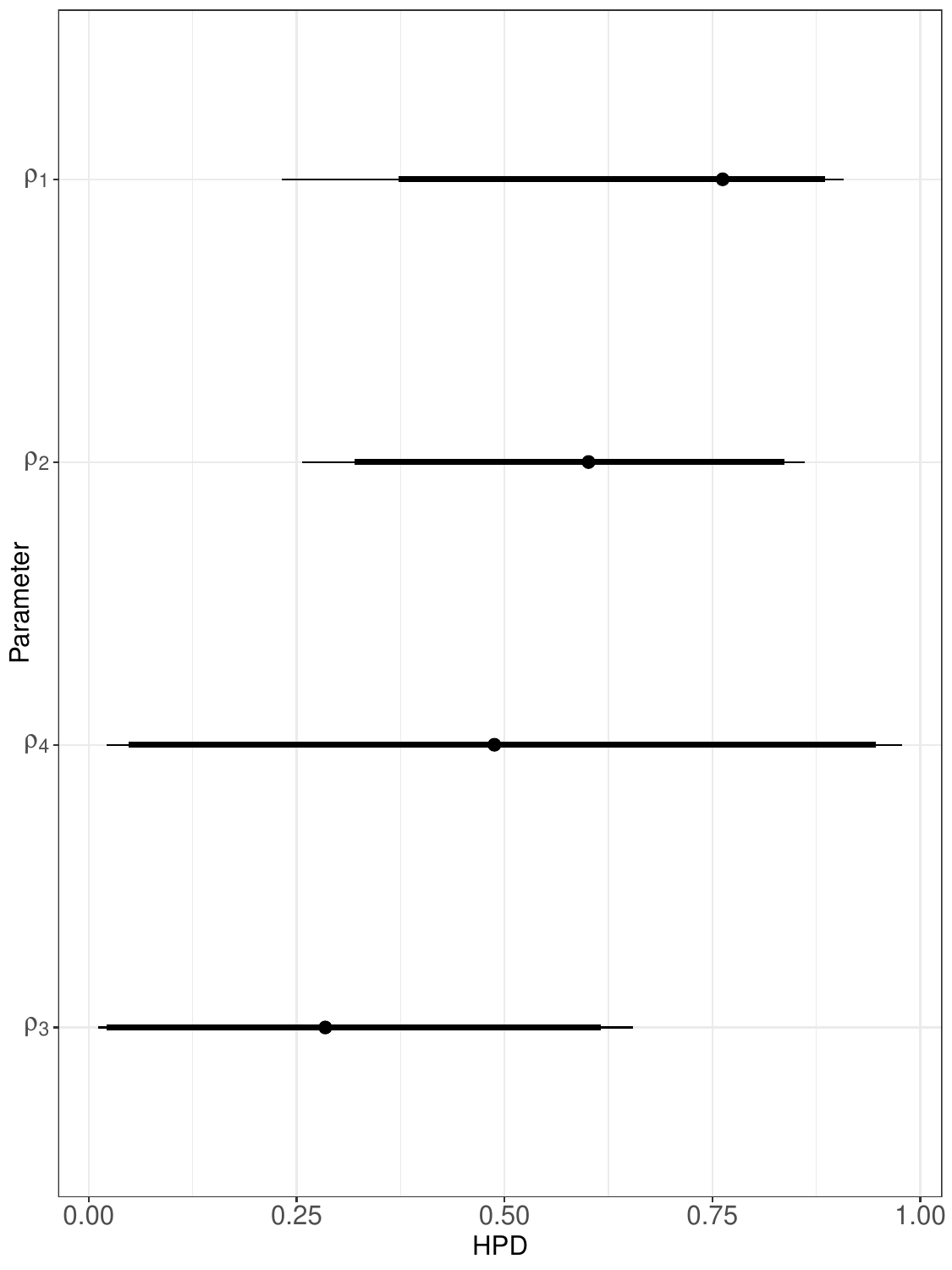}
    \caption{\color{orange} Highest posterior density (HPD) intervals at 90\% (thick lines) and 95\% (thin lines) levels for $\bV$ (left panel), which consists of the 15 stick-breaking weights used in Eq.~\eqref{eq:DP}, and for $\brho$ (right panel), which consists of the 4 spatial autocorrelation parameters (one for each disease) in the MDAGAR model. These intervals are arranged in increasing order of posterior medians. The variations among these posterior intervals indicate substantial learning from the data that informs the joint distribution of the spatial effects through the stick-breaking probabilities and the baseline MDAGAR distribution in the DP.}
    \label{fig:post_v_rho}
\end{figure}

\begin{figure}[h!]
    \centering
    \includegraphics[width=0.49\linewidth]{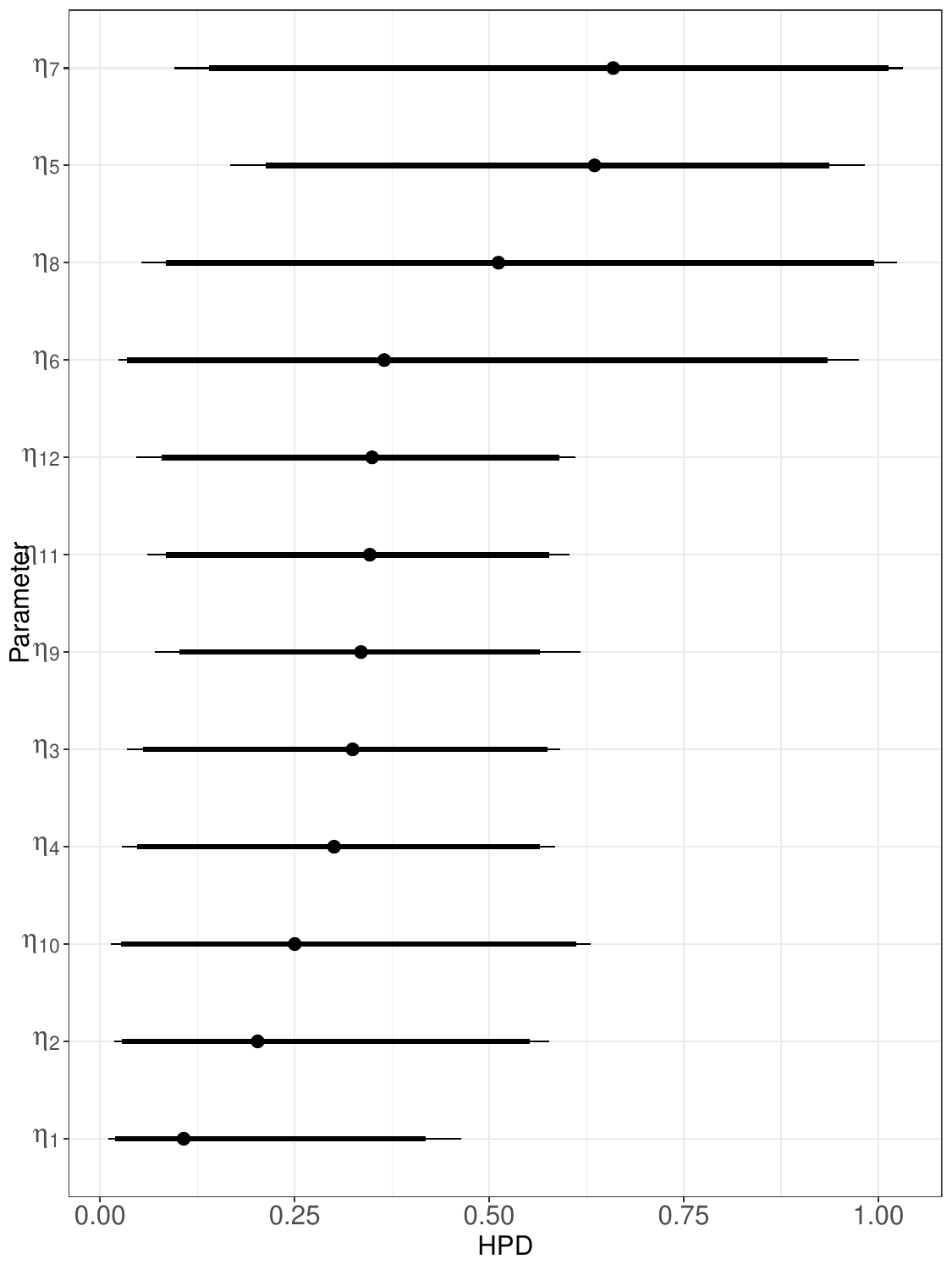}
    \includegraphics[width=0.49\linewidth]{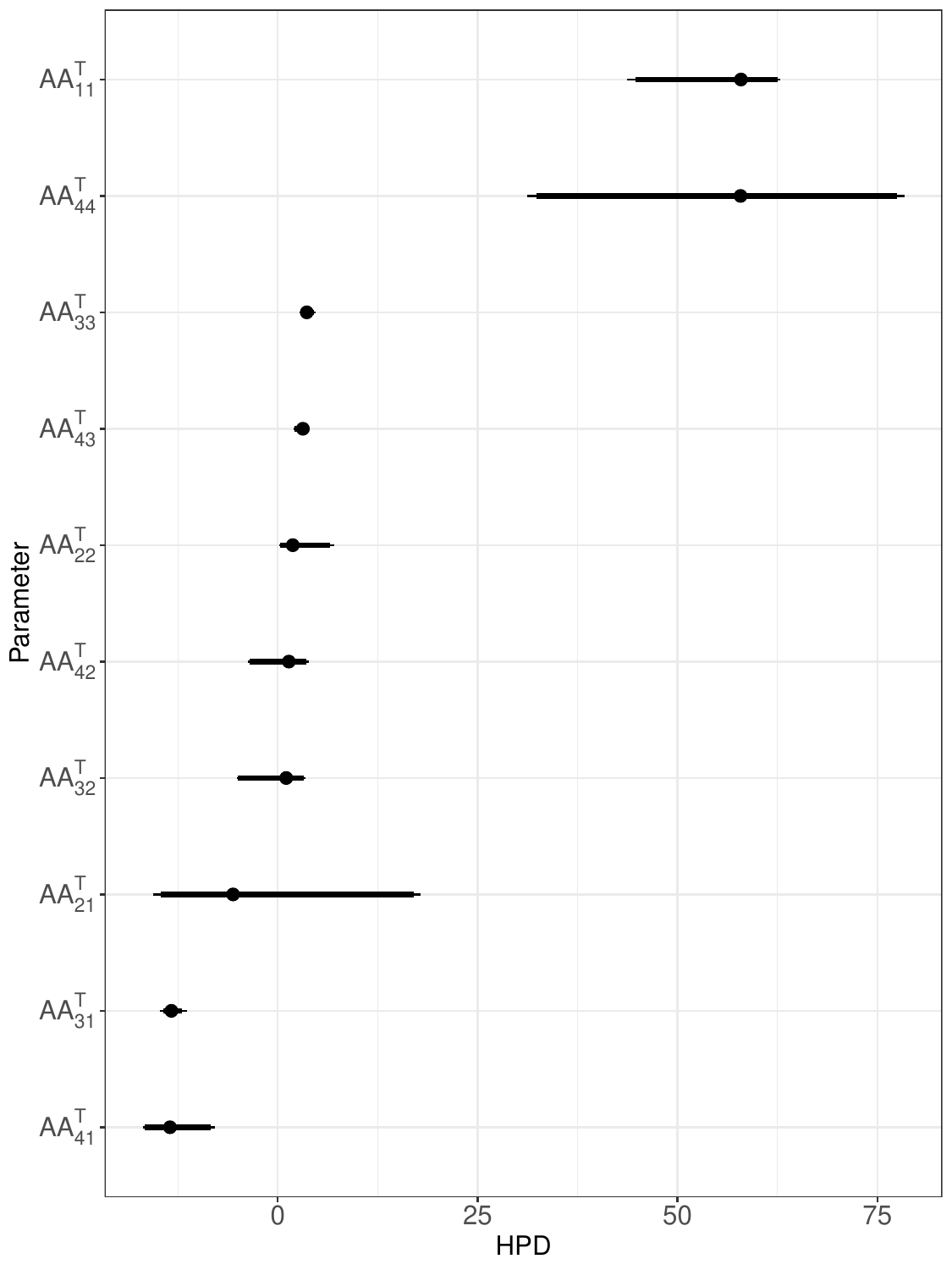}
    \caption{\color{orange} Highest posterior density (HPD) intervals at 90\% (thick lines) and 95\% (thin lines) levels for $\bet$ (left panel) and $\bA$ (right panel). The 12 coefficients in $\bet$ include 3 slopes for each of the 4 cancer types corresponding to the covariates shown in Figure~\ref{fig:covariates_plot}. For each of the 4 cancers, these parameters appear in \eqref{eq:w_binary} as coefficients of the $3\times 1$ vector $\bz_{ij}$, where the elements of $\bz_{ij}$ are $|z_{ki}-z_{kj}|$ for $k=1,2,3$ corresponding to the 3 covariates. The four slopes corresponding to smoking are denoted as $\eta_{1}$, $\eta_{4}$, $\eta_{7}$, and $\eta_{10}$. For the variable 'population over 65', the corresponding slopes are $\eta_{2}$, $\eta_{5}$, $\eta_{8}$, and $\eta_{11}$. The slopes related to poverty levels are $\eta_{3}$, $\eta_{6}$, $\eta_{9}$, and $\eta_{12}$. For all cancers considered, these coefficients correspond respectively to lung, esophageal, laryngeal, and colorectal cancers. Here, the greater the $\eta$'s, the more likely two regions are not considered to be adjacent by the model. For instance, the smoking covariate plays a significant role in determining adjacencies for larynx cancer ($\eta_7$), while the over 65 covariate has a similar impact for both esophageal and larynx cancers ($\eta_5$ and $\eta_8$). Additionally, the below poverty covariate appears to have a notable effect on the adjacencies related to esophageal cancer ($\eta_6$). In the right panel, the HPD intervals for the elements of $\bA\bA^{\T}$ are shown, representing the non-spatial covariance among the cancers. Lung and larynx cancers exhibit the highest variances $(\bA\bA^{\T})_{11}$ and $(\bA\bA^{\T}_{44})$, with larynx showing the widest interval. The non-spatial association between lung and larynx captured by $(\bA\bA^{\T})_{41}$ and that between lung and colorectal captured by $(\bA\bA^{\T})_{31}$ are negative with the $95\%$ intervals situated entirely below $0$. It is important to interpret the non-spatial component of these associations with caution. What can, perhaps, be gleaned from these negative non-spatial associations is a phenomenon of spatial confounding at the residual scale, where much of the positive associations among these aggregated data are absorbed by the spatial associations arising from geographic proximity.}
    \label{fig:post_eta_A}
\end{figure}

We present further analysis for the model used in Section~\ref{sec: cancer_analysis} of the main article. In this section the regression model $\bx_{ij}^{\T}\bbeta_{j}$ consists of only cancer-specific intercepts while $\bz_{ij}$ in the adjacency model is built from the absolute differences between geographic neighbors for the three covariates described in Section~\ref{sec: data} of the main article. We present posterior means and standard deviations in Table~\ref{tab:post_inf} and credible intervals in Fig.~\ref{fig:post_beta_theta},~\ref{fig:post_gamma},~\ref{fig:post_v_rho}~and~\ref{fig:post_eta_A}. 


Monte Carlo estimates, whether for Markov chain Monte Carlo output or any other purpose, should be accompanied by Monte Carlo standard error as a measure of reliability of posterior inference based on the estimates from the Monte Carlo samples. If the central limit theorem applies, then the corresponding covariance matrices' estimates must be reported, either directly or indirectly. Due to the correlation between the obtained samples, these quantities necessitate more advanced tools compared to typical sample estimators. The validity of a central limit theorem is not guaranteed in all cases because it depends on the convergence rate of the underlying Markov chain. Nonetheless, this assumption is also made when employing various convergence diagnostics. We use the \texttt{mcmcse R} package \citep{mcmcse} to compute Monte Carlo standard errors and the effective sample size, based on \cite{gong2016practical} to demonstrate that we can maintain control over the Monte Carlo error by applying a predefined threshold \cite{vats2019multivariate}.

Table~\ref{tab:post_inf} presents the estimates of the model described in Section~\ref{sec: cancer_analysis} including only cancer-specific intercepts in the regression model and the 3 covariates described in Section~\ref{sec: data} of the main article in the adjacency model. The last column of Table~\ref{tab:post_inf} presents Monte Carlo standard errors with the only value higher than 0.05, the usual threshold, being 0.061. We compute the multivariate effective sample size of the chain to be $1847.068$ using the formula $B \frac{\lvert \Lambda \rvert^{1/p}}{\lvert \Sigma \rvert^{1/p}}$. Here, $B$ is the total number of iterations post burn-in, $\Lambda$ represents the sample covariance matrix, and $\Sigma$ is an estimate of the Monte Carlo standard error. This computation accounts for the multivariate dependence structure of the process. Comparing our effective sample size to the minimum requirement of $7,619$ for achieving a precision level relative to the variability in the target distribution of $0.05$, we observe that ours is significantly lower. However, it is important to note that we computed the precision level of our estimate, obtained with an estimated effective sample size of $1847.068$, which amounts approximately to $0.01$ and, while higher than precision levels used in simpler models, is still deemed acceptable for the inference we draw using these complex hierarchical models.

\begin{table}[htbp!]
    \centering
    \small
    \caption{\color{orange} Posterior estimates, posterior standard deviations and Monte Carlo standard errors for the regression coefficients, atoms and their precision in the hierarchical model described \eqref{eq:post} using the Poisson regression model described in Section~\ref{sec: cancer_analysis} of the main article. Here, $\beta_1$ through $\beta_4$ denote the intercepts corresponding to lung, esophageal, larynx and colorectal, respectively. The $\theta$'s denote the atoms and $\tau$ is the precision in the prior of the $\theta_k$'s as described in Section~\ref{sec: sre}. While none of the intercepts are significant, we note that lung and esophageal cancers reveal positive estimates indicating higher expected counts than the standardized rates, $E_{id}$ defined in Section~\ref{sec: data} and introduced in the model in Section~\ref{sec: cancer_analysis}, while larynx and colorectal exhibit negative estimates indicating lower expected counts than their standardized rates.}
    \label{tab:post_inf}
    \begin{tabular}{cccc}
        \hline
        Variable & Estimate & Standard deviation & Monte Carlo Standard Errors\\
        \hline
        $\beta_1$  & $0.022$ & $0.027$ & $0.006$ \\ 
        $\beta_2$  & $0.075$ & $0.033$ & $0.005$\\ 
        $\beta_3$  & $-0.026$ & $0.037$ & $0.005$\\ 
        $\beta_4$  & $-0.045$ & $0.030$ & $0.007$\\ 
        $\theta_1$ & $0.072$ & $0.030$ & $0.007$\\ 
        $\theta_2$ & $-0.283$ & $0.043$ & $0.007$\\ 
        $\theta_3$ & $-0.098$ & $0.029$ & $0.006$\\ 
        $\theta_4$ & $-0.020$ & $0.031$ & $0.007$\\ 
        $\theta_5$ & $0.156$ & $0.162$ & $0.027$\\ 
        $\theta_6$ & $-0.072$ & $0.175$ & $0.061$\\ 
        $\theta_7$ & $0.365$ & $0.041$ & $0.009$\\ 
        $\theta_8$ & $-0.115$ & $0.085$ & $0.020$\\ 
        $\theta_9$ & $0.312$ & $0.086$ & $0.023$\\ 
        $\theta_{10}$ & $0.178$ & $0.062$ & $0.019$\\ 
        $\theta_{11}$ & $-0.149$ & $0.170$ & $0.029$\\ 
        $\theta_{12}$ & $-0.158$ & $0.028$ & $0.006$\\ 
        $\theta_{13}$ & $0.184$ & $0.037$ & $0.007$\\ 
        $\theta_{14}$ & $0.095$ & $0.139$ & $0.032$\\ 
        $\theta_{15}$ & $0.010$ & $0.029$ & $0.006$\\ 
        $\tau$  & $7.246$ & $2.368$ & $0.024$\\ 
        \hline
    \end{tabular}
\end{table}

\clearpage

\section{Additional figures and tables}\label{sec: additional_analysis}

\begin{figure}[htbp!]
    \centering
    \includegraphics[width = 0.75\textwidth]{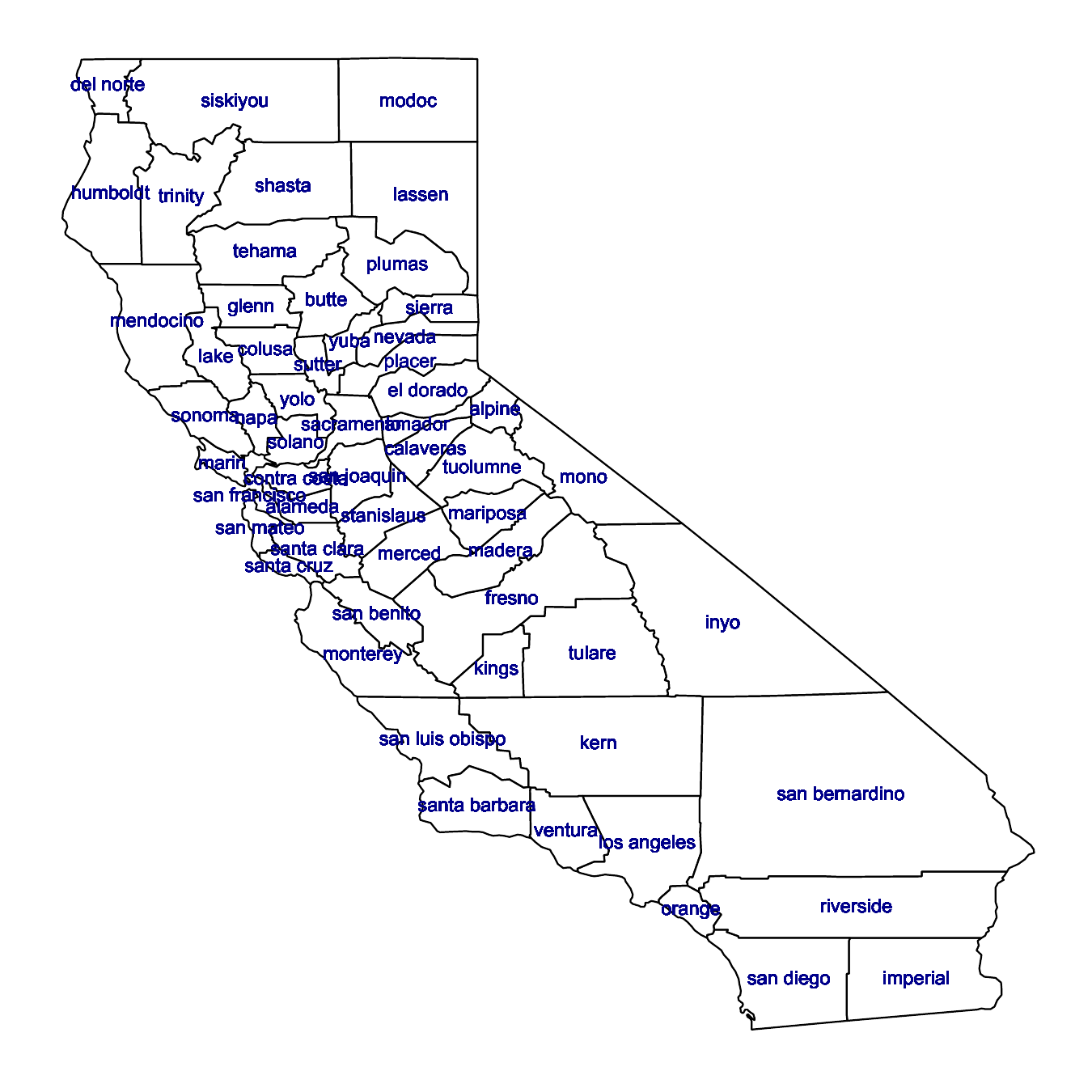}
    \caption{California map with county names. This will be helpful in detecting the boundaries.}
    \label{fig:named_map}
\end{figure}

\subsection{Analysis with covariates in the mean and adjacency model}\label{subsec: meanadj_model}

We present inference from the model used in Section~\ref{sec: cancer_analysis} of the main article, where we include cancer-specific intercepts and slopes for the three covariates described in Section~\ref{sec: data} of the main article and also include the covariates in the adjacency model. Table~\ref{tab:post_inf_meanadj} presents the estimates of the model described in Section~\ref{sec: cancer_analysis} including only cancer-specific intercepts in the regression model and the 3 covariates described in Section~\ref{sec: data} of the main article in the adjacency model. The last column of Table~\ref{tab:post_inf_meanadj} presents Monte Carlo standard errors corresponding to the parameter estimates.

\begin{table}[htbp!]
    \centering
    \small
    \caption{\color{orange} Posterior estimates, standard deviations and their Monte Carlo standard errors for the regression coefficients, atoms and their precision in the hierarchical model in \eqref{eq:post}. Here, $\beta_1$ through $\beta_4$ denote the regression coefficient corresponding to lung, $\beta_5$ through $\beta_8$ to esophageal, $\beta_9$ through $\beta_{12}$ to larynx and $\beta_{13}$ through $\beta_{16}$ colorectal, respectively. The $\theta$'s denote the atoms and $\tau$ is the precision in the prior of the $\theta_k$'s as described in Section~\ref{sec: sre}. While the intercepts regarding esophageal, larynx and colorectal ($\beta_5$, $\beta_9$ and $\beta_{13}$) are significant, we note that lung intercept is not. Moreover, across cancers the smoking covariate always reveal positive estimates ($\beta_2$, $\beta_6$, $\beta_{10}$ and $\beta_{14}$) while over 65 and below poverty covariates do not show such a consistent behaviour. This indicates that the smoking covariate has a positive impact on the count means across the cancers.}
    \label{tab:post_inf_meanadj}
    \begin{tabular}{cccc}
        \hline
        Variable & Estimate & Standard deviation & Monte Carlo Standard Error \\
        \hline
        $\beta_1$ & $-0.068$ & $0.037$ & $0.003$\\ 
        $\beta_2$ & $0.029$ & $0.002$ & $< 10^{-3}$\\ 
        $\beta_3$ & $-0.011$ & $0.002$ & $< 10^{-3}$\\ 
        $\beta_4$ & $-0.009$ & $0.002$ & $< 10^{-3}$ \\ 
        $\beta_5$ & $-0.361$ & $0.109$ & $0.009$ \\ 
        $\beta_6$ & $0.034$ & $0.007$ & $< 10^{-3}$ \\ 
        $\beta_7$ & $0.010$ & $0.007$ & $< 10^{-3}$ \\ 
        $\beta_8$ & $-0.010$ & $0.006$ & $< 10^{-3}$\\ 
        $\beta_9$ & $-0.394$ & $0.132$ & $0.008$\\ 
        $\beta_{10}$ & $0.032$ & $0.008$ &  $0.001$\\ 
        $\beta_{11}$ & $0.000$ & $0.009$ &  $< 10^{-3}$\\ 
        $\beta_{12}$ & $0.004$ & $0.007$ &  $< 10^{-3}$\\ 
        $\beta_{13}$ & $0.156$ & $0.052$ & $0.007$\\ 
        $\beta_{14}$ & $0.013$ & $0.002$ & $< 10^{-3}$\\ 
        $\beta_{15}$ & $-0.018$ & $0.004$ & $< 10^{-3}$\\ 
        $\beta_{16}$ & $-0.006$ & $0.002$ & $< 10^{-3}$\\ 
        $\theta_1$ & $0.021$ & $0.024$ & $0.005$\\ 
        $\theta_2$ & $0.190$ & $0.077$ & $0.011$\\ 
        $\theta_3$ & $-0.223$ & $0.060$ & $0.010$\\ 
        $\theta_4$ & $0.168$ & $0.198$ & $0.029$ \\ 
        $\theta_5$ & $-0.082$ & $0.234$ & $0.030$ \\ 
        $\theta_6$ & $0.209$ & $0.136$ & $0.018$ \\ 
        $\theta_7$ & $-0.081$ & $0.290$ & $0.030$ \\ 
        $\theta_8$ & $0.259$ & $0.139$ & $0.015$ \\ 
        $\theta_9$ & $0.061$ & $0.307$ & $0.036$ \\ 
        $\theta_{10}$ & $-0.071$ & $0.241$ & $0.034$ \\ 
        $\theta_{11}$ & $0.115$ & $0.034$ & $0.007$\\ 
        $\theta_{12}$ & $-0.144$ & $0.029$ & $0.007$\\ 
        $\theta_{13}$ & $0.175$ & $0.048$ & $0.008$\\ 
        $\theta_{14}$ & $0.132$ & $0.145$ & $0.018$\\ 
        $\theta_{15}$ & $-0.082$ & $0.025$ & $0.005$\\ 
        $\tau$ & $6.979$ & $2.336$ & $0.026$\\ 
        \hline
    \end{tabular}
\end{table}

\begin{figure}[htbp!]
    \centering
    \includegraphics[width=0.49\linewidth]{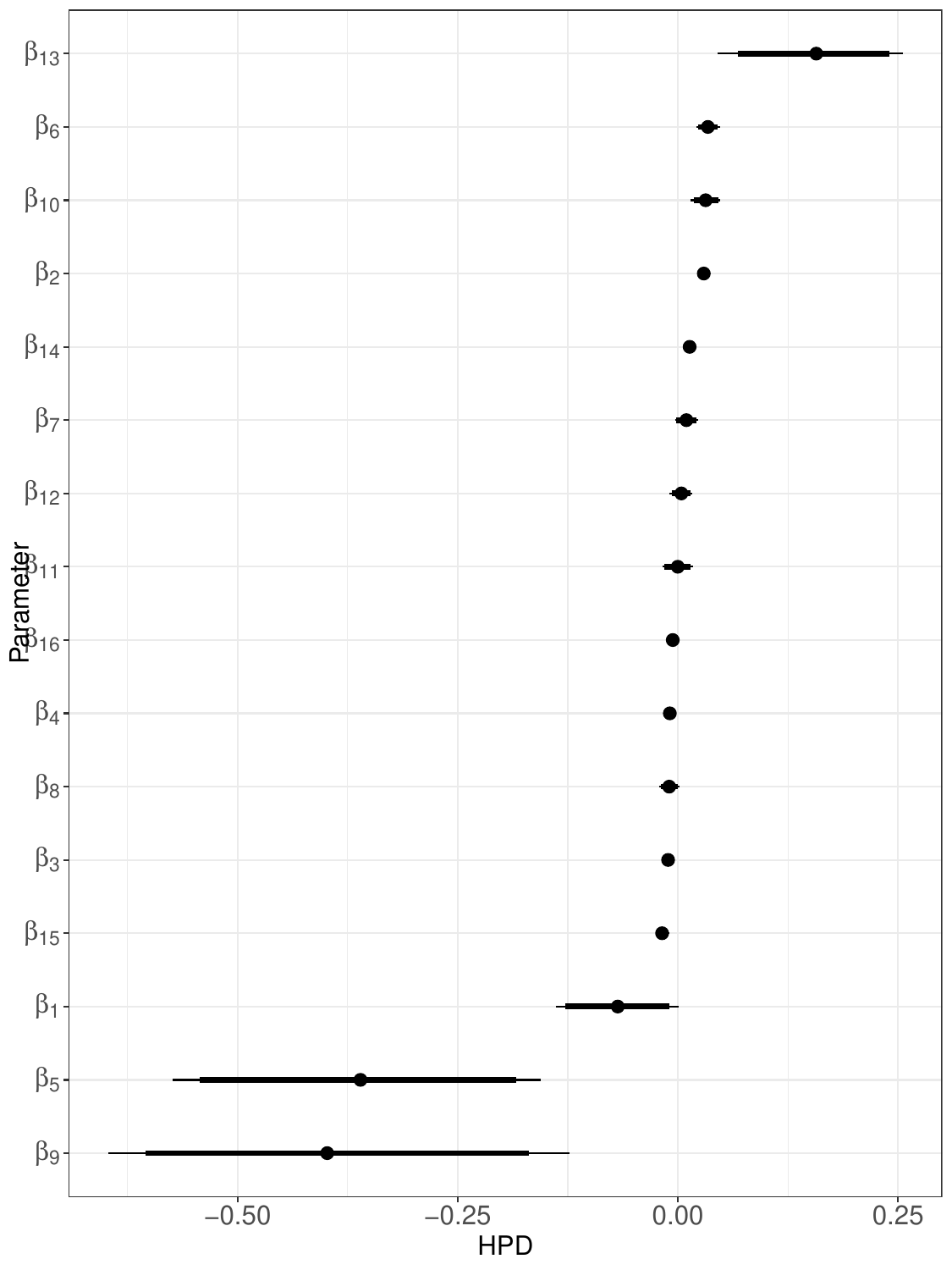}
    \includegraphics[width=0.49\linewidth]{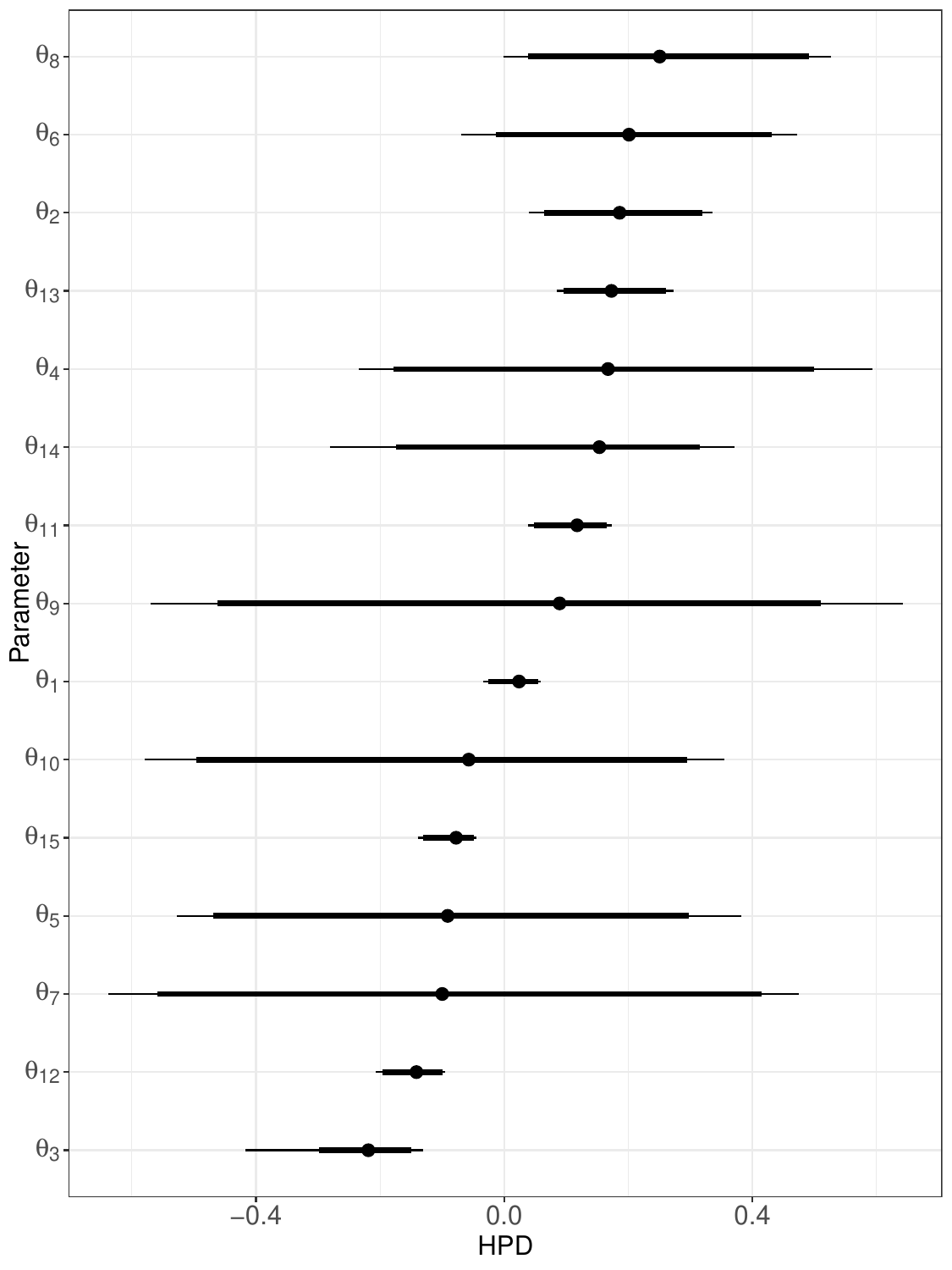}
    \caption{\color{orange} Highest posterior density (HPD) intervals at 90\% (thick lines) and 95\% (thin lines) levels for $\bbeta$ (left panel) and $\btheta$ (right panel). These are ordered according to their posterior medians. As with the posterior means in Table~\ref{tab:post_inf_meanadj}, we see the intercepts regarding esophageal and larynx ($\beta_5$ and $\beta_9$) having negative posterior medians with the $95\%$ interval entirely to the left of $0$, colorectal ($\beta_{13}$) having positive posterior medians with the $95\%$ interval entirely to the right of $0$ and lung ($\beta_1$) with a much smaller negative effect with the $95\%$ interval almost including the $0$. All of the others regression coefficients have much smaller intervals with the median estimates very close to $0$. Turning to the right panel, we see about 3 atoms ($\theta_3$, $\theta_{12}$ and $\theta_{15}$) to have their HPD intervals almost entirely to the left of $0$, while 4 atoms ($\theta_{2}$, $\theta_{8}$, $\theta_{11}$ and $\theta_{13}$) have intervals situated entirely to the right of $0$. The remaining 8 atoms (out of a total of $K=15$) are not significantly different from $0$. The variation in the posterior distribution of these atoms reflect substantial posterior learning and information from the data that propagates to the spatial random effects.}
    \label{fig:post_beta_theta_meanadj}
\end{figure}

\begin{figure}[htbp!]
    \centering
    \includegraphics[width=0.49\linewidth]{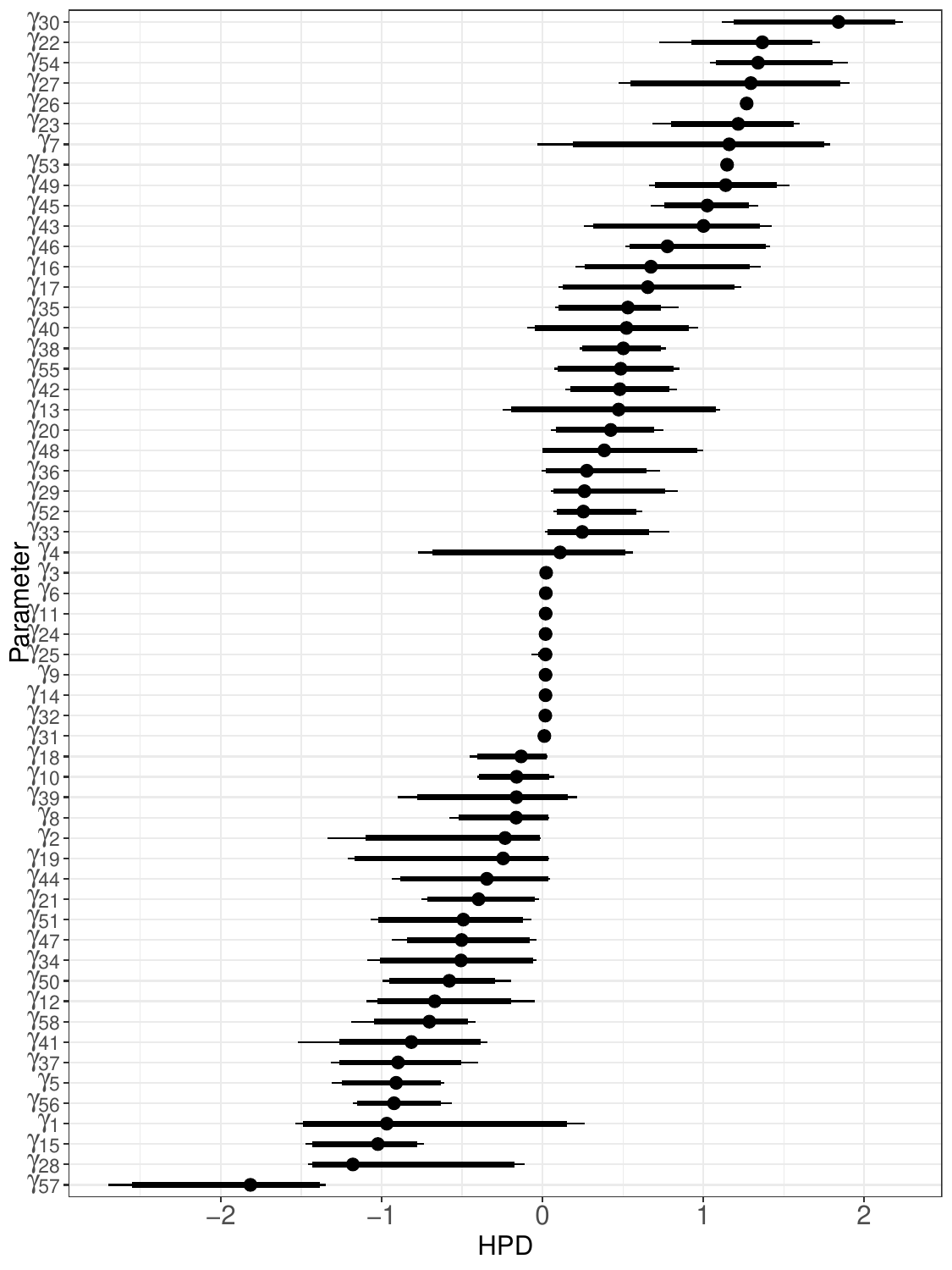}
    \includegraphics[width=0.49\linewidth]{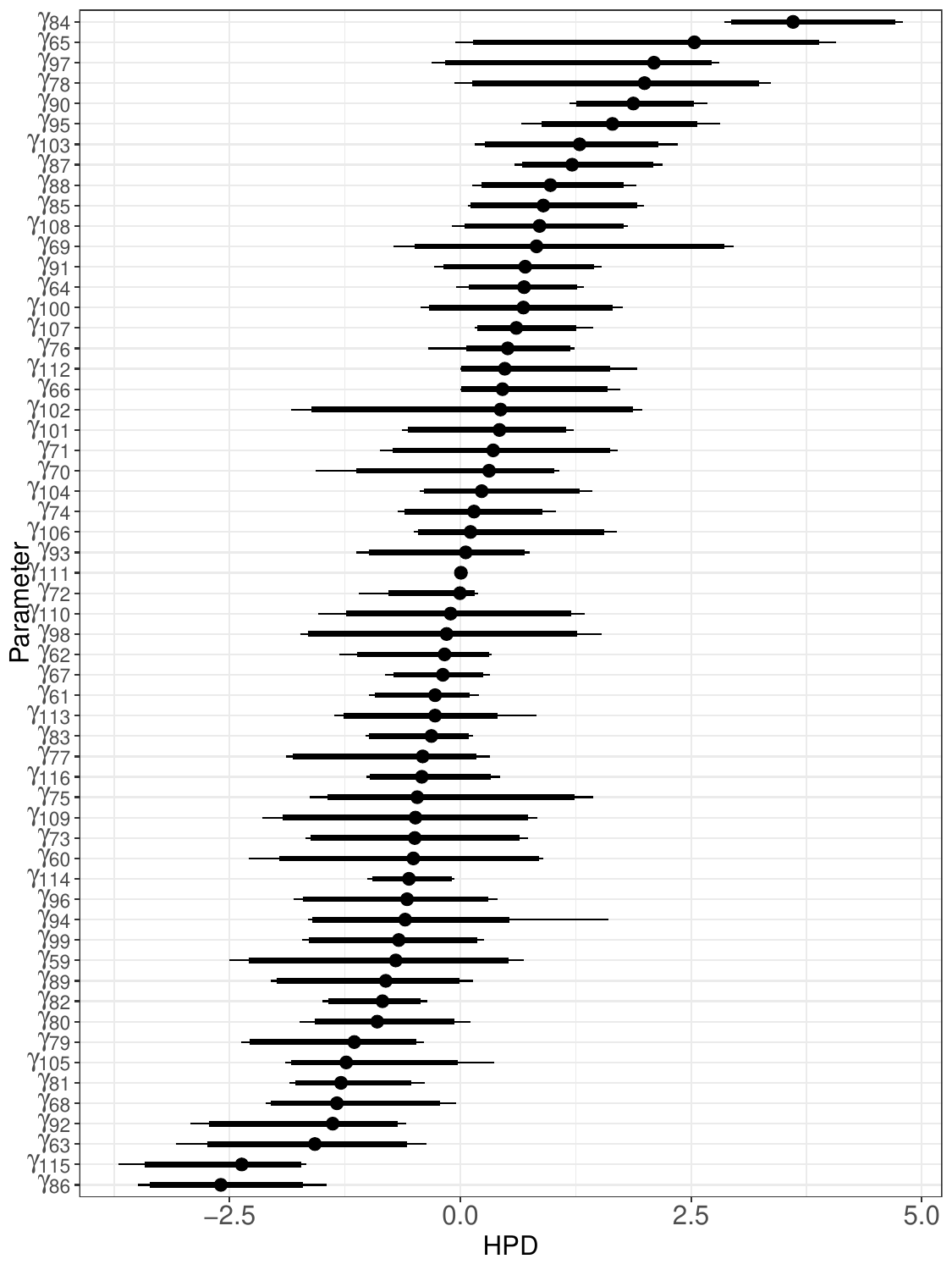}\\
    \includegraphics[width=0.49\linewidth]{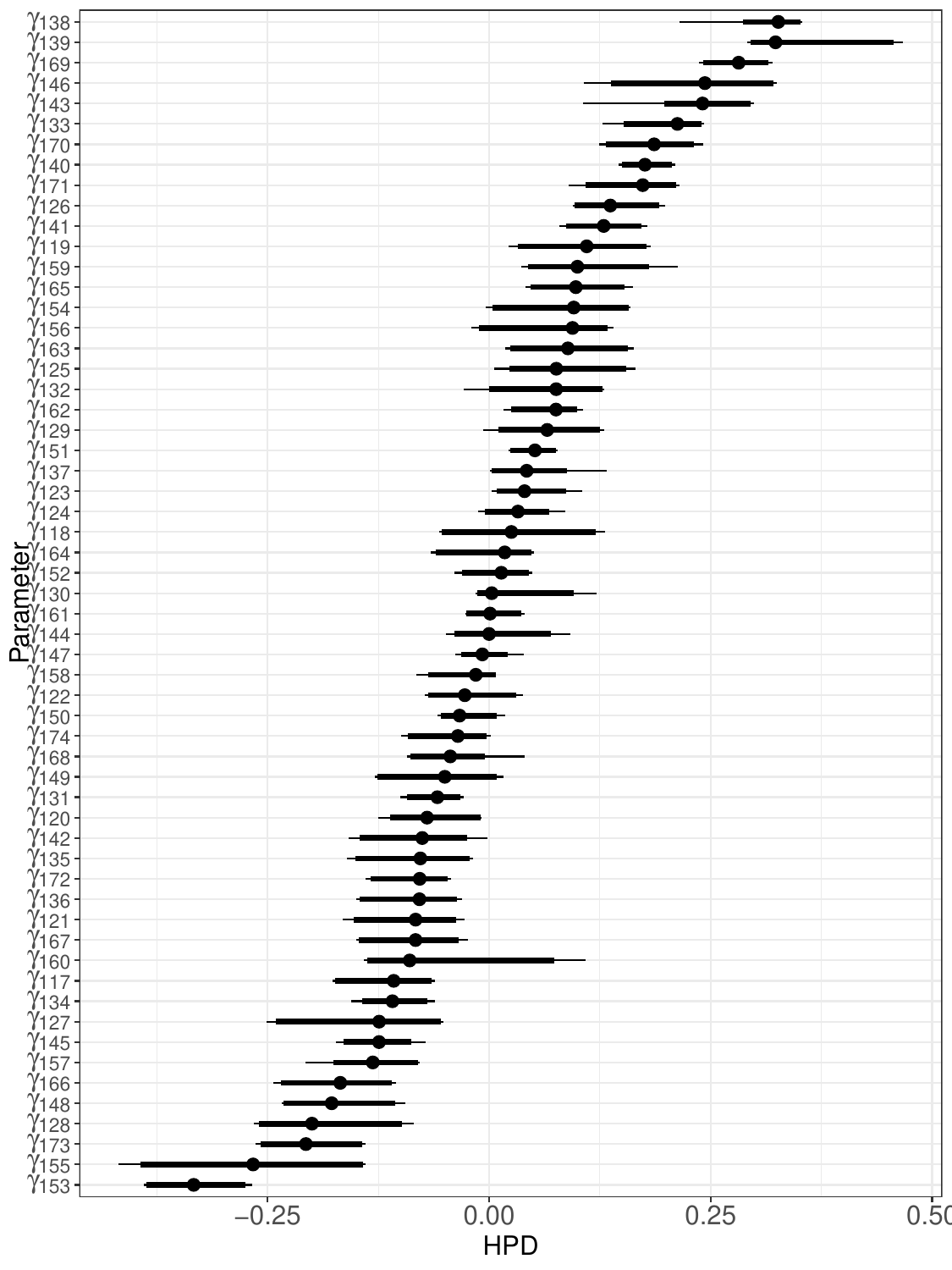}
    \includegraphics[width=0.49\linewidth]{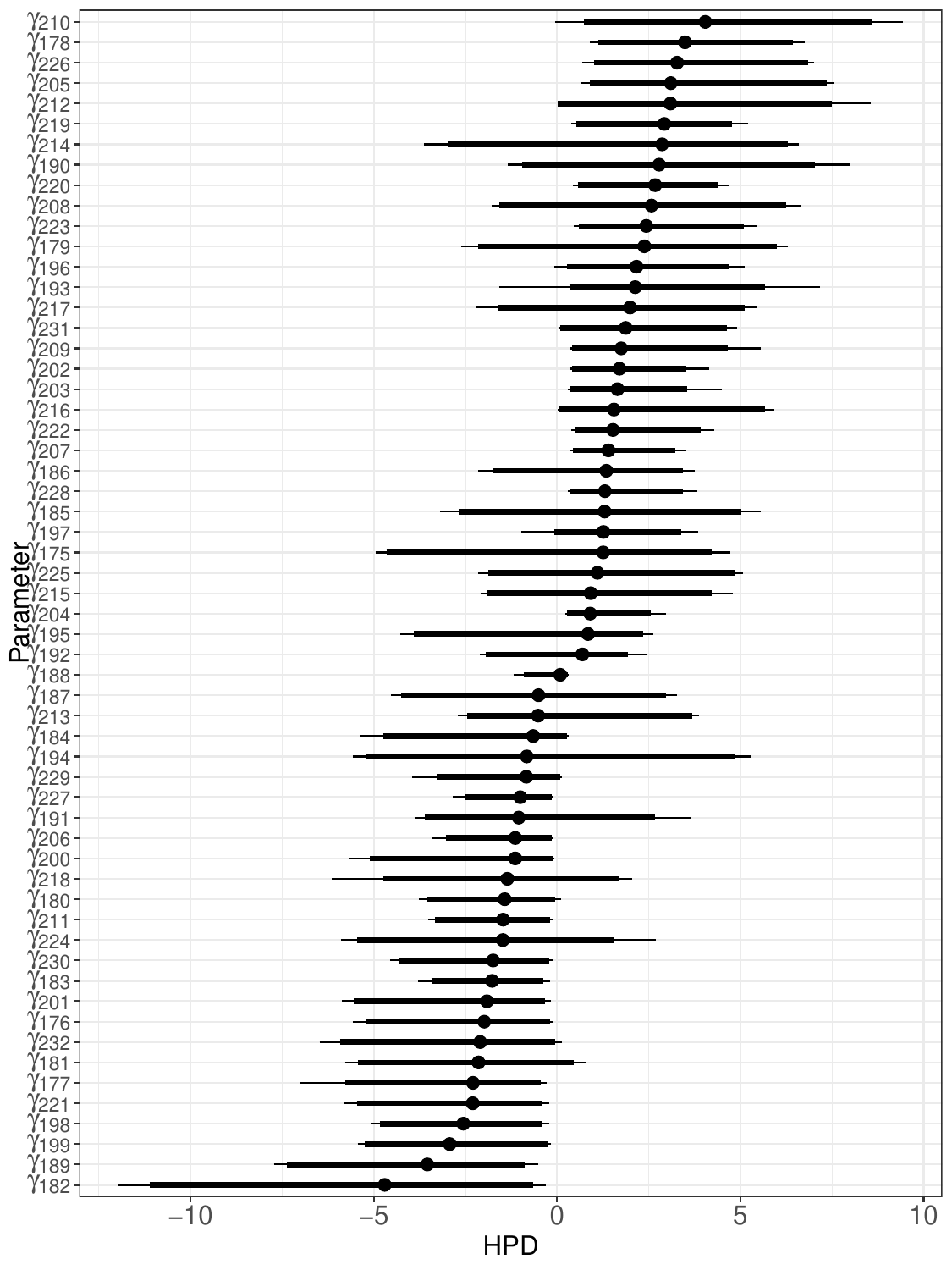}
    \caption{\color{orange} Highest posterior density (HPD) intervals at 90\% (thick lines) and 95\% (thin lines) levels for the elements of the disease-specific spatial effects in $\bgamma$ using the MDAGAR specification in Section~\ref{sec: unstructured_graph} and the precision matrix given in Eq.~\ref{eq:sp_cov}. The four panels correspond to the 4 cancers: lung (top left), esophageal (top right), larynx (bottom left) and colorectal (bottom right). The intervals are arranged in increasing order of value for posterior medians. It is expected that the intervals for these random effects will mostly not be significantly different from $0$, but they reveal substantial variation indicating enough information in the data to discern spatial patterns. It is worth pointing out that the degree of dispersion varies significantly among different cancers. For instance, the intervals for lung and esophageal cancers cover similar ranges, while those for laryngeal cancer are much narrower, and those for colorectal cancer are much wider.}
    \label{fig:post_gamma_meanadj}
\end{figure}

\begin{figure}[htbp!]
    \centering
    \includegraphics[width=0.49\linewidth]{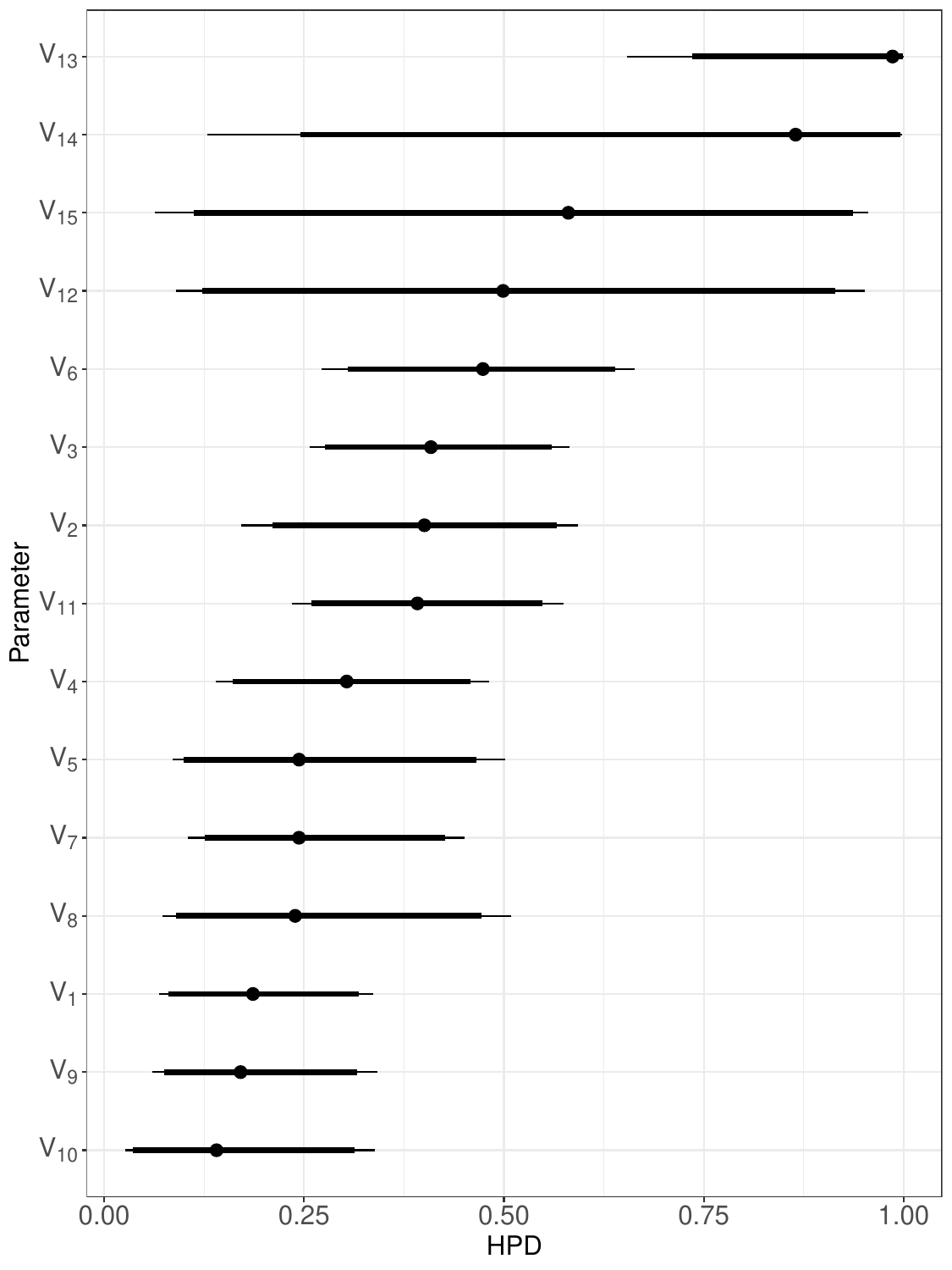}
    \includegraphics[width=0.49\linewidth]{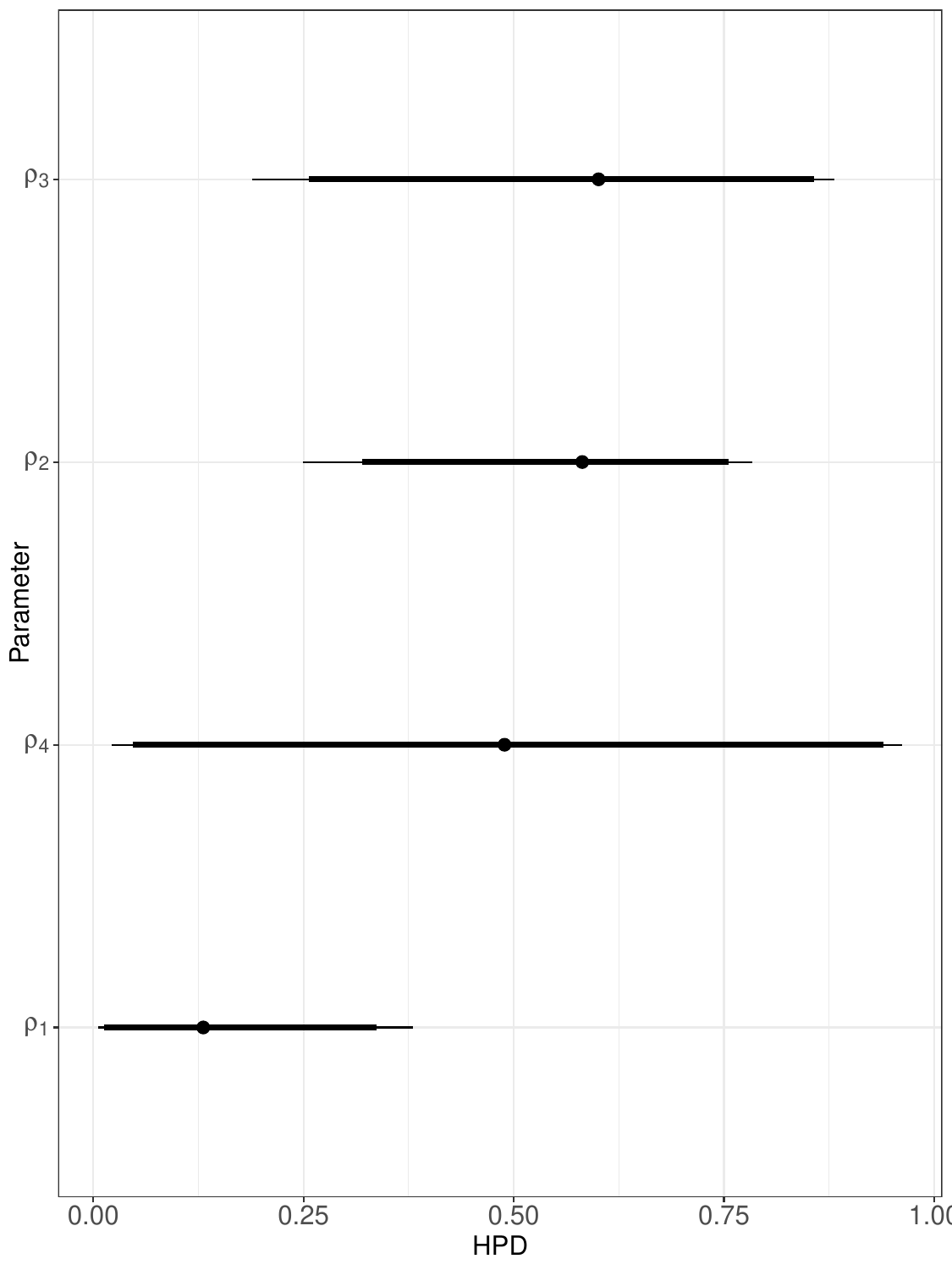}
    \caption{\color{orange} Highest posterior density (HPD) intervals at 90\% (thick lines) and 95\% (thin lines) levels for $\bV$ (left panel), which consists of the 15 stick-breaking weights used in Eq.~\eqref{eq:DP}, and for $\brho$ (right panel), which consists of the 4 spatial autocorrelation parameters (one for each disease) in the MDAGAR model. These intervals are arranged in increasing order of posterior medians. The variations among these posterior intervals indicate substantial learning from the data that informs the joint distribution of the spatial effects through the stick-breaking probabilities and the baseline MDAGAR distribution in the DP.}
    \label{fig:post_v_rho_meanadj}
\end{figure}

\begin{figure}[h!]
    \centering
    \includegraphics[width=0.49\linewidth]{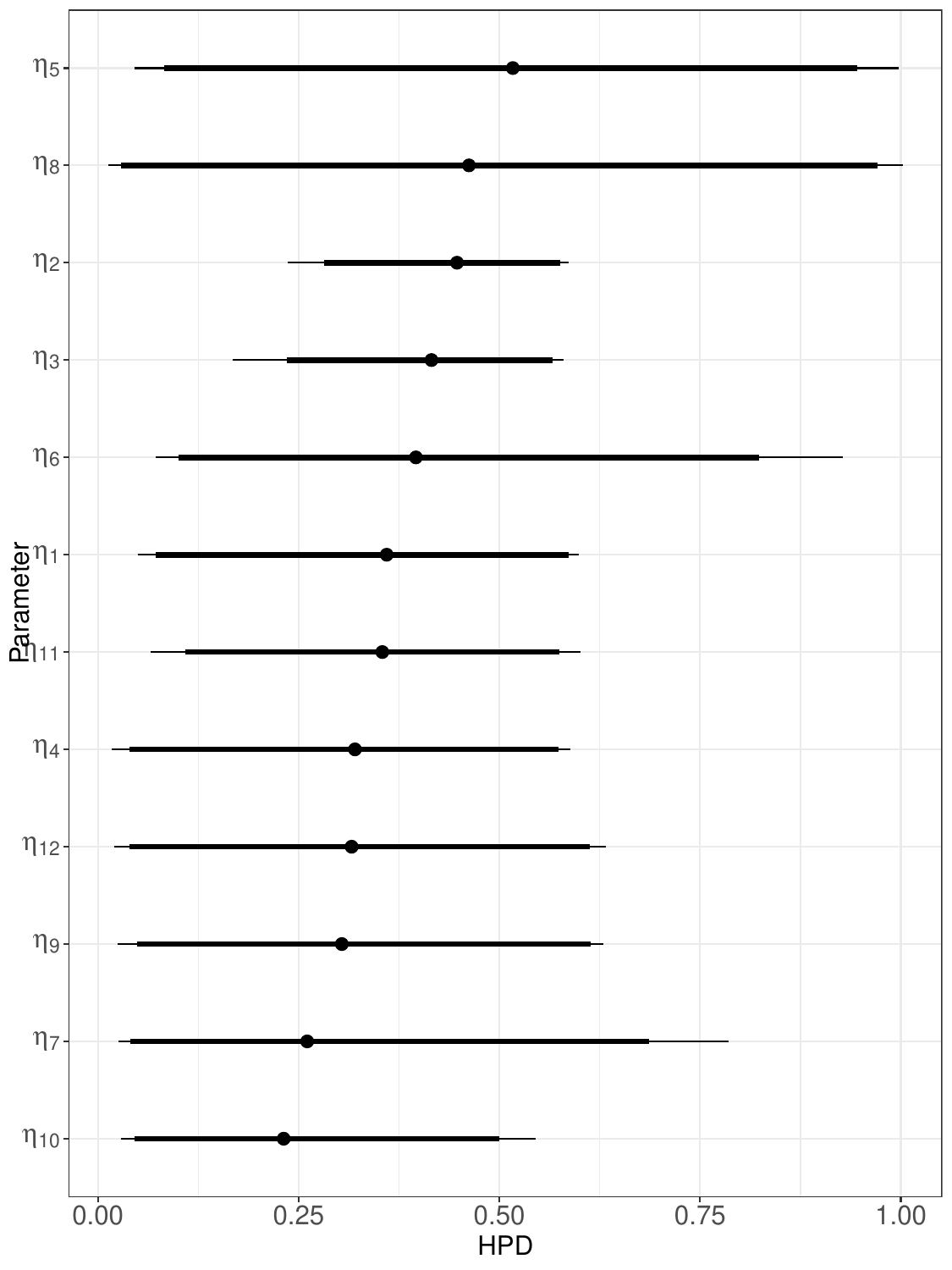}
    \includegraphics[width=0.49\linewidth]{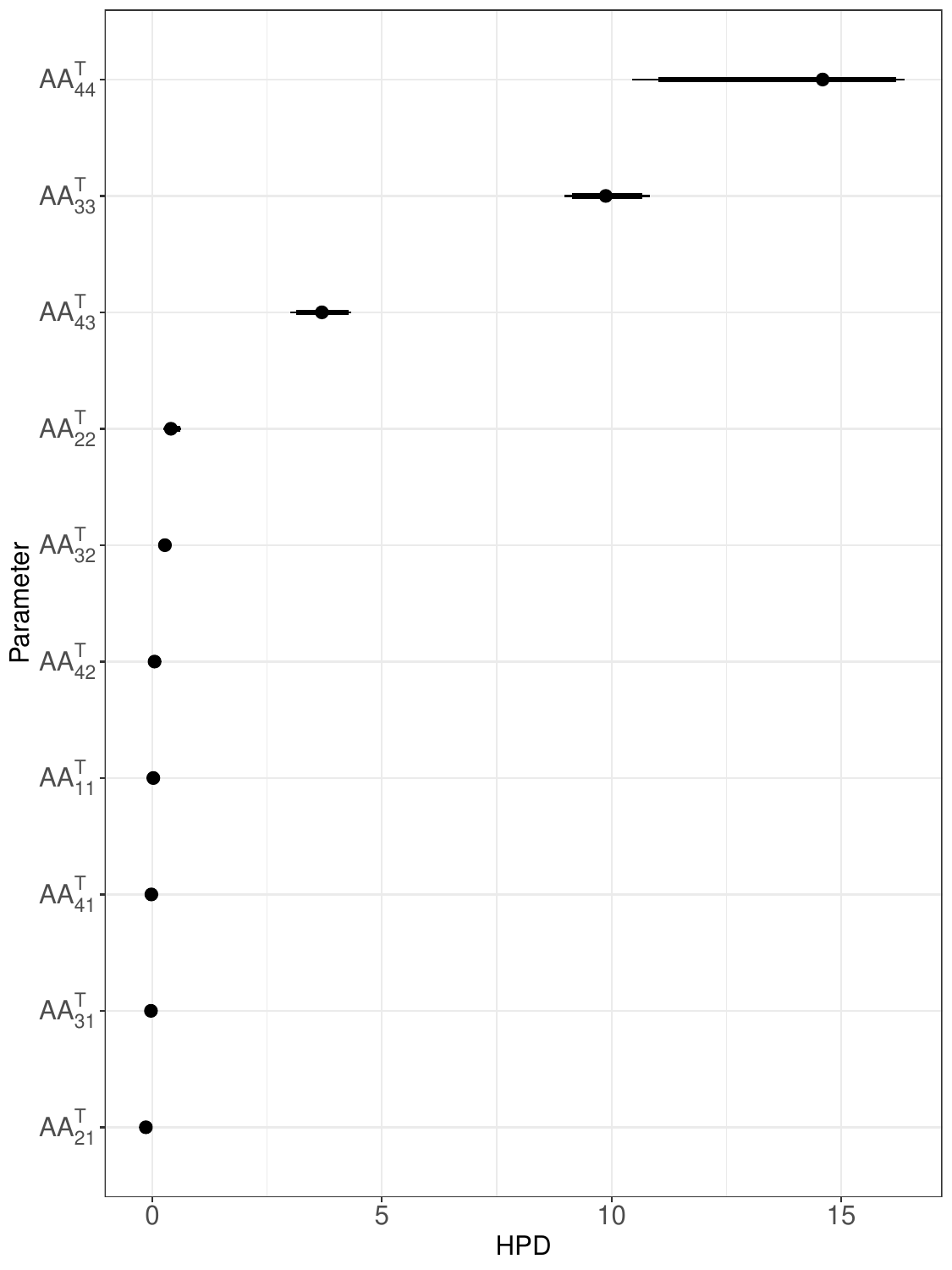}
    \caption{\color{orange} Highest posterior density (HPD) intervals at 90\% (thick lines) and 95\% (thin lines) levels for $\bet$ (left panel) and $\bA$ (right panel). The 12 coefficients in $\bet$ include 3 slopes for each of the 4 cancer types corresponding to the covariates shown in Figure~\ref{fig:covariates_plot}. For each of the 4 cancers, these parameters appear in \eqref{eq:w_binary} as coefficients of the $3\times 1$ vector $\bz_{ij}$, where the elements of $\bz_{ij}$ are $|z_{ki}-z_{kj}|$ for $k=1,2,3$ corresponding to the 3 covariates. The four slopes corresponding to smoking are denoted as $\eta_{1}$, $\eta_{4}$, $\eta_{7}$, and $\eta_{10}$. For the variable 'population over 65', the corresponding slopes are $\eta_{2}$, $\eta_{5}$, $\eta_{8}$, and $\eta_{11}$. The slopes related to poverty levels are $\eta_{3}$, $\eta_{6}$, $\eta_{9}$, and $\eta_{12}$. For all cancers considered, these coefficients correspond respectively to lung, esophageal, laryngeal, and colorectal cancers. Here, the greater the $\eta$'s, the more likely two regions are not considered to be adjacent by the model. For example, the over 65 covariate influences the adjacency assessment similarly for both esophageal and larynx cancers ($\eta_5$ and $\eta_8$), while it plays a more significant role in determining adjacencies for lung cancer ($\eta_2$). Additionally, the below poverty covariate appears to have a considerable impact on the adjacencies associated with lung cancer as well ($\eta_3$). In the right panel, the HPD intervals for the elements of $\bA\bA^{\T}$ are shown, representing the non-spatial covariance among the cancers. Colorectal and larynx cancers exhibit the highest variances ($(\bA\bA^{\T})_{44}$ and $(\bA\bA^{\T})_{33}$), with colorectal showing the widest interval. Colorectal and larynx reveal a significantly positive non-spatial component of their association ($(\bA\bA^{\T}_{43})$) but, as in Fig.~\ref{fig:post_eta_A}, we recommend caution in interpreting the value of these associations since much of the positive associations among these aggregated data are explained by shared risk factors that are also absorbed into geographic neighbors.}
    \label{fig:post_eta_A_meanadj}
\end{figure}

\begin{figure}[htbp!]
    \centering
    \includegraphics[width = 0.85\textwidth]{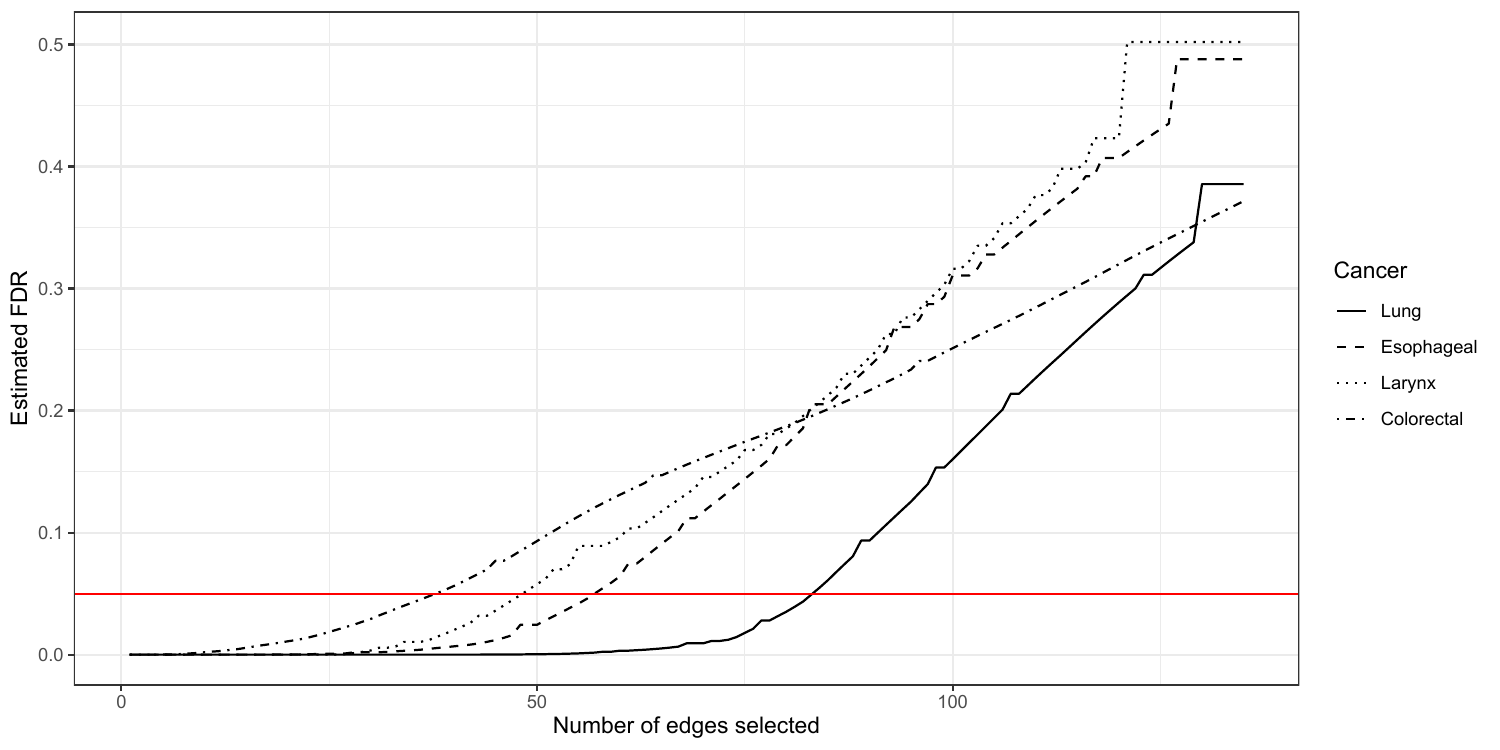}
    \caption{Estimated FDR curves plotted against the number of selected difference boundaries for the four cancers}
    \label{fig:FDR_meanadj}
\end{figure}

\begin{figure}[htbp!]
    \centering
    \includegraphics[width = 0.66 \textwidth]{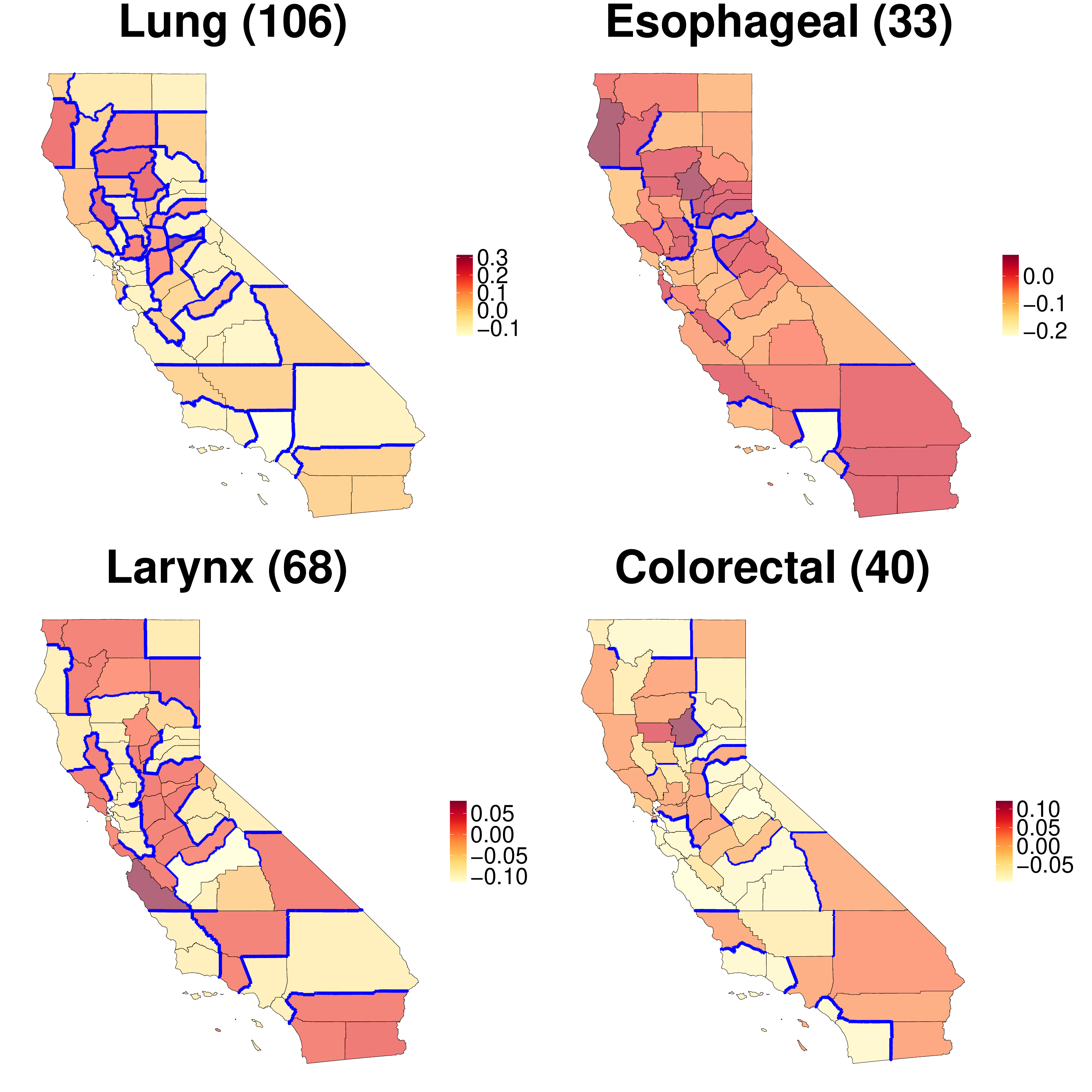}
    \caption{Difference boundaries (highlighted in red) detected by the model in the California map, colored according to the posterior mean of the corresponding $\phi_{id}$, for four cancers individually when $\zeta = 0.05$. The values in brackets are the number of difference boundaries detected.}
    \label{fig:bd_meanadj}
\end{figure}

\begin{figure}[htbp!]
    \centering
    \includegraphics[width = \textwidth]{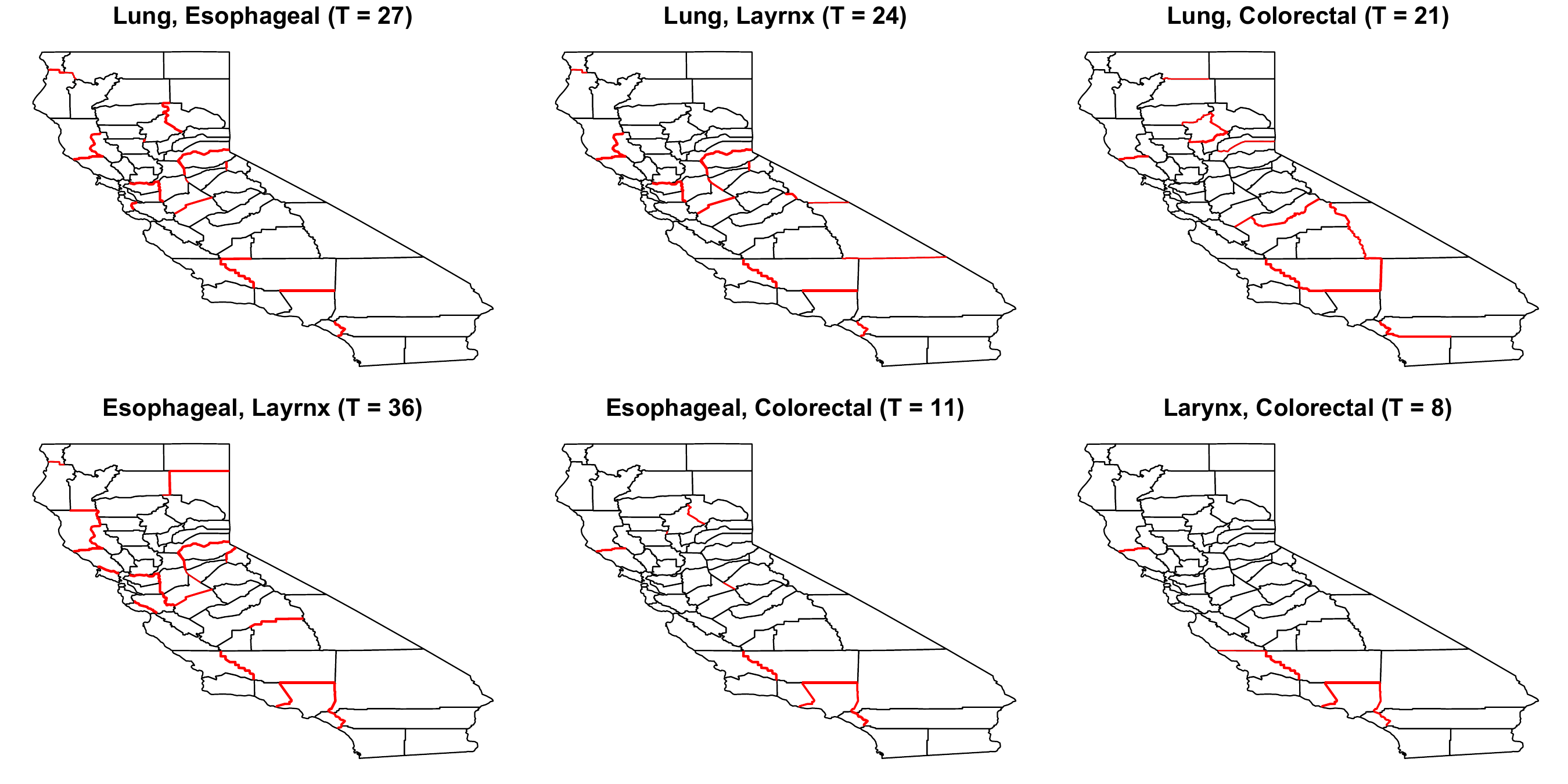}
    \caption{Shared difference boundaries (highlighted in red) detected by the model for each pair of cancers in California map when $\zeta = 0.05$. The values in brackets are the number of difference boundaries detected.}
    \label{fig:sbd_meanadj}
\end{figure}

\begin{figure}[htbp!]
    \centering
    \includegraphics[width = \textwidth]{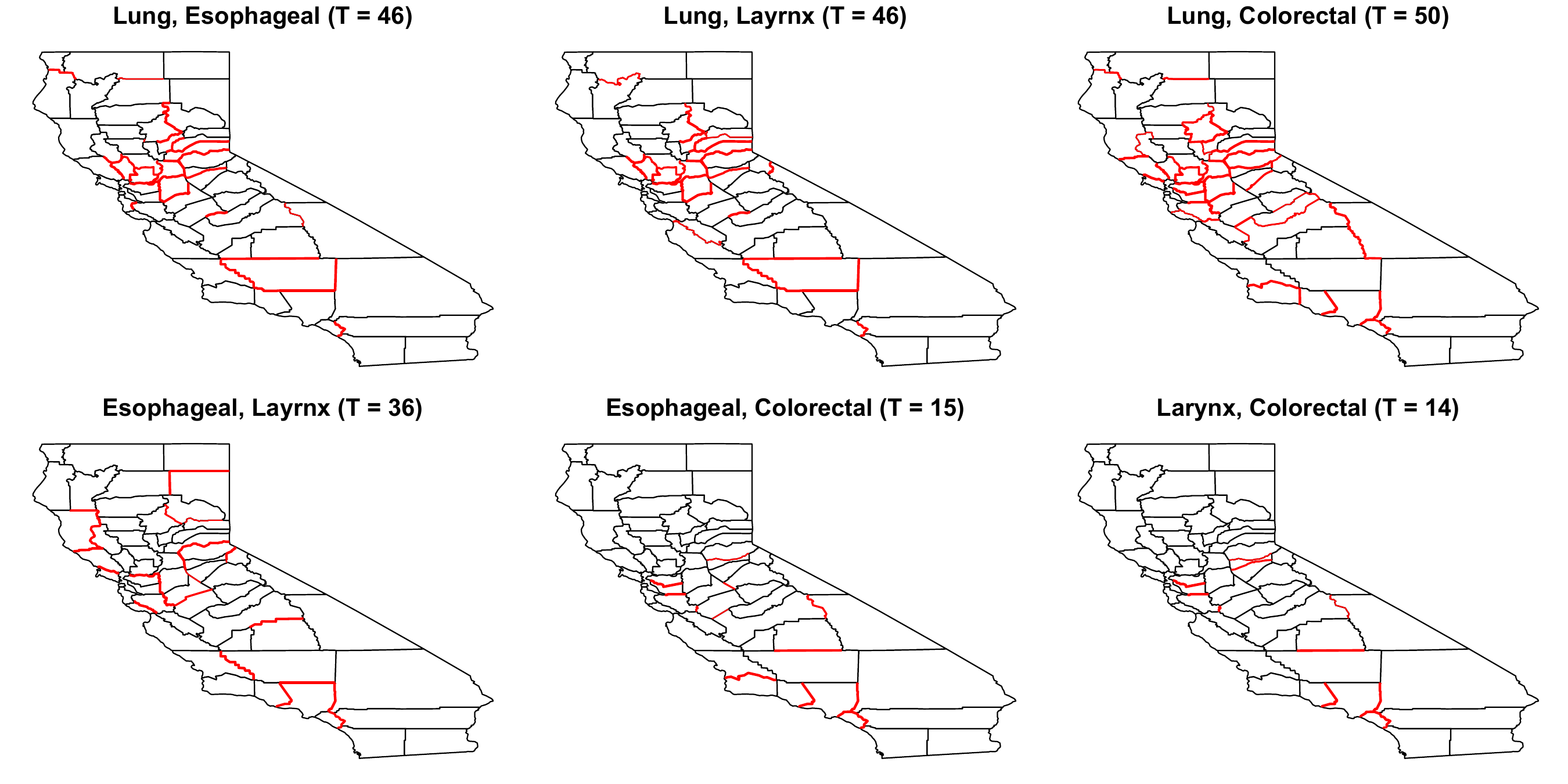}
    \caption{Mutual cross-difference boundaries (highlighted in red) detected by the model for each pair of cancers in California map when $\zeta = 0.05$. The values in brackets are the number of difference boundaries detected.}
    \label{fig:mbd_meanadj}
\end{figure}

\begin{figure}[htbp!]
    \centering
    \includegraphics[width = 0.66 \textwidth]{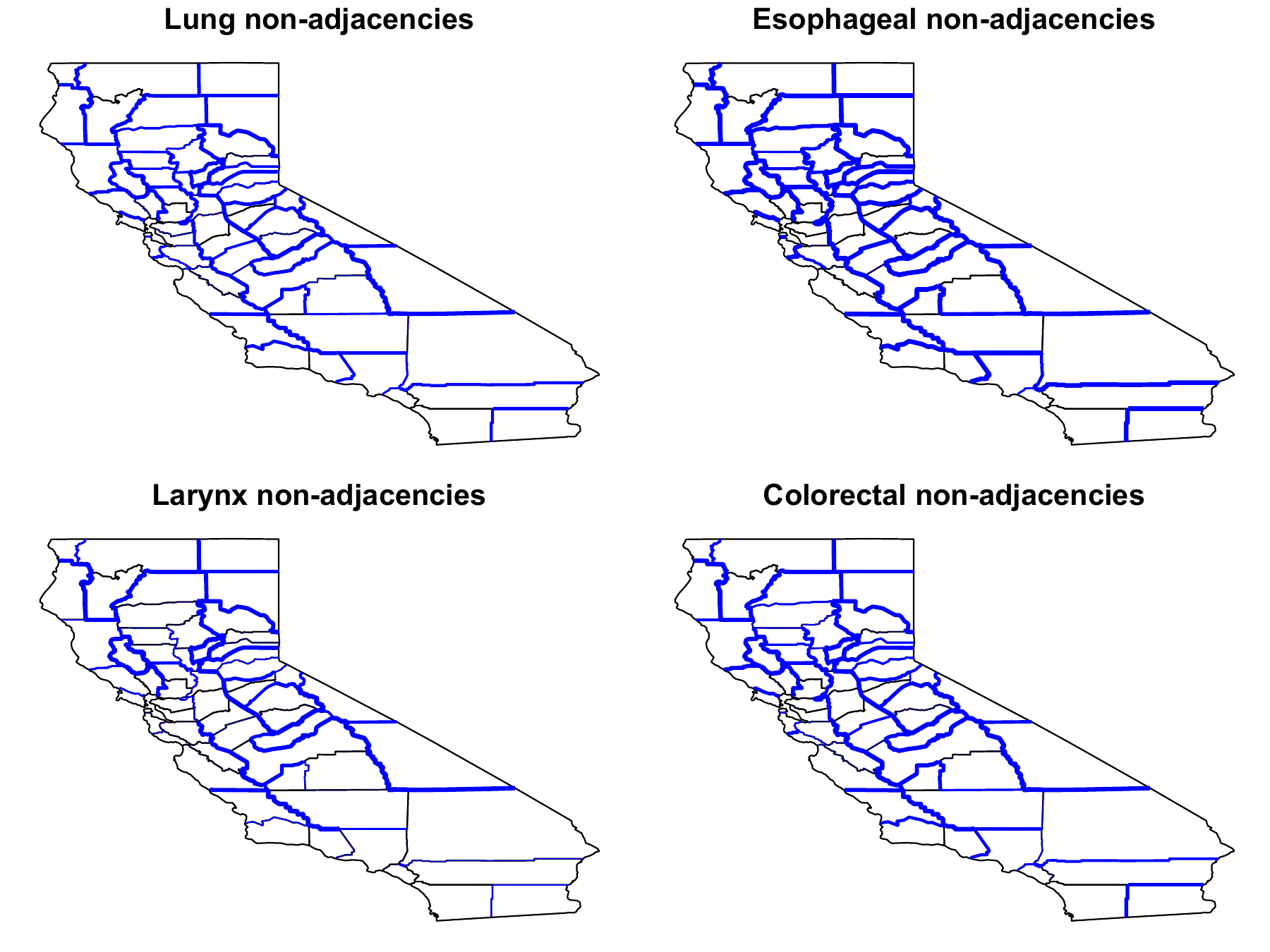}
    \caption{Non-adjacencies (shown in blue) over the California map. The thickness of the lines is proportional to the probability of being considered as a non-adjacency}
    \label{fig:adj_meanadj}
\end{figure}

\clearpage

\subsection{Analysis with covariates in the mean with fixed adjacencies}\label{subsec: mean_model}

We present inference from the model used in Section~\ref{sec: cancer_analysis} of the main article, where we include cancer-specific intercepts and slopes for the three covariates described in Section~\ref{sec: data} of the main article in the regression $\bx_{ij}^{\T}\bbeta_{j}$, but exclude modeling adjacencies and keep them fixed. Table~\ref{tab:post_inf_mean} presents the estimates of the model described in Section~\ref{sec: cancer_analysis} including only cancer-specific intercepts in the regression model and the 3 covariates described in Section~\ref{sec: data} of the main article in the adjacency model. The last column of Table~\ref{tab:post_inf_mean} presents Monte Carlo standard errors corresponding to the parameter estimates.

\begin{table}[htbp!]
    \centering
    \small
    \caption{\color{orange} Posterior estimates and standard deviations for the regression coefficients, atoms and their precision in the hierarchical model in \eqref{eq:post}. Here, $\beta_1$ through $\beta_4$ denote the regression coefficient corresponding to lung, $\beta_5$ through $\beta_8$ to esophageal, $\beta_9$ through $\beta_{12}$ to larynx and $\beta_{13}$ through $\beta_{16}$ colorectal, respectively. The $\theta$'s denote the atoms and $\tau$ is the precision in the prior of the $\theta_k$'s as described in Section~\ref{sec: sre}. While the intercepts regarding esophageal, larynx and colorectal ($\beta_5$, $\beta_9$ and $\beta_{13}$) are significant, we note that lung intercept is not. Moreover, across cancers the smoking covariate always reveal positive estimates ($\beta_2$, $\beta_6$, $\beta_{10}$ and $\beta_{14}$) while over 65 and below poverty covariates do not show such a consistent behavior. This indicates that the smoking covariate has a positive impact on the count means across the cancers.}
    \label{tab:post_inf_mean}
    \begin{tabular}{cccc}
        \hline
        Variable & Estimate & Standard deviation & Monte Carlo Standard Errors\\
        \hline
        $\beta_1$ & $-0.060$ & $0.040$ & $0.005$\\ 
        $\beta_2$ & $0.031$ & $0.003$ & $< 10^{-3}$\\ 
        $\beta_3$ & $-0.009$ & $0.003$ & $< 10^{-3}$\\ 
        $\beta_4$ & $-0.010$ & $0.003$ & $< 10^{-3}$\\ 
        $\beta_5$ & $-0.304$ & $0.104$ & $0.009$\\ 
        $\beta_6$ & $0.028$ & $0.010$ & $0.001$\\ 
        $\beta_7$ & $0.014$ & $0.008$ & $0.001$\\ 
        $\beta_8$ & $-0.009$ & $0.008$ & $0.001$\\ 
        $\beta_9$ & $-0.407$ & $0.150$ & $0.011$\\ 
        $\beta_{10}$ & $0.026$ & $0.010$ & $0.001$\\ 
        $\beta_{11}$ & $0.005$ & $0.010$ & $0.001$\\ 
        $\beta_{12}$ & $0.008$ & $0.008$ & $0.001$\\ 
        $\beta_{13}$ & $0.236$ & $0.061$ & $0.009$\\ 
        $\beta_{14}$ & $0.011$ & $0.004$ &  $< 10^{-3}$\\ 
        $\beta_{15}$ & $-0.019$ & $0.004$ &  $< 10^{-3}$\\ 
        $\beta_{16}$ & $-0.005$ & $0.002$ &  $< 10^{-3}$\\ 
        $\theta_1$ & $-0.122$ & $0.038$ & $0.014$\\ 
        $\theta_2$ & $-0.019$ & $0.045$ & $0.013$\\ 
        $\theta_3$ & $0.001$ & $0.036$ & $0.009$\\ 
        $\theta_4$ & $-0.016$ & $0.067$ & $0.012$\\ 
        $\theta_5$ & $-0.033$ & $0.061$ & $0.013$\\ 
        $\theta_6$ & $-0.026$ & $0.046$ & $0.016$ \\ 
        $\theta_7$ & $-0.089$ & $0.294$ & $0.046$ \\ 
        $\theta_8$ & $-0.087$ & $0.120$ & $0.029$ \\ 
        $\theta_9$ & $-0.231$ & $0.104$ & $0.023$ \\ 
        $\theta_{10}$ & $0.038$ & $0.137$ & $0.054$ \\ 
        $\theta_{11}$ & $0.054$ & $0.125$ & $0.036$ \\ 
        $\theta_{12}$ & $-0.181$ & $0.043$ & $0.017$ \\ 
        $\theta_{13}$ & $-0.034$ & $0.095$ & $0.030$ \\ 
        $\theta_{14}$ & $-0.042$ & $0.241$ & $0.082$\\ 
        $\theta_{15}$ & $0.082$ & $0.052$ & $0.021$\\ 
        $\tau$ & $8.042$ & $2.676$ & $0.028$ \\  
        \hline
    \end{tabular}
\end{table}

\begin{figure}[htbp!]
    \centering
    \includegraphics[width=0.49\linewidth]{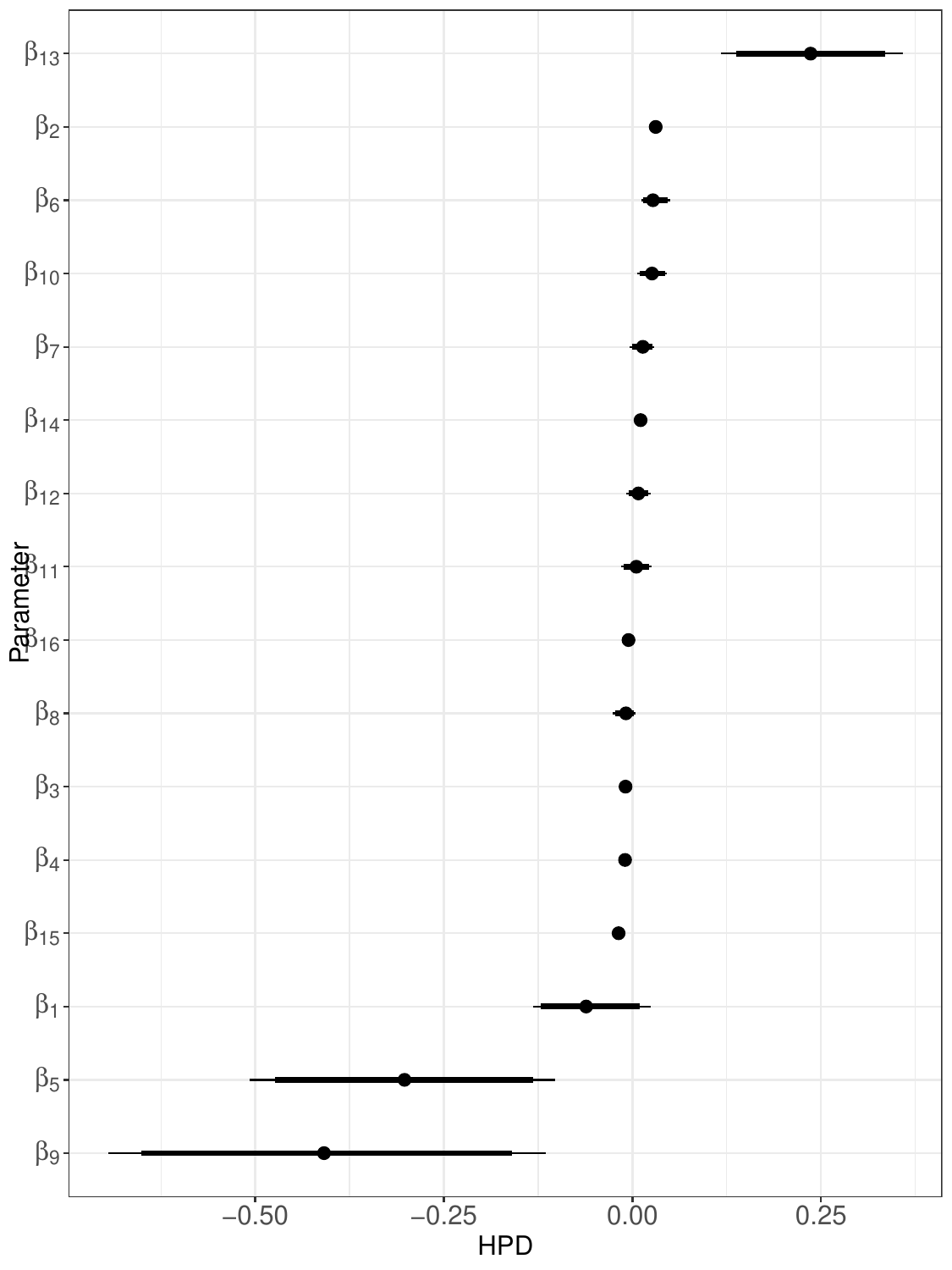}
    \includegraphics[width=0.49\linewidth]{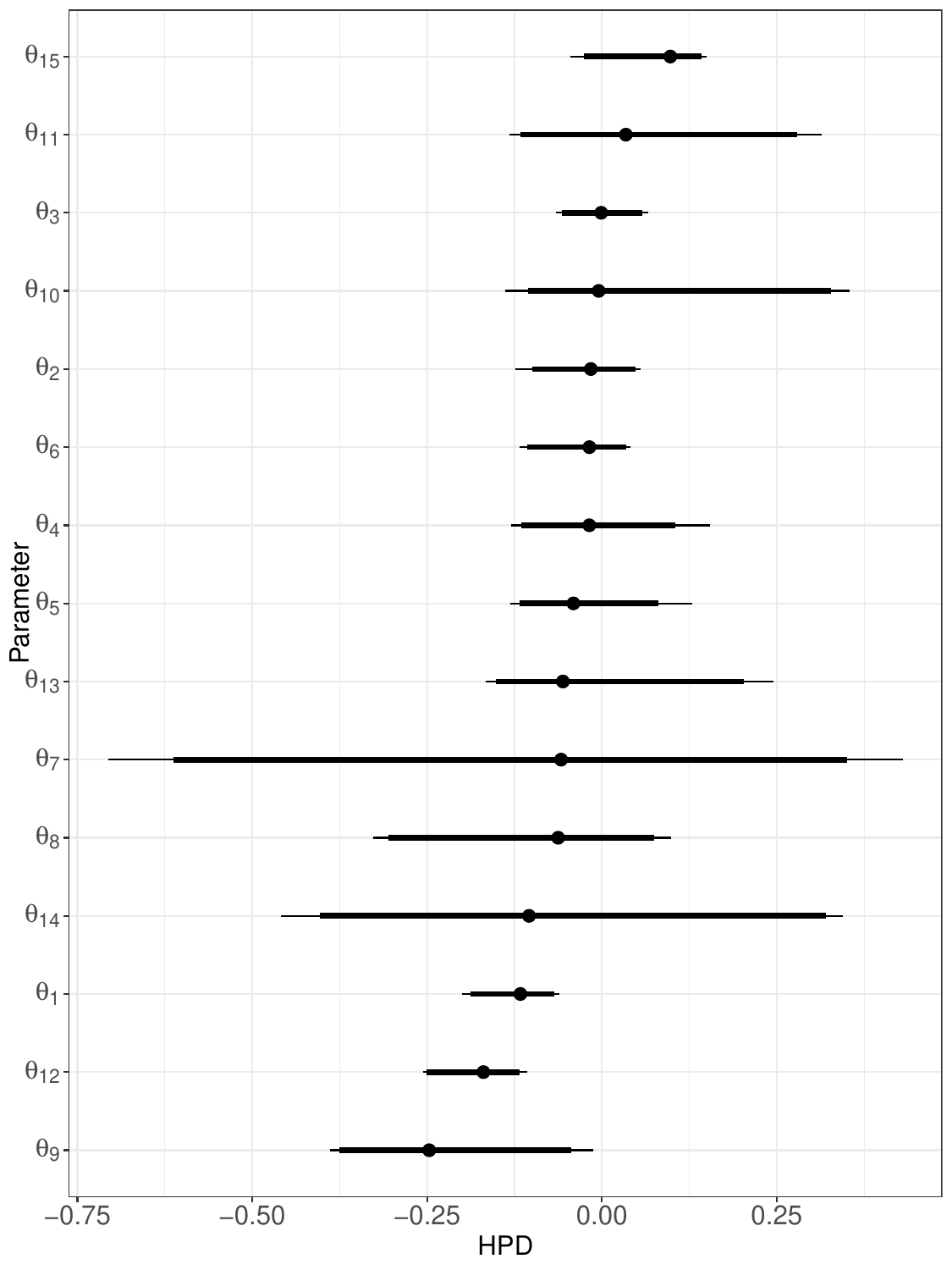}
    \caption{\color{orange} Highest posterior density (HPD) intervals at 90\% (thick lines) and 95\% (thin lines) levels for $\bbeta$ (left panel) and $\btheta$ (right panel). These are ordered according to their posterior medians. As with the posterior means in Table~\ref{tab:post_inf_mean}, we see the intercepts regarding esophageal and larynx ($\beta_5$ and $\beta_9$) having negative posterior medians with the $95\%$ interval entirely to the left of $0$, colorectal ($\beta_{13}$) having positive posterior medians with the $95\%$ interval entirely to the right of $0$ and lung ($\beta_1$) with a much smaller negative effect with the $90\%$ interval almost including the $0$. All of the others regression coefficients have much smaller intervals with the median estimates very close to $0$. Turning to the right panel, we see about 3 atoms ($\theta_1$, $\theta_9$ and $\theta_{12}$) to have their HPD intervals almost entirely to the left of $0$, while none of the remaining 12 atoms (out of a total of $K=15$) are significantly different from $0$ since their $95\%$ interval include $0$. The variation in the posterior distribution of these atoms reflect substantial posterior learning and information from the data that propagates to the spatial random effects.}
    \label{fig:post_beta_theta_mean}
\end{figure}

\begin{figure}[htbp!]
    \centering
    \includegraphics[width=0.49\linewidth]{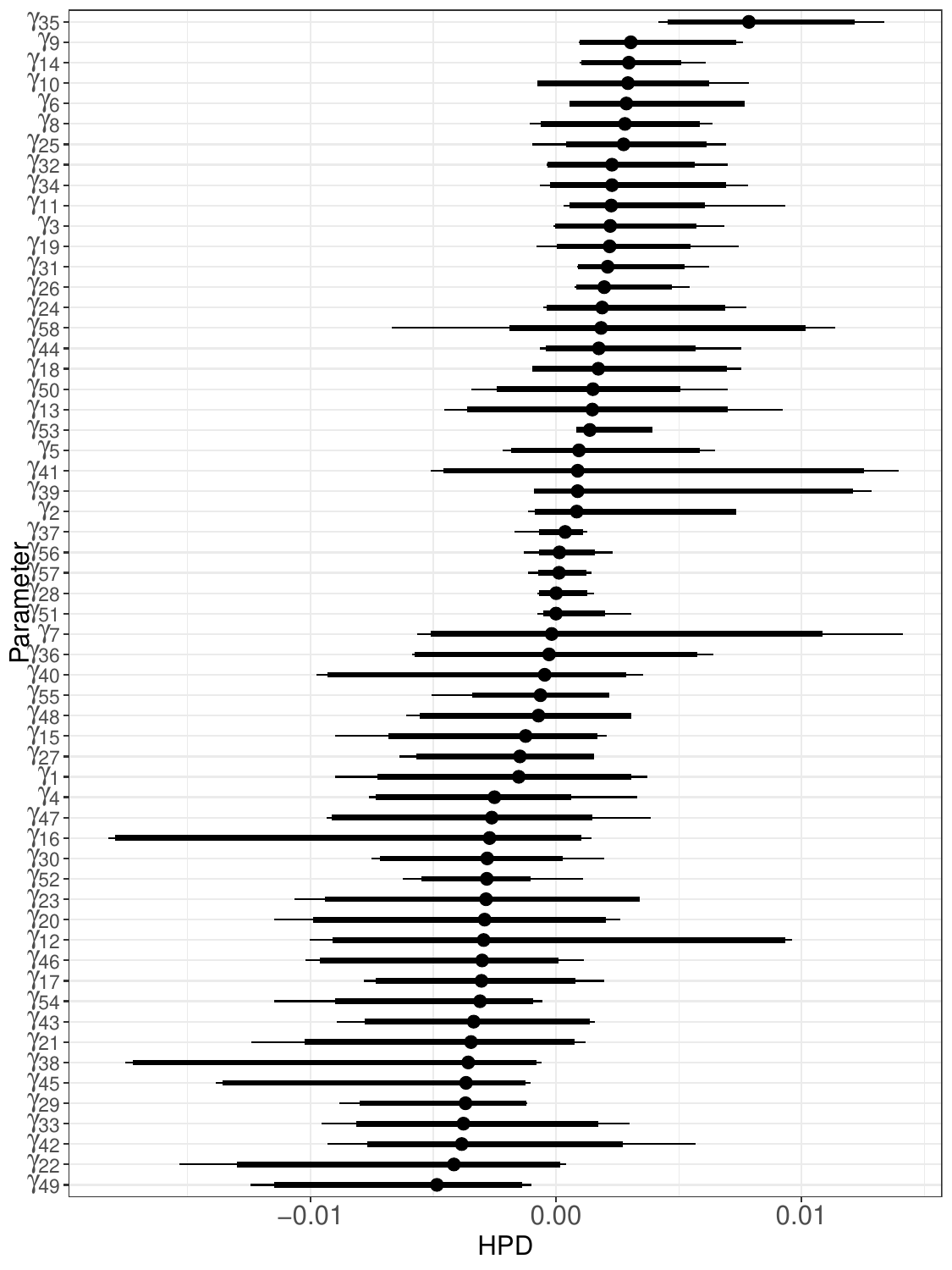}
    \includegraphics[width=0.49\linewidth]{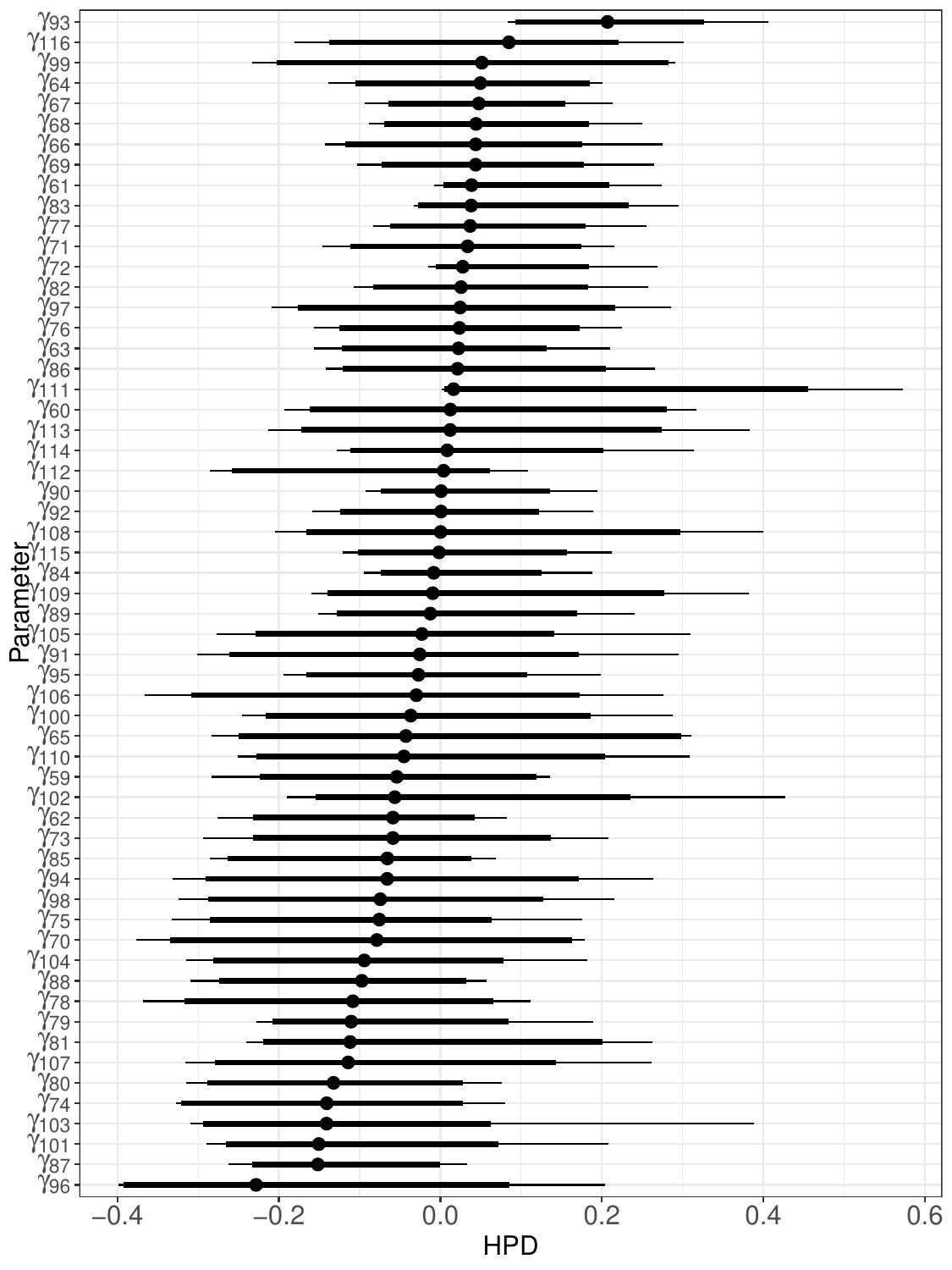}\\
    \includegraphics[width=0.49\linewidth]{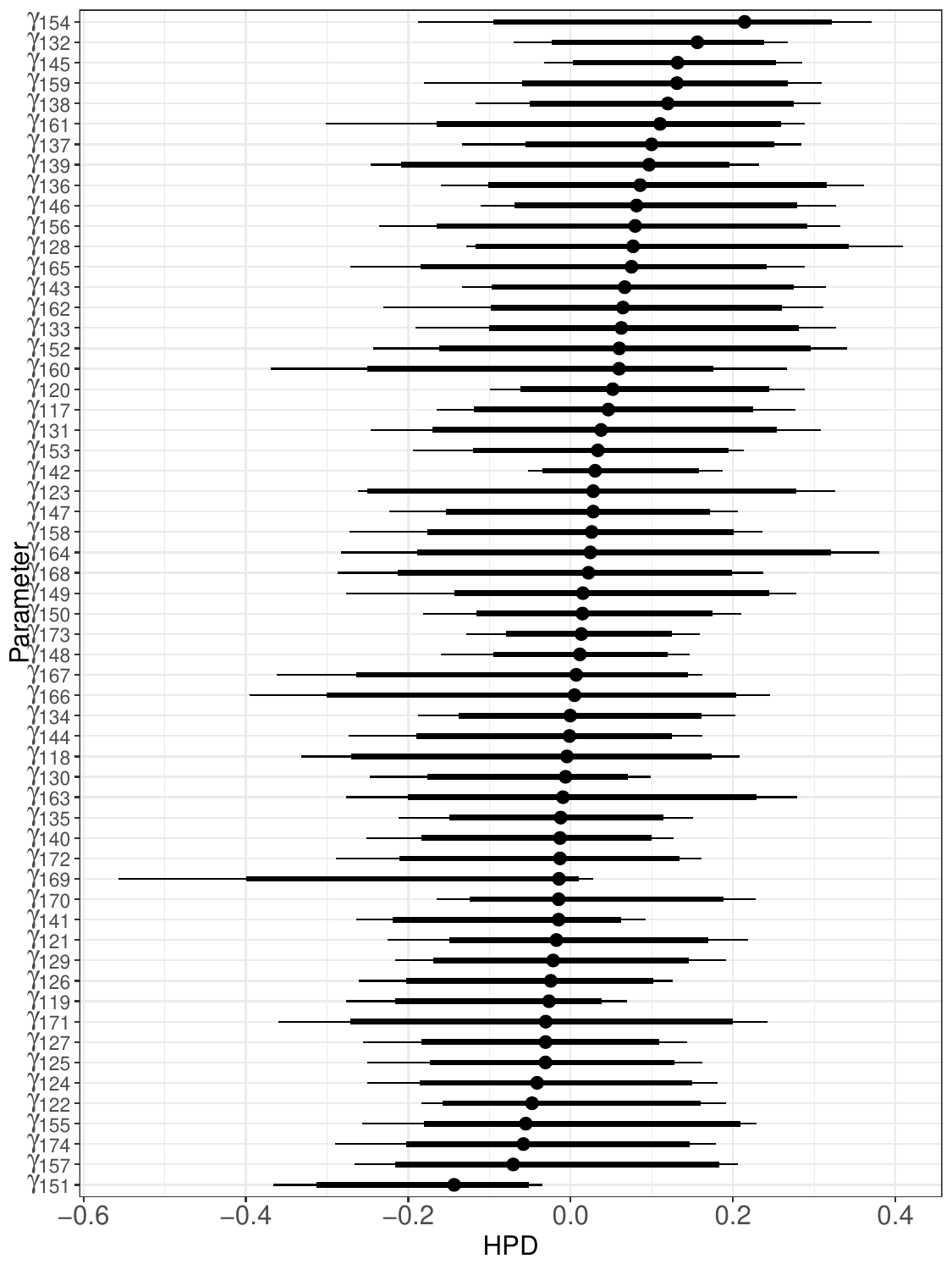}
    \includegraphics[width=0.49\linewidth]{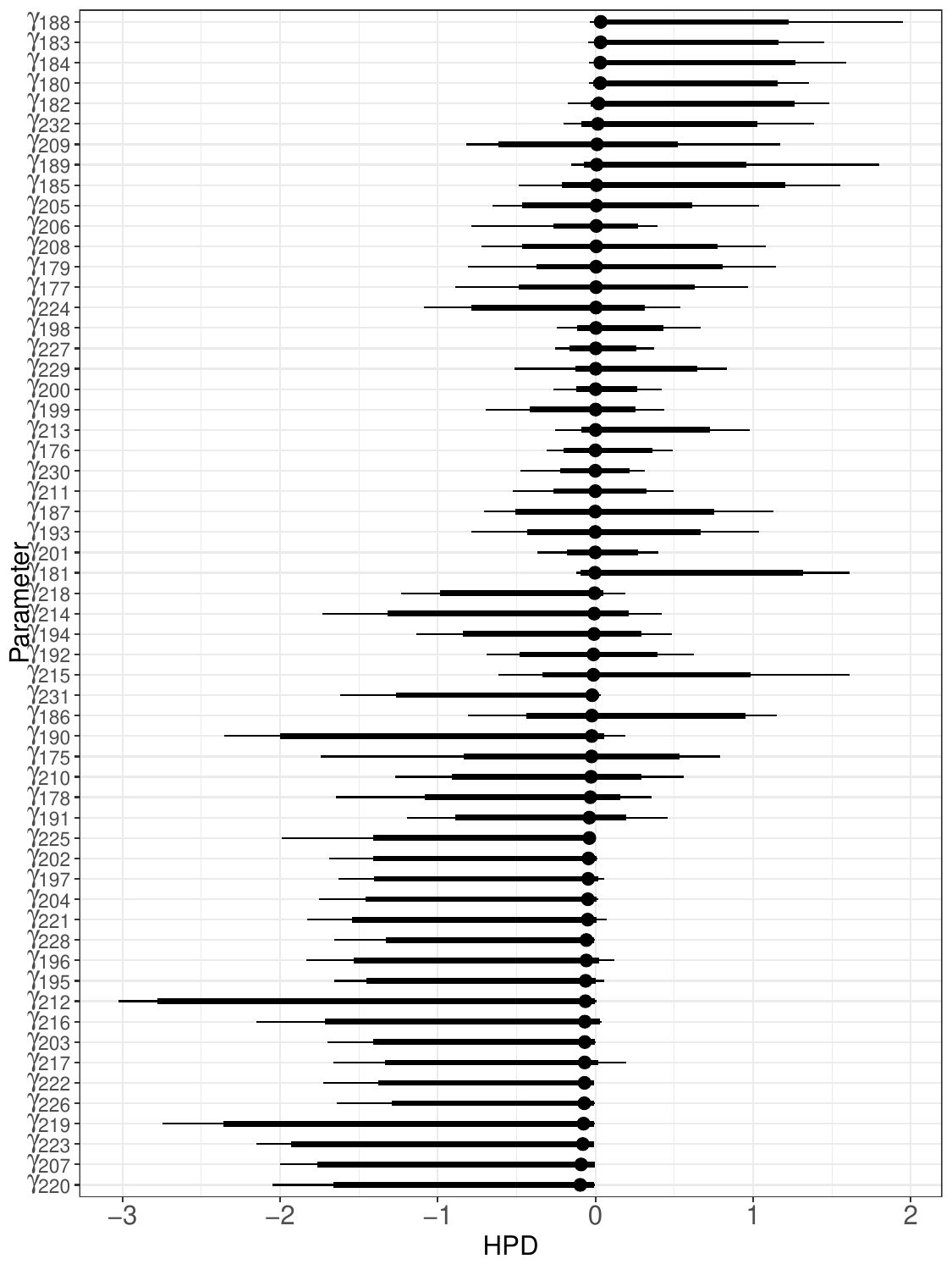}
    \caption{\color{orange} Highest posterior density (HPD) intervals at 90\% (thick lines) and 95\% (thin lines) levels for the elements of the disease-specific spatial effects in $\bgamma$ using the MDAGAR specification in Section~\ref{sec: unstructured_graph} and the precision matrix given in Eq.~\ref{eq:sp_cov}. The four panels correspond to the 4 cancers: lung (top left), esophageal (top right), larynx (bottom left) and colorectal (bottom right). The intervals are arranged in increasing order of value for posterior medians. It is expected that the intervals for these random effects will mostly not be significantly different from $0$, but they reveal substantial variation indicating enough information in the data to discern spatial patterns. It is important to note that the degree of dispersion varies significantly among different types of cancer. For example, lung cancer has the narrowest intervals, while esophageal and laryngeal cancers exhibit similar ranges. In contrast, colorectal cancer shows much wider intervals, with its medians being very close to $0$.}
    \label{fig:post_gamma_mean}
\end{figure}

\begin{figure}[htbp!]
    \centering
    \includegraphics[width=0.49\linewidth]{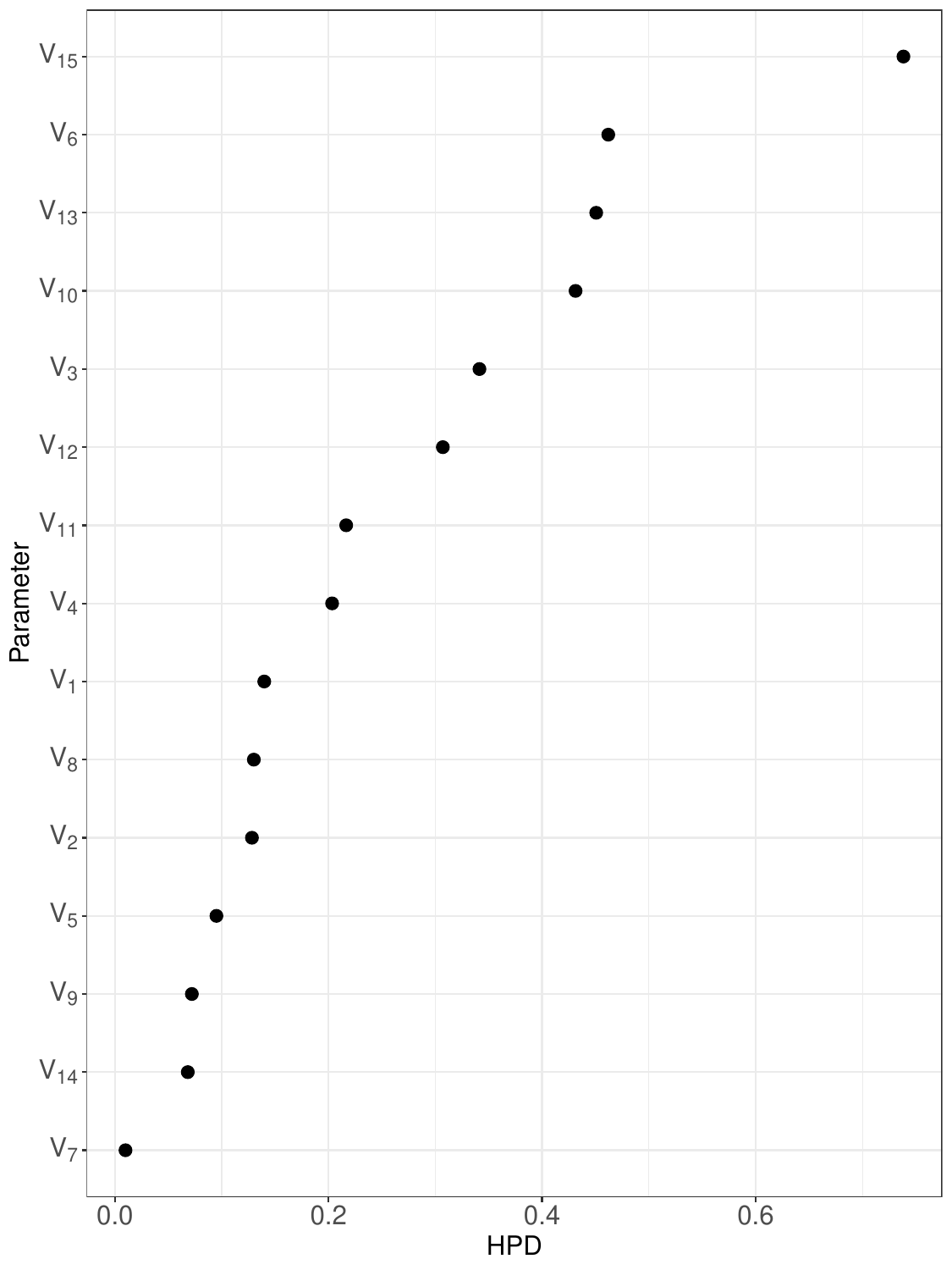}
    \includegraphics[width=0.49\linewidth]{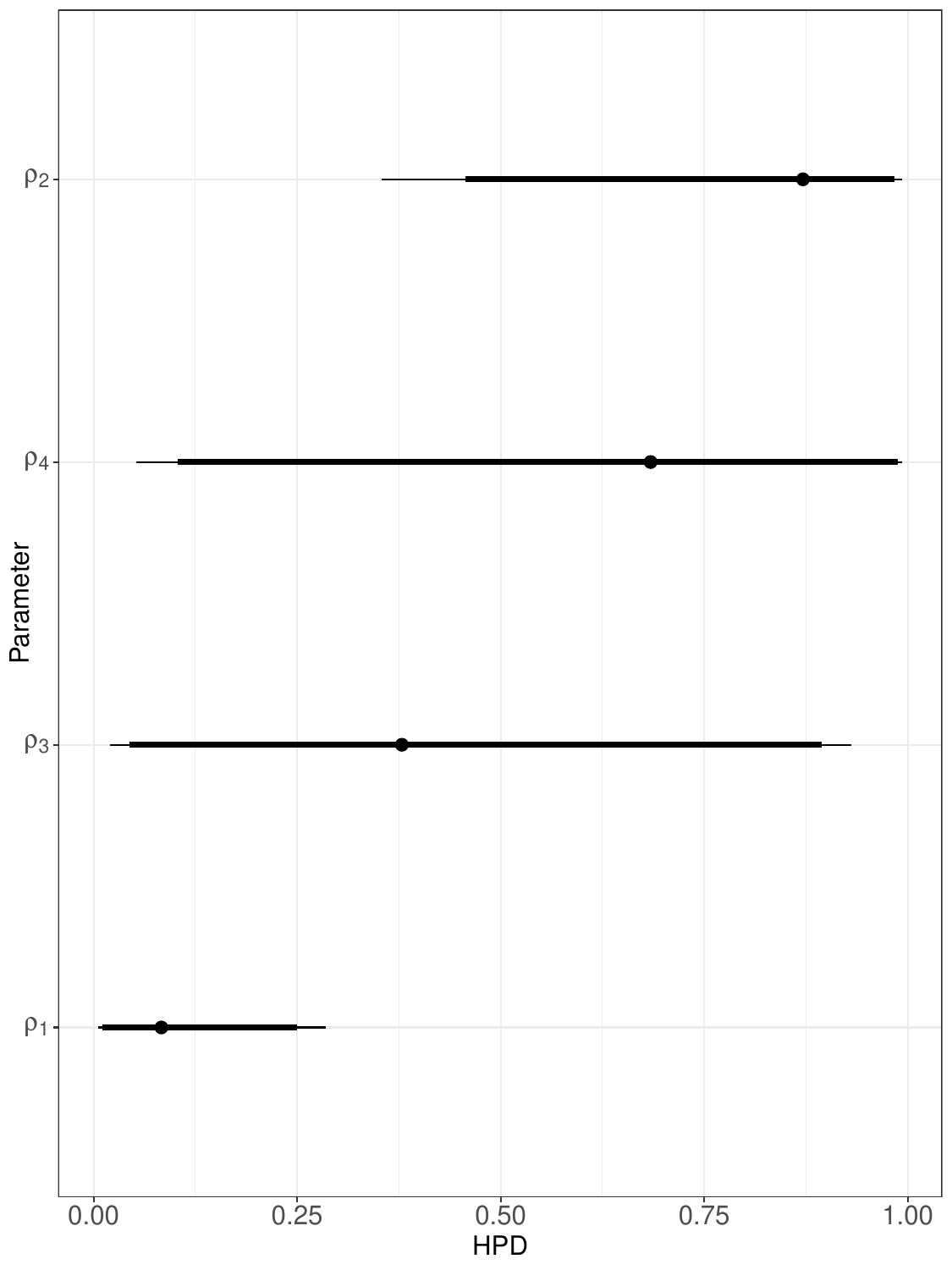}
    \caption{\color{orange} Highest posterior density (HPD) intervals at 90\% (thick lines) and 95\% (thin lines) levels for $\bV$ (left panel), which consists of the 15 stick-breaking weights used in Eq.~\eqref{eq:DP}, and for $\brho$ (right panel), which consists of the 4 spatial autocorrelation parameters (one for each disease) in the MDAGAR model. These intervals are arranged in increasing order of posterior medians. The variations among these posterior intervals indicate substantial learning from the data that informs the joint distribution of the spatial effects through the stick-breaking probabilities and the baseline MDAGAR distribution in the DP. We notice that the intervals regarding $\bV$ are very narrow.}
    \label{fig:post_v_rho_mean}
\end{figure}

\begin{figure}[h!]
    \centering
    \includegraphics[width=0.49\linewidth]{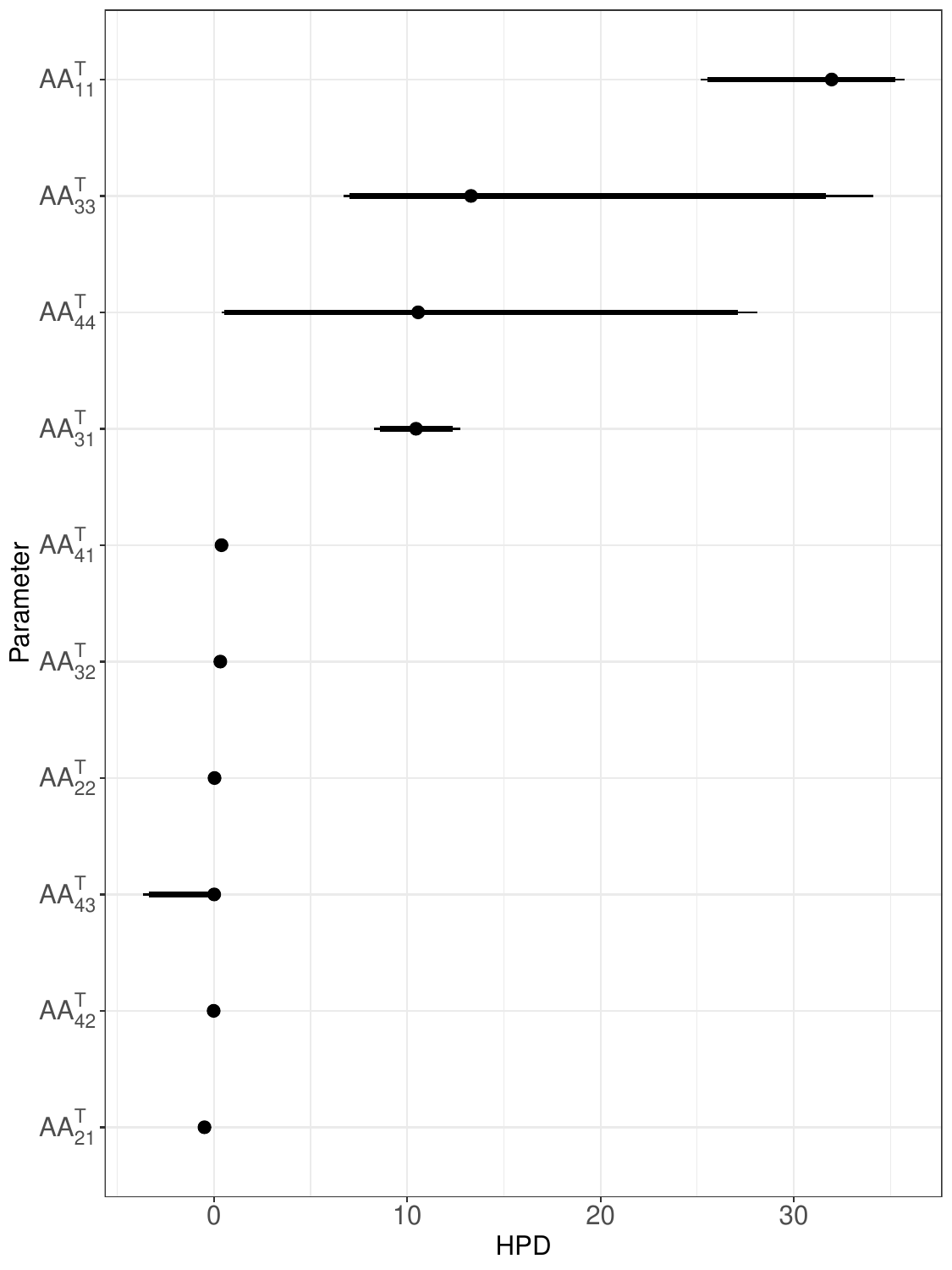}
    \caption{\color{orange} We provide the HPD intervals for the elements of $\bA\bA^{\T}$ representing the non-spatial component of associations among the cancers. Lung, larynx and colorectal cancers reveal the highest variances ($(\bA\bA^{\T}_{11}$, $(\bA\bA^{\T}_{33}$ and $(\bA\bA^{\T})_{44}$) with the last two having the widest intervals. Lung and larynx reveal significantly positive associations with the $95\%$ interval for $(\bA\bA^{\T})_{31}$ situated entirely above $0$. As emphasized earlier, it is important to interpret the non-spatial component of these associations with caution, where much of the positive associations among these aggregated data are absorbed by the spatial associations arising from geographic proximity.}
    \label{fig:post_A_mean}
\end{figure}

\begin{figure}[htbp!]
    \centering
    \includegraphics[width = 0.85\textwidth]{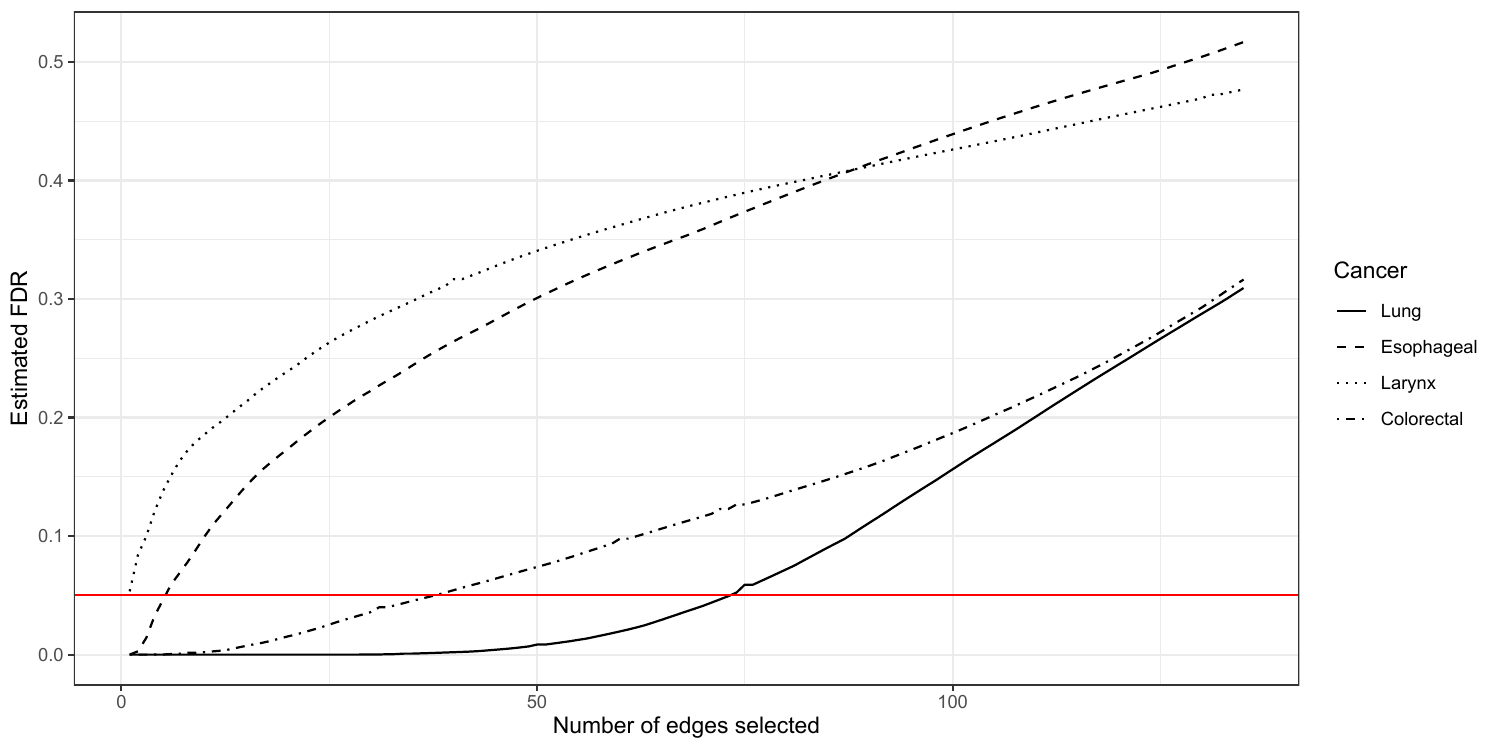}
    \caption{Estimated FDR curves plotted against the number of selected difference boundaries for four cancers}
    \label{fig:FDR_mean}
\end{figure}

\begin{figure}[htbp!]
    \centering
    \includegraphics[width = 0.66 \textwidth]{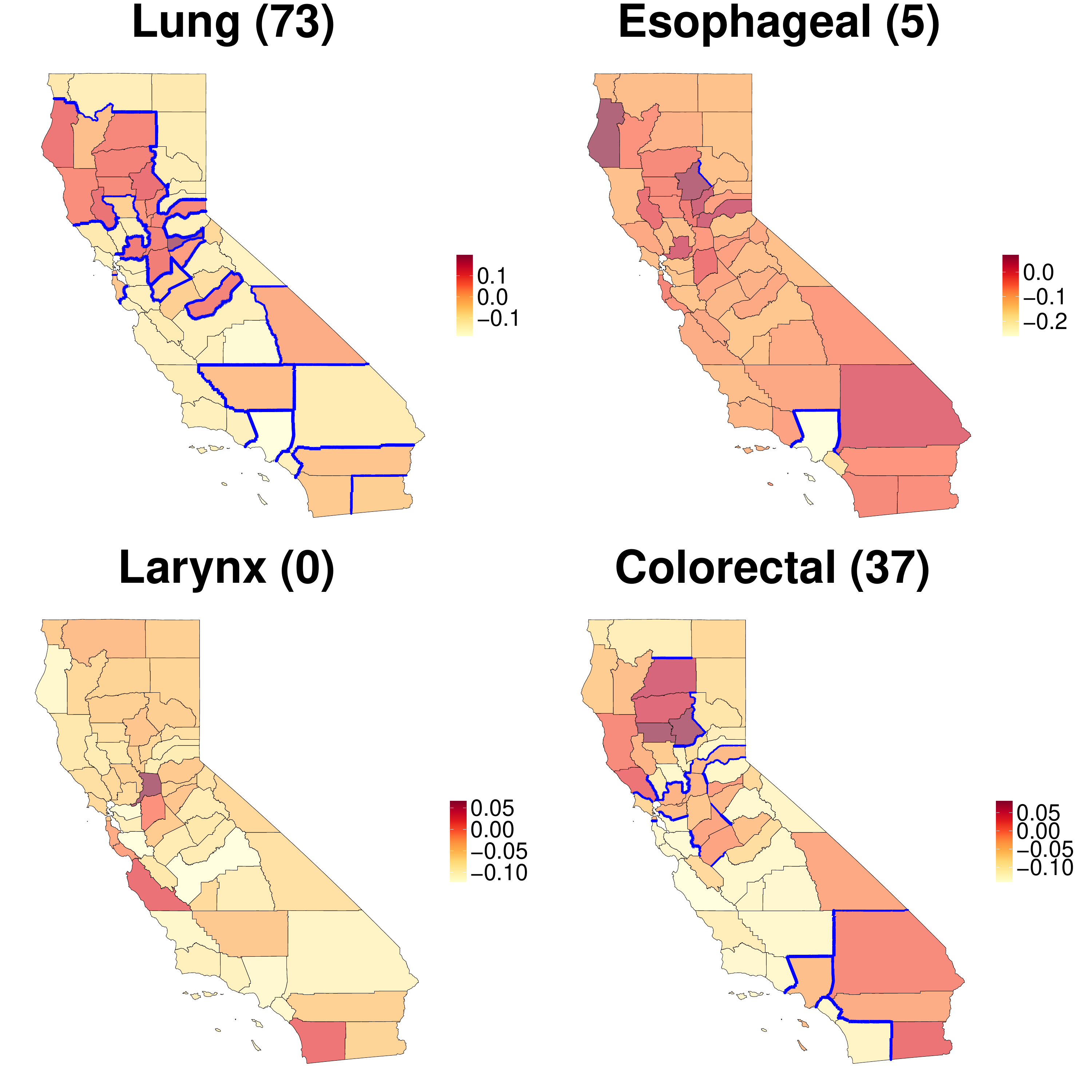}
    \caption{Difference boundaries (highlighted in red) detected by the model in the California map, colored according to the posterior mean of the corresponding $\phi_{id}$, for four cancers individually when $\zeta = 0.05$. The values in brackets are the number of difference boundaries detected.}
    \label{fig:bd_mean}
\end{figure}

\begin{figure}[htbp!]
    \centering
    \includegraphics[width = \textwidth]{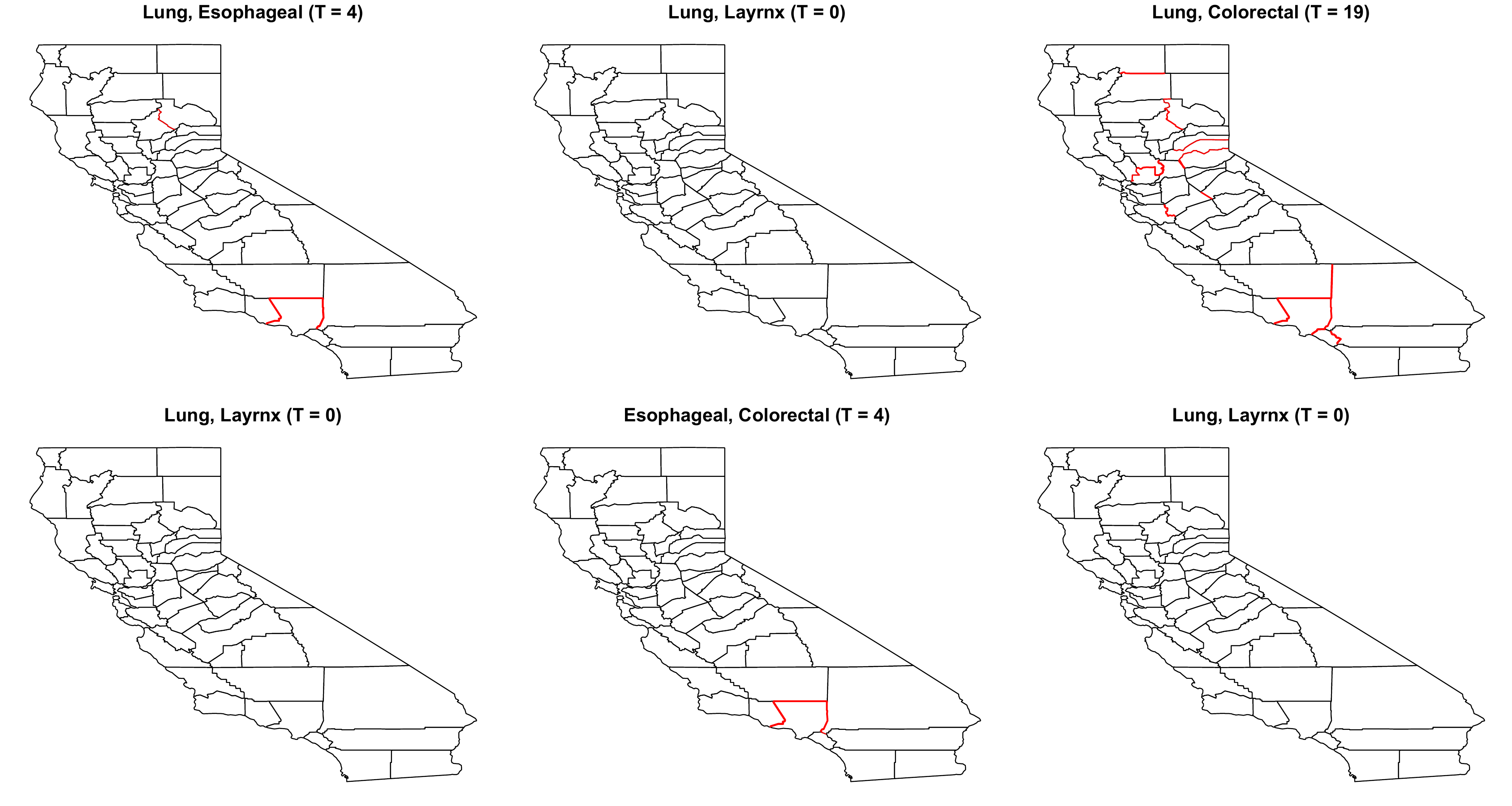}
    \caption{Shared difference boundaries (highlighted in red) detected by the model for each pair of cancers in California map when $\zeta = 0.05$. The values in brackets are the number of difference boundaries detected.}
    \label{fig:sbd_mean}
\end{figure}

\begin{figure}[htbp!]
    \centering
    \includegraphics[width = \textwidth]{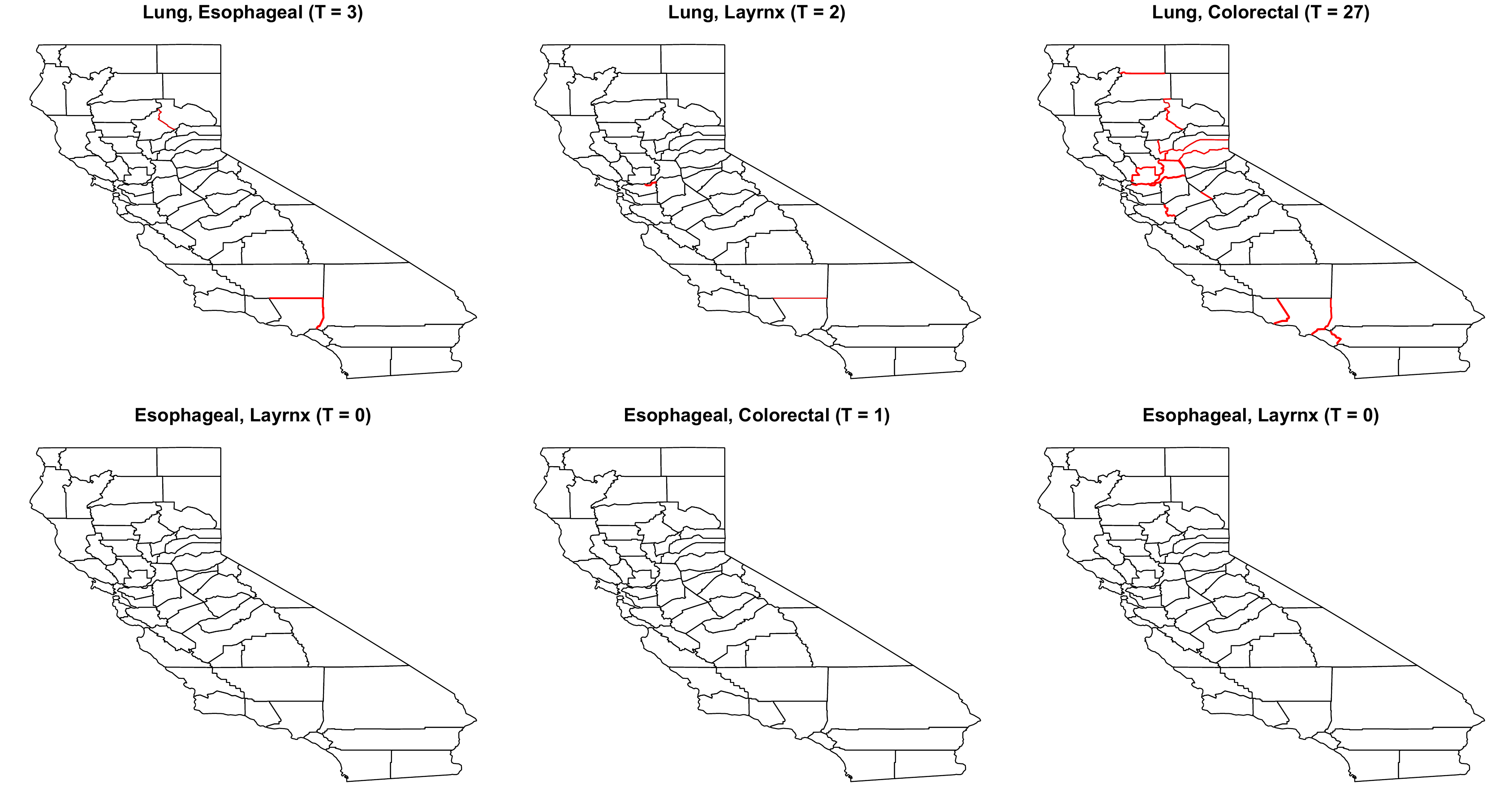}
    \caption{Mutual cross-difference boundaries (highlighted in red) detected by the model for each pair of cancers in California map when $\zeta = 0.05$. The values in brackets are the number of difference boundaries detected.}
    \label{fig:mbd_mean}
\end{figure}

\end{appendix}

\clearpage

\bibliography{biblio}
\bibliographystyle{apalike}